\documentclass[a4paper,UKenglish,cleveref, autoref, thm-restate]{lipics-v2021}

\usepackage{paralist}

\DeclareMathOperator*{\Asterisk}{Asterisk}
\renewcommand{\iff}{\hspace*{2pt}\Leftrightarrow\hspace*{2pt}}

\usepackage[createShortEnv,conf={end,no link to proof,restate}]{proof-at-the-end}

\newenvironment{myFiveTextE}{\begin{textAtEnd}[category=five]}{\end{textAtEnd}}
\newenvironment{mySixTextE}{\begin{textAtEnd}[category=six]}{\end{textAtEnd}}

\renewcommand{\paragraph}[1]{\noindent{\textbf{#1}}}

\newcommand{\nat}{\mathbb{N}}
\newcommand{\arityof}[1]{\#{#1}}

\newcommand{\universeGeneral}{U}

\newcommand{\vars}{\mathbb{V}_1}
\newcommand{\Vars}{\mathbb{V}_2}

\newcommand{\preds}{\mathbb{A}}
\newcommand{\isdef}{\stackrel{\scriptscriptstyle{\mathsf{def}}}{=}}
\newcommand{\interv}[2]{[{#1},{#2}]}
\newcommand{\tuple}[1]{\langle {#1} \rangle}

\newcommand{\set}[1]{\{ {#1} \}}

\newcommand{\pow}[1]{\mathrm{pow}({#1})}

\newcommand{\dom}[1]{\mathrm{dom}({#1})}

\newcommand{\cardof}[1]{|\!| {#1} |\!|}

\newcommand{\np}{$\mathsf{NP}$}
\newcommand{\ptime}{$\mathsf{PTIME}$}

\newcommand{\pspace}{$\mathsf{PSPACE}$}

\newcommand{\signature}{\Sigma}

\newcommand{\arel}{\mathsf{R}}

\newcommand{\acst}{\mathsf{c}}

\newcommand{\nrel}{N}
\newcommand{\ncst}{M}
\newcommand{\nrule}{R}
\newcommand{\npred}{P}

\newcommand{\ports}{\Pi}

\makeatletter
\newcommand*{\da@rightarrow}{\mathchar"0\hexnumber@\symAMSa 4B }
\newcommand*{\da@leftarrow}{\mathchar"0\hexnumber@\symAMSa 4C }
\newcommand*{\xdashrightarrow}[2][]{%
  \mathrel{%
    \mathpalette{\da@xarrow{#1}{#2}{}\da@rightarrow{\,}{}}{}%
  }%
}
\newcommand{\xdashleftarrow}[2][]{%
  \mathrel{%
    \mathpalette{\da@xarrow{#1}{#2}\da@leftarrow{}{}{\,}}{}%
  }%
}
\newcommand*{\da@xarrow}[7]{%
  \sbox0{$\ifx#7\scriptstyle\scriptscriptstyle\else\scriptstyle\fi#5#1#6\m@th$}%
  \sbox2{$\ifx#7\scriptstyle\scriptscriptstyle\else\scriptstyle\fi#5#2#6\m@th$}%
  \sbox4{$#7\dabar@\m@th$}%
  \dimen@=\wd0 %
  \ifdim\wd2 >\dimen@
    \dimen@=\wd2 %
  \fi
  \count@=2 %
  \def\da@bars{\dabar@\dabar@}%
  \@whiledim\count@\wd4<\dimen@\do{%
    \advance\count@\@ne
    \expandafter\def\expandafter\da@bars\expandafter{%
      \da@bars
      \dabar@
    }%
  }%
  \mathrel{#3}%
  \mathrel{%
    \mathop{\da@bars}\limits
    \ifx\\#1\\%
    \else
      _{\copy0}%
    \fi
    \ifx\\#2\\%
    \else
      ^{\copy2}%
    \fi
  }%
  \mathrel{#4}%
}
\makeatother

\newcommand{\store}{\nu}
\newcommand{\struc}{\sigma}
\newcommand{\strucof}[1]{\mathit{Str}({#1})}
\newcommand{\astruc}{S}
\newcommand{\adomof}[1]{\universeGeneral_{#1}}
\newcommand{\interpof}[1]{\struc_{#1}}
\newcommand{\glue}{\mathit{glue}}
\newcommand{\absglue}{\glue^\sharp}
\newcommand{\glueof}[2]{\glue({#1},{#2})}
\newcommand{\absglueof}[2]{\absglue({#1},{#2})}
\newcommand{\fgcst}[1]{\mathit{fgcst}_{#1}}
\newcommand{\absfgcst}[1]{\mathit{fgcst}^\sharp_{#1}}
\newcommand{\qrof}[1]{\mathrm{qr}({#1})}
\newcommand{\twsid}[1]{\asid({#1})}
\newcommand{\twformsid}[2]{\asid({#1},{#2})}
\newcommand{\typesof}[1]{\mathbb{F}^{#1}_{\scriptscriptstyle{\mathsf{MSO}}}}
\newcommand{\atype}{\tau}
\newcommand{\typeof}[2]{\mathit{type}^{#1}({#2})}
\newcommand{\absof}[1]{{#1}^\sharp}
\newcommand{\tweq}[1]{\mathit{Tw}({#1})}
\newcommand{\abstweq}[1]{\mathit{Tw}^\sharp({#1})}
\newcommand{\Dom}[1]{\mathrm{Supp}({#1})}
\newcommand{\Rel}[1]{\mathrm{Rel}({#1})}

\newcommand{\encof}[3]{\encode({#2},{#3})}

\newcommand{\comp}{\uplus}

\newcommand{\bigscomp}{\scalebox{2}{$\scomp$}}

\newcommand{\scomp}{\bullet}

\newcommand{\predname}[1]{\mathsf{#1}}
\newcommand{\apred}{\predname{A}}
\newcommand{\bpred}{\predname{B}}

\newcommand{\emp}{\predname{emp}}

\let\Asterisk\undefined
\newcommand{\Asterisk}{\mathop{\scalebox{1.9}{\raisebox{-0.2ex}{$\ast$}}}\hspace*{1pt}}%

\newcommand{\fv}[1]{\mathrm{fv}({#1})}

\newcommand{\cmso}[1]{$\mathsf{MSO}_{#1}$}
\newcommand{\sol}{\textsf{SO}}

\newcommand{\mso}{\textsf{MSO}}
\newcommand{\eso}{\textsf{ESO}}
\newcommand{\ccmso}{\textsf{CMSO}}

\newcommand{\solmso}{\textsf{(M)SO}}
\newcommand{\Models}{\Vdash}

\newcommand{\seplog}{\textsf{SL}}

\newcommand{\slr}{\textsf{SLR}}
\newcommand{\gl}{\textsf{GL}}

\newcommand{\asid}{\Delta}
\newcommand{\imodels}[3]{\models^{\scalebox{.7}{\raisebox{3pt}{\hspace*{-7pt}$\bullet$}}\hspace*{3pt}{#1}}_{#2}}

\newcommand{\arule}{\mathsf{r}}

\newcommand{\predtrue}{\predname{true}}

\newcommand{\rbr}{{\bf ]\!]}}
\newcommand{\lbr}{{\bf [\![}}
\newcommand{\sem}[1]{{\lbr #1 \rbr}}
\newcommand{\gsem}[2]{\sem{#1}^{\domsymb,{#2}}}
\newcommand{\sidsem}[2]{\sem{#1}_{#2}}
\newcommand{\gsidsem}[3]{{\sidsem{#1}{#2}}^{\!\!\!\!\domsymb,{#3}}}

\newcommand{\proj}[2]{{#1}\!\!\downharpoonleft_{\scriptscriptstyle{#2}}}

\newcounter{index}

\newcommand{\vertices}{\mathcal{V}}
\newcommand{\edges}{\mathcal{E}}
\newcommand{\graph}{\mathcal{G}}
\newcommand{\univgraph}{\universeGeneral_\graph}
\newcommand{\strucgraph}{\struc_\graph}
\newcommand{\signagraph}{\Gamma}
\newcommand{\vertset}{\mathfrak{V}}
\newcommand{\edgerel}{\mathfrak{E}}
\newcommand{\tnodes}{\mathcal{N}}
\newcommand{\tedges}{\mathcal{F}}
\newcommand{\labels}{\Lambda}
\newcommand{\alabel}{\lambda}
\newcommand{\tree}{\mathcal{T}}
\newcommand{\subtree}[2]{{#1}[{#2}]}

\newcommand{\clique}[1]{\mathcal{K}_{#1}}

\newcommand{\twof}[1]{\mathrm{tw}({#1})}
\newcommand{\charform}[3]{\Theta({#2},{#3})}

\newcommand{\domsymb}{\mathfrak{D}}
\newcommand{\heapsymb}{\mathfrak{H}}
\newcommand{\vertex}{v}
\newcommand{\vertexSet}{V}

\newcommand{\encode}{\mathit{encode}}

\hideLIPIcs

\title{Expressiveness Results for an Inductive Logic of Separated Relations}

\author{Radu Iosif}{Univ. Grenoble Alpes, CNRS, Grenoble INP, VERIMAG, 38000, France}{Radu.Iosif@univ-grenoble-alpes.fr}{https://orcid.org/0000-0003-3204-3294}{}
\author{Florian Zuleger}{Institute of Logic and Computation, Technische Universit\"{a}t Wien, Austria}{Florian.Zuleger@tuwien.ac.at}{https://orcid.org/0000-0003-1468-8398}{}

\authorrunning{R. Iosif and F. Zuleger} 

\Copyright{Radu Iosif and Florian Zuleger} 



\relatedversion{} 

\acknowledgements{The first author wishes to acknowledge the support of the French National Research Agency project Non-Aggregative Resource COmpositions (NARCO) under grant number ANR-21-CE48-0011.}

\begin{document}

\nolinenumbers

\maketitle

\begin{abstract}
  In this paper we study a Separation Logic of Relations (\slr) and
  compare its expressiveness to (Monadic) Second Order Logic
  [\solmso].  \slr\ is based on the well-known Symbolic Heap fragment
  of Separation Logic, whose formul\ae\ are composed of points-to
  assertions, inductively defined predicates, with the separating
  conjunction as the only logical connective. \slr\ generalizes the
  Symbolic Heap fragment by supporting general relational atoms,
  instead of only points-to assertions.  In this paper, we restrict
  ourselves to finite relational structures, and hence only consider
  Weak \solmso, where quantification ranges over finite sets.
  Our main results are that \slr\ and \mso\ are incomparable on
  structures of unbounded treewidth, while \slr\ can be embedded in
  \sol\ in general.  Furthermore, \mso\ becomes a strict subset of
  \slr, when the treewidth of the models is bounded by a parameter and
  all vertices attached to some hyperedge belong to the interpretation
  of a fixed unary relation symbol. We also discuss the problem of
  identifying a fragment of \slr\ that is equivalent to \mso\ over
  models of bounded treewidth.
\end{abstract}

\section{Introduction}
\label{sec:intro}

Relational structures are interpretations of relation symbols that
define the standard semantics of first and second order logic
\cite{DBLP:books/daglib/0080654}. They provide a unifying framework
for reasoning about a multitude of graph types e.g., graphs with
multiple edges, labeled graphs, colored graphs, hypergraphs,
etc. Graphs are, in turn, important for many areas of computing, e.g.,
static analysis \cite{10.1145/582153.582161}, databases and knowledge
representation \cite{AbitebouldBunemanSuciu00} and concurrency
\cite{DBLP:conf/birthday/2008montanari}.

A well-established language for specifying graph properties is Monadic
Second Order Logic (\mso), where quantification is over vertices only,
or both vertices and edges, and sets thereof
\cite{courcelle_engelfriet_2012}. Other graph description logics use
formal language theory (e.g., regular expressions, context-free
grammars) to check for paths with certain patterns
\cite{DBLP:conf/lics/FigueiraL15}.

Another way of describing graphs is by an algebra of operations, such
as vertex/hyperedge replacement, i.e., substitution of a
vertex/hyperedge in a graph by another graph. Graph algebras come with
robust notions of \emph{recognizable sets} (i.e., unions of
equivalence classes of a finite index congruence) and \emph{inductive
  sets} (i.e., least solutions of recursive sets of equations,
sometimes also called \emph{equational} or \emph{context-free} sets
\cite{courcelle_engelfriet_2012}). The relation between the
expressivity of \mso-definable, recognizable and inductive sets is
well-understood: all definable sets are recognizable, but there are
recognizable sets that are not definable \cite{CourcelleI}. The
equivalence between definability and recognizability has been
established for those sets in which the \emph{treewidth} (a positive
integer that indicates how close the graph is to a tree) is bounded by
a fixed constant \cite{10.1145/2933575.2934508}. Moreover, it is known
that the set of graphs of treewidth bounded by a constant is inductive
\cite[Theorem 2.83]{courcelle_engelfriet_2012}.

From a system designer's point of view, logical specification is
declarative (i.e., it describes required properties, such as
acyclicity, hamiltonicity, etc.), whereas algebraic specification is
operational (i.e., describes the way graphs are built from pieces),
relying on low-level details (e.g., designated source
vertices). Because of this, system provers (e.g., model checkers or
deductive verifiers) tend to use logic both for requirement
specification and internal representation of configuration sets.
However, algebraic theories (e.g., automata theory) are used to obtain
algorithms for discharging the generated logical verification
conditions, e.g., satisfiability of formul{\ae} or validity of
entailments between formul{\ae}.

Separation Logic (\seplog)
\cite{Ishtiaq00bias,Reynolds02,Cardelli2002Spatial} is a first order
substructural logic with a \emph{separating conjunction} $*$ that
decomposes structures. For reasons related to its applications in the
deductive verification of pointer-manipulating programs, the models of
\seplog\ are finite graphs of fixed outdegree, described by partial
functions, called \emph{heaps}. The separating conjunction is
interpreted in \seplog\ as the union of heaps with disjoint domains.

Since their early days, substructural logics have had (abstract)
algebraic semantics \cite{PymOHearn99}, yet their relation with graph
algebras has received scant attention. However, as we argue in this
paper, the standard interpretation of the separating conjunction has
the flavor of certain graph-algebraic operations, such as the disjoint
union with fusion of designated nodes \cite{CourcelleVII}.

The benefits of \seplog\ over purely boolean graph logics (e.g., \mso)
are two-fold: \begin{compactenum}[I.]
\item The separating conjunction in combination with inductive
  definitions \cite{ACZEL1977739} provide concise descriptions of
  datastructures in the heap memory of a program. For instance, the
  rules (1) $\mathsf{ls}(x,y) \leftarrow x = y$ and (2)
  $\mathsf{ls}(x,y) \leftarrow \exists z ~.~ x \mapsto z *
  \mathsf{ls}(z,y)$ define finite singly-linked list segments, that
  are either (1) empty with equal endpoints, or (2) consist of a
  single cell $x$ \emph{separated from} the rest of the list segment
  $\mathsf{ls}(z,y)$. Most recursive datastructures (singly- and
  doubly-linked lists, trees, etc.) can be defined using only
  existentially quantified spatial conjunctions of atoms, that are
  (dis-)equalities and points-to atoms. This simple subset of
  \seplog\ is referred to as the \emph{Symbolic Heap} fragment. The
  problems of model checking \cite{DBLP:conf/popl/BrotherstonGKR16},
  satisfiability \cite{DBLP:conf/csl/BrotherstonFPG14}, robustness
  properties \cite{JansenKatelaanMathejaNollZuleger17} and entailment
  \cite{DBLP:conf/concur/CookHOPW11,DBLP:conf/lpar/KatelaanZ20,EchenimIosifPeltier21b,EchenimIosifPeltier21,tocl-guarded-separation-logic}
  for this fragment have been studied extensively.
  \item The separating conjunction is a powerful tool for
    reasoning about mutations of heaps. In fact, the built-in
    separating conjunction allows to describe actions \emph{locally},
    i.e., only with respect to the resources (e.g., memory cells,
    network nodes) involved, while framing out the part of the state
    that is irrelevant for that particular action. This principle of
    describing mutations, known as \emph{local reasoning}
    \cite{CalcagnoOHearnYan07}, is at the heart of very powerful
    compositional proof techniques for pointer programs using
    \seplog\ \cite{CalcagnoDistefanoOHearnYang11}.
\end{compactenum}

The extension of \seplog\ from heaps to relational structures, called
\emph{Separation Logic of Relations} (\slr), has been first considered
for relational databases and type systems of object-oriented
languages, known as role logic
\cite{10.1007/978-3-540-27864-1_26}. Our motivation for studying the
expressivity of \slr\ arose from several
works: \begin{inparaenum}[(1)]
\item deductive verification of self-adapting distributed systems,
  where Hoare-style local reasoning is applied to write correctness
  proofs for systems with dynamically changing network architectures
  \cite{AhrensBozgaIosifKatoen21,DBLP:conf/cade/BozgaBI22,DBLP:conf/concur/BozgaBI22},
  and
\item model-checking such systems for absence of deadlocks and
  critical section violations \cite{DBLP:journals/tcs/BozgaIS23}.
\end{inparaenum}
Another possible application of \slr\ is reasoning about programs with
overlaid datastructures
\cite{10.1007/978-3-642-38856-9_10,DBLP:conf/cade/KatelaanJW18}, using
variants of \seplog\ with a per-field composition of heaps, naturally
expressed in \slr.

The \slr\ separating conjunction is understood as splitting the
interpretation of each relation symbol from the signature into
disjoint parts.  For instance, the formula $\arel(x_1, \ldots, x_n)$
describes a structure in which all relations are empty and $\arel$
consists of a single tuple of values $x_1, \ldots, x_n$, whereas
$\arel(x_1, \ldots, x_n) * \arel(y_1, \ldots, y_n)$ says that $\arel$
consists of two distinct tuples, i.e., the values of $x_i$ and $y_i$
differ for at least one index $1 \leq i \leq n$. In contrast to the
Courcelle-style composition of disjoint structures with fusion of
nodes that interpret the common constants (i.e., function symbols of
arity zero) \cite{CourcelleVII}, the \slr-style composition (i.e., the
pointwise disjoint union of the interpretations of each relation
symbol) is more fine-grained. For instance, if structures are used to
encode graphs, \slr\ allows to specify (hyper-)edges that have no
connected vertices, isolated vertices, or both. The same style of
composition is found in other spatial logics for graphs, such as the
\gl\ logic of Cardelli, Gardner and Ghelli \cite{Cardelli2002Spatial}.

In particular, \slr\ is strictly more expressive than standard
\seplog\ interpreted over heaps. For instance, the previous definition
of a list segment can be written in a relational signature having at
least a unary relation $\domsymb$ and a binary relation $\heapsymb$,
as (1) $\mathsf{rls}(x,y) \leftarrow x = y$ and (2) $\mathsf{rls}(x,y)
\leftarrow \exists z ~.~ \domsymb(x) * \heapsymb(x,z) *
\mathsf{rls}(z,y)$.  Note that the $\domsymb(x)$ atoms joined by
separating conjunction ensure that all the nodes are pairwise
different, except for the last one denoted by $y$.  We will later
generalize this use of $\domsymb$ for the definition of a
Courcelle-style composition operator~\cite{CourcelleVII}, where
$\domsymb$ ensures that all but a bounded number of nodes are pairwise
different.  Further, \slr\ can describe graphs of unbounded degree,
e.g., stars with a central vertex and outgoing binary edges $E$ to
frontier vertices e.g., (1) $\mathsf{star}(x) \leftarrow
\mathfrak{N}(x) * \mathsf{node}(x)$ (2) $\mathsf{node}(x) \leftarrow x
= x$ and (3) $\mathsf{node}(x) \leftarrow \exists y ~.~
\mathfrak{E}(x,y) * \mathfrak{N}(y) * \mathsf{node}(x)$. The
definition of stars is not possible with \seplog\ interpreted over
heaps, because of their bounded out-degree.

\paragraph{Our contributions}
We compare the expressiveness of \slr\ with (monadic) second-order
logic \solmso. We are interested in finite relational structures, and
hence only consider \emph{weak} \solmso, where relations are
interpreted as finite sets.

For a logic $\mathcal{L} \in \set{\text{\slr}, \text{\mso},
  \text{\sol}}$ using a finite set $\signature$ of relation and
constant symbols, we denote by $\sem{\mathcal{L}}$ the set of sets of models for all formul\ae\ $\phi \in \mathcal{L}$. For a unary relation
symbol $\domsymb$ not in $\signature$, considered fixed in the rest of
the paper, we say that a graph is \emph{guarded} if all elements from
a tuple in the interpretation of a relation symbol belong to the
interpretation of $\domsymb$. Then $\gsem{\mathcal{L}}{k}$ is the set
of sets of guarded models of treewidth at most $k$ of a formula from
$\mathcal{L}$, where the signature of $\mathcal{L}$ is extended with
$\domsymb$, and $\gsem{\mathcal{L}_1}{k} \subseteq
\sem{\mathcal{L}_2}$ means that $\mathcal{L}_2$ is at least as
expressive as $\mathcal{L}_1$, when only guarded models of treewidth
at most $k$ are considered. Note that $\gsem{\mathcal{L}}{k} \subseteq
\sem{\mathcal{L}}$ is not a trivial statement, in general, because it
asserts the existence of a formula of $\mathcal{L}$ that defines the
set of guarded structures of treewidth at most $k$.

\begin{table}[t!]
  \vspace*{-\baselineskip}
  \begin{center}
    \scalebox{.9}{
  \begin{tabular}{|c|c||c|c||c|c||c|c|}
    \hline
    \multicolumn{2}{|c||}{}
    & \multicolumn{2}{|c||}{\slr} & \multicolumn{2}{|c||}{\mso} & \multicolumn{2}{|c|}{\sol} \\
    \hline
    \hline
    \multicolumn{2}{|c||}{\slr} & \checkmark & ? & $\times$ (\S\ref{sec:mso-slr}) & $\times$ (\S\ref{sec:mso-slr}) & \checkmark (\S\ref{sec:slr-so}) & \checkmark (\S\ref{sec:remaining}) \\
    \hline
    \multicolumn{2}{|c||}{\mso} & $\times$ (\S\ref{sec:mso-slr}) & \checkmark (\S\ref{sec:k-mso-slr}) & \checkmark & \checkmark (\S\ref{sec:remaining}) & \checkmark & \checkmark (\S\ref{sec:remaining}) \\
    \hline
    \multicolumn{2}{|c||}{\sol} & $\times$ (\S\ref{sec:mso-slr}) & ? & $\times$ (\S\ref{sec:remaining}) & $\times$ (\S\ref{sec:remaining}) & \checkmark & \checkmark (\S\ref{sec:remaining}) \\
    \hline
  \end{tabular}
    }
  \end{center}
  \caption{}
  \label{tab:intro:expressiveness}
  \vspace*{-2.5\baselineskip}
\end{table}

Each cell of Table \ref{tab:intro:expressiveness} shows
$\sem{\mathcal{L}_1} \subseteq \sem{\mathcal{L}_2}$ (left) and
$\gsem{\mathcal{L}_1}{k} \subseteq \sem{\mathcal{L}_2}$ (right). Here
$\checkmark$ means that the inclusion holds, $\times$ means it does
not and $?$ denotes an open problem, with reference to the sections
where (non-trivial) proofs are given. The most interesting cases are:
\begin{compactenum}
\item \slr\ and \mso\ are incomparable on unguarded structures of
  unbounded treewidth, i.e., there are formul{\ae} in each of the
  logics that do not have an equivalent in the other,
\item \sol\ is strictly more expressive than \slr, when considering unguarded
  structures of unbounded treewidth,
  and at least as expressive as \slr, when considering guarded
  structures of bounded treewidth,
\item \slr\ is strictly more expressive than \mso, when considering
  guarded structures of bounded treewidth; this shows the expressive
  power of \slr, emphasizing (once more) the model-theoretic
  importance of the treewidth parameter.
\end{compactenum}
Note that, when considering \slr-definable sets of bounded treewidth,
we systematically assume these structures to be guarded. We state as
an open problem and conjecture that every infinite \slr-definable set
of structures of bounded treewidth is necessarily guarded, in a hope
that the guardedness condition can actually be lifted. So far, similar
conditions have been used to, e.g., obtain decidability of entailments
between \seplog\ symbolic heaps
\cite{DBLP:conf/cade/IosifRS13,DBLP:conf/lpar/KatelaanZ20} and of
invariance for assertions written in a fragment of \slr\ for verifying
distributed networks \cite{DBLP:conf/concur/BozgaBI22}. Moreover, the
problem of checking if a given set of inductive definitions defines a
guarded set of structures is decidable for these logics
\cite{JansenKatelaanMathejaNollZuleger17,DBLP:journals/corr/abs-2202-09637}.

A further natural question asks for a fragment of \slr\ with the same
expressive power as \mso, over structures of bounded treewidth. This
is also motivated by the need for a general fragment of \slr\ with a
decidable entailment problem, that is instrumental in designing
automated verification systems.  Unfortunately, such a definition is
challenging because the \mso-definability of the sets defined by
\slr\ is an undecidable problem, whereas treewidth boundedness of such
sets remains an open problem, conjectured to be decidable.

Technical material from Sections \ref{sec:mso-slr}, \ref{sec:slr-so}
and \ref{sec:k-mso-slr} is given in Appendix \ref{app:slr-mso},
\ref{app:slr-so} and \ref{app:k-mso-slr}, respectively.

\paragraph{Related work}
Treewidth is a cornerstone of algorithmic tractability. For instance,
many \np-complete graph problems such as Hamiltonicity and 3-Coloring
become \ptime, when restricted to inputs whose treewidth is bounded by
a constant, see, e.g., \cite[Chapter 11]{DBLP:series/txtcs/FlumG06}.
Moreover, bounding the treewidth by a constant sets the frontier between the decidability and undecidability of monadic second order (\mso) logical theories. A result of Courcelle \cite{CourcelleI} proves that \mso\ is decidable over bounded treewidth structures, by reduction to the emptiness problem of tree
automata. A dual result of Seese \cite{Seese91} proves that each class
of structures with a decidable \mso\ theory necessarily has bounded
treewidth.

Comparing the expressiveness of \seplog\ \cite{Reynolds02} with
classical logics received a fair amount of attention. A first proof of
undecidability of the satisfiability problem for \seplog, with first
order quantification, negation and separating implication, but without
inductive definitions \cite{DBLP:conf/fsttcs/CalcagnoYO01}, is based
on a reduction to Trakhtenbrot's undecidability result for first order
logic on finite models \cite{DBLP:books/daglib/0082516}. This proof
uses heaps of outdegree two to encode arbitrary binary relations as
$\arel(x,y) \isdef \exists z ~.~ z \mapsto (x,y) * \predtrue$. A more
refined proof for heaps of outdegree one was given in
\cite{DBLP:journals/iandc/BrocheninDL12}, where it was shown that
\sol\ has the same expressivity as \seplog, when negation and
separating implication is allowed, which is not the case for our
fragment of \slr.

A related line of work, pioneered by Lozes~\cite{PhD-lozes}, is the
translation of quantifier-free \seplog\ formul{\ae} into boolean
combinations of \emph{core formul{\ae}}, belonging to a small set of
very simple patterns. This enables a straightforward translation of
the quantifier-free fragment of \seplog\ into first order logic, over
unrestricted signatures with both relation and function symbols,
subsequently extended to two quantified variables
\cite{DBLP:journals/tocl/DemriD16} and restricted quantifier prefixes
\cite{DBLP:journals/tocl/EchenimIP20}. Moreover, a translation of
quantifier-free \seplog\ into first order logic, based on the small
model property of the former, has been described in
\cite{CalcagnoGardnerHague05}. These are fragments of \seplog\ without
inductive definitions, but with arbitrary combinations of boolean
(conjunction, negation) and spatial (separating conjunction, magic
wand) connectives. A non-trivial attempt of generalizing the technique
of core formul{\ae} to \emph{reachability} and \emph{list segment}
predicates is given in \cite{DBLP:journals/tocl/DemriLM21}. Moreover,
an in-depth comparison between the expressiveness of various models of
separation, i.e., spatial, as in \seplog, and contextual
(subtree-like), as in Ambient Logic \cite{10.1145/325694.325742}, can
be found in \cite{DBLP:phd/hal/Mansutti20}. The restriction of
\slr\ on trees is, however, out of the scope of this paper.

An early combination of spatial connective for graph decomposition
with (least fixpoint) recursion is Graph Logic (\gl)
\cite{Cardelli2002Spatial}, whose expressiveness is compared to that
of \cmso{2}, i.e., \mso\ interpreted over graphs, with quantification
over both vertices and edges \cite{Dawar2007Expressiveness}. For
reasons related to its applications, \gl\ quantifies over the vertices
and edge labels of a graph, unlike \cmso{2} that quantifies over
vertices, edges and sets thereof. Another fairly subtle difference is
that \gl\ can describe graphs with multiple edges that involve the
same vertices and same label, whereas the models of \cmso{2} are
simple graphs. Without recursion, \gl\ can be translated into \cmso{2}
and it has been shown that \cmso{2} is strictly more expressive than
\gl\ without edge label quantification
\cite{10.1007/978-3-642-00596-1_6}. Little is known for \gl\ with
recursion, besides that it can express \pspace-complete model checking
problems \cite{Dawar2007Expressiveness}, whereas model checking is
\pspace-complete for \mso\ \cite{DBLP:conf/stoc/Vardi82}.

The separating conjunction used in \slr\ has been first introduced in
role logic \cite{10.1007/978-3-540-27864-1_26}, a logic designed to
reason about properties of record fields in object-oriented
programs. This logic uses separating conjunction in combination with
boolean connectives and first order quantifier (ranging over vertices)
and has no recursive constructs (least fixpoints or inductive
definitions). A bothways translation between role logic and \sol\ has
been described in \cite{DBLP:journals/corr/cs-LO-0410073}. These
translations rely on boolean connectives and first order quantifiers,
instead of least fixpoint recursion, which is the case in our work.

To complete the picture, a substructural logic with separating
conjunction and implication, based on a layered decomposition of
graphs has been developped in
\cite{DBLP:journals/logcom/CollinsonMP14}. However, the relation
between this logic and \solmso\ remains unexplored, to the best of our
knowledge.

\section{Definitions}
\label{sec:defs}

For a set $A$, we denote by $\pow{A}$ its powerset, $A^1 \isdef A$,
$A^{i+1} \isdef A^i \times A$, for all $i \geq 1$, where $\times$ is
the Cartesian product, and $A^+ \isdef \bigcup_{i\geq1} A^i$. The
cardinality of a finite set $A$ is denoted by $\cardof{A}$. Given
integers $i$ and $j$, we write $\interv{i}{j}$ for the set
$\set{i,i+1,\ldots,j}$, empty if $i>j$. For a partial function $f : A
\rightarrow B$, we denote by $\dom{f}$ its domain and by $\proj{f}{S}$
its restriction to $S \subseteq \dom{f}$. $f$ is \emph{locally
  co-finite} iff the set $\set{a \in A \mid f(a)=b}$ is finite, for
all $b \in B$. $f$ is \emph{effectively computable} iff there exists a
Turing machine $\mathcal{M}$, such that, for any $a \in \dom{f}$,
$\mathcal{M}$ outputs $f(a)$ in finitely many steps and diverges for
$a\not\in\dom{f}$.

\paragraph{Signatures and Structures}
Let $\signature = \set{\arel_1, \ldots, \arel_\nrel, \acst_1, \ldots,
  \acst_\ncst}$ be a finite \emph{signature}, where $\arel_i$ are
relation symbols of arity $\arityof{\arel_i} \geq 1$ and $\acst_j$ are
constant symbols, i.e., function symbols of arity zero. Additionally, we
assume the existence of a unary relation symbol $\domsymb$, not in
$\signature$. Unless stated otherwise, we consider $\signature$ and
$\domsymb$ to be fixed in the following.

A \emph{structure} is a pair $(\universeGeneral, \struc)$, where
$\universeGeneral$ is an \emph{infinite} set, called \emph{universe},
and $\struc : \signature \rightarrow \universeGeneral \cup
\pow{\universeGeneral^+}$ is an \emph{interpretation} that maps each
relation symbol $\arel$ to a relation $\struc(\arel) \subseteq
\universeGeneral^{\arityof{\arel}}$ and each constant $\acst$ to an
element $\struc(\acst) \in \universeGeneral$. Two structures are
\emph{isomorphic} iff they differ only by a renaming of their elements
(a formal definition is given in, e.g., \cite[\S
  A3]{DBLP:books/daglib/0082516}). We write $\Rel{\struc}$ for the set
of elements that belong to $\struc(\arel)$, for some relation symbol
$\arel \in \signature$ and $\Dom{\struc} \isdef \Rel{\struc} \cup
\set{\struc(\acst_1), \ldots, \struc(\acst_\ncst)}$ for the
\emph{support} of the structure, that includes the interpretation of
constants. We denote by $\strucof{\signature}$
(resp. $\strucof{\signature,\domsymb}$) the set of structures over the
signature $\signature$ (resp. $\signature\cup\set{\domsymb}$).

A structure is \emph{guarded} iff all nodes that occur in some tuple
from the denotation of a relation symbol sit also inside the
denotation of the unary relation $\domsymb$:

\vspace*{-.2\baselineskip}
\begin{definition}\label{def:guarded}
  A structure $(\universeGeneral,\struc) \in
  \strucof{\signature,\domsymb}$ is \emph{guarded} iff $\Rel{\struc} =
  \struc(\domsymb)$.
\end{definition}
Two interpretations $\struc_1$ and $\struc_2$ are \emph{compatible}
iff $\struc_1(\acst)=\struc_2(\acst)$, for all constant symbols
$\acst\in\signature$. Two structures $(\universeGeneral_1,\struc_1)$
and $(\universeGeneral_2,\struc_2)$ are \emph{locally disjoint} iff
$\struc_1(\arel)\cap\struc_2(\arel)=\emptyset$, for all relation
symbols $\arel\in\signature$.
The (spatial) \emph{composition} of structures is defined below:
\begin{definition}\label{def:composition}
  The \emph{composition} of two compatible and locally disjoint
  structures $(\universeGeneral_1,\struc_1)$ and
  $(\universeGeneral_2,\struc_2)$ is $(\universeGeneral_1,\struc_1)
  \scomp (\universeGeneral_2,\struc_2) \isdef
  (\universeGeneral_1\cup\universeGeneral_2,\struc_1 \comp \struc_2)$,
  where $(\struc_1 \comp \struc_2)(\arel_i) \isdef \struc_1(\arel_i) \cup
  \struc_2(\arel_i)$ and $(\struc_1 \comp \struc_2)(\acst_j) \isdef
  \struc_1(\acst_j) = \struc_2(\acst_j)$, for all $i \in
  \interv{1}{\nrel}$ and $j \in \interv{1}{\ncst}$. The composition is
  undefined for structures that are not compatible or not locally
  disjoint.
\end{definition}

\paragraph{Graphs and Treewidth}
A graph is a pair $\graph = (\vertices,\edges)$, such that $\vertices$
is a set of \emph{vertices} and $\edges \subseteq \vertices \times
\vertices$ is a set of \emph{edges}. All graphs considered in this
paper are finite and directed, i.e., $\edges$ is not necessarily a symmetric
relation. Graphs are naturally encoded as structures:
\begin{definition}\label{def:graph-struc} A graph
 $\graph = (\vertices,\edges)$ is encoded by the structure
  $(\univgraph,\strucgraph)$ over the signature $\signagraph \isdef
  \set{\vertset, \edgerel}$, where $\arityof{\vertset}=1$ and
  $\arityof{\edgerel}=2$, such that $\univgraph=\vertices$,
  $\strucgraph(\vertset) = \vertices$ and $\strucgraph(\edgerel) =
  \edges$.
\end{definition}
A \emph{path} in $\graph$ is a sequence of pairwise distinct vertices
$v_1, \dots, v_n$, such that $(v_i,v_{i+1}) \in \edges$, for all $i
\in \interv{1}{n-1}$. We say that $v_1, \dots, v_n$ is an
\emph{undirected path} if $\set{(v_i,v_{i+1}),(v_{i+1},v_i)} \cap
\edges \neq \emptyset$ instead, for all $i \in \interv{1}{n-1}$. A set
of vertices $V \subseteq \vertices$ is \emph{connected in $\graph$}
iff there is an undirected path in $\graph$ between any two vertices
in $V$. A graph $\graph$ is \emph{connected} iff $\vertices$ is
connected in $\graph$. A \emph{clique} is a graph such that each two
distinct nodes are the endpoints of an edge, the direction of which is
not important. We denote by $\clique{n}$ the set of cliques with $n$
vertices.

Given a set $\labels$ of labels, a \emph{$\labels$-labeled tree} is a
tuple $\tree = (\tnodes,\tedges,r,\alabel)$, where $(\tnodes,\tedges)$
is a graph, $r \in \tnodes$ is a designated vertex called the
\emph{root}, such that there exists a unique path in
$(\tnodes,\tedges)$ from $r$ to any other vertex $v \in
\tnodes\setminus\set{r}$ and $r$ has no incoming edges $(p,r) \in
\tedges$. The mapping $\alabel : \tnodes \rightarrow \labels$
associates each vertex of the tree a \emph{label} from $\labels$.

\begin{definition}\label{def:treewidth}
  A \emph{tree decomposition} of a structure
  $(\universeGeneral,\struc)$ over the signature $\signature$ is a
  $\pow{\universeGeneral}$-labeled tree
  $\tree=(\tnodes,\tedges,r,\alabel)$, such that the following
  hold: \begin{compactenum}
  \item\label{it1:treewidth} for each relation symbol $\arel \in
    \signature$ and each tuple $\tuple{u_1, \ldots,
      u_{\arityof{\arel}}} \in \struc(\arel)$ there exists $n \in
    \tnodes$, such that $\set{u_1, \ldots, u_{\arityof{\arel}}}
    \subseteq \alabel(n)$, and
  \item\label{it2:treewidth} for each $u \in \Dom{\struc}$, the set
    $\set{n \in \tnodes \mid u \in \alabel(n)}$ is nonempty and
    connected in $(\tnodes,\tedges)$.
  \end{compactenum}
  The \emph{width} of the tree decomposition is $\twof{\tree} \isdef
  \max_{n \in \tnodes} \cardof{\alabel(n)}-1$. The \emph{treewidth} of
  the structure $(\universeGeneral,\struc)$ is $\twof{\universeGeneral,\struc} \isdef \min
  \set{\twof{\tree} \mid \tree \text{ is a tree decomposition of }
    \struc}$.
\end{definition}
A set of structures is \emph{treewidth-bounded} iff the set of
corresponding treewidths is finite and \emph{treewidth-unbounded}
otherwise. A set is \emph{strictly treewidth-unbounded} iff it is
treewidth-unbounded and any of its infinite subsets is
treewidth-unbounded. The following result can be found in
\cite[Theorem 12.3.9]{DBLP:books/daglib/0030488} and is restated here
for self-containment:

\begin{proposition}\label{prop:strictly-unbounded}
  The set of cliques $\set{\clique{n} \mid n \in \nat}$ is strictly treewidth-unbounded.
\end{proposition}

\section{Logics}
\label{sec:logics}

We introduce two logics over a relational signature $\signature =
\set{\arel_1, \ldots, \arel_\nrel, \acst_1, \ldots,
  \acst_\ncst}$. First, the \emph{Separation Logic of Relations}
(\slr) uses a set of \emph{first order variables} $\vars =
\set{x,\ldots}$ and a set of \emph{predicates} $\preds = \set{\apred,
  \ldots}$ (also called \emph{recursion variables} in the literature,
e.g., \cite{Cardelli2002Spatial}) of arities $\arityof{\apred}\geq0$.
We use the symbols $\xi, \chi \in \vars \cup \set{\acst_1, \ldots,
  \acst_\ncst}$ to denote \emph{terms}, i.e., either first order
variables or constants. The formul{\ae} of \slr\ are defined by the
following syntax:
\begin{flalign*}
\phi := \emp \mid \xi=\chi \mid \xi\neq\chi \mid \arel(\xi_1, \ldots,
\xi_{\arityof{\arel}}) \mid \apred(\xi_1, \ldots,
\xi_{\arityof{\apred}}) \mid \phi * \phi \mid \exists x ~.~ \phi
\end{flalign*}
The formul{\ae} $\xi=\chi$ and $\xi\neq\chi$ are called
\emph{equalities} and \emph{disequalities}, $\arel(\xi_1,
\ldots,\xi_{\arityof{\arel}})$ and
$\apred(\xi_1,\ldots,\xi_{\arityof{\apred}})$ are called
\emph{relation} and \emph{predicate atoms}, respectively. A formula
with no occurrences of predicate atoms (resp. existential quantifiers)
is called \emph{predicate-free} (resp. \emph{quantifier-free}). A
variable is \emph{free} if it does not occur within the scope of an
existential quantifier and \emph{bound} otherwise. We denote by
$\fv{\phi}$ be the set of free variables of $\phi$. A \emph{sentence}
is a formula with no free variables. A \emph{substitution}
$\phi[x_1/\xi_1 \ldots x_n/\xi_n]$ replaces simultaneously every
occurrence of the free variable $x_i$ by the term $\xi_i$ in $\phi$,
for all $i \in \interv{1}{n}$. As a convention, the bound variables in
$\phi$ are renamed to avoid clashes with $\xi_1, \ldots, \xi_n$.

The predicates from $\preds$ are interpreted as sets of structures,
defined inductively:

\begin{definition}\label{def:sid}
  A \emph{set of inductive definitions (SID)} $\asid$ is a
  \emph{finite} set of \emph{rules} of the form $\apred(x_1, \ldots,
  x_{\arityof{\apred}}) \leftarrow \phi$, where $x_1, \ldots,
  x_{\arityof{\apred}}$ are pairwise distinct variables, called
  \emph{parameters}, such that $\fv{\phi} \subseteq \set{x_1, \ldots,
    x_{\arityof{\apred}}}$.
  A rule $\apred(x_1, \ldots, x_{\arityof{\apred}}) \leftarrow \phi$
  is said to \emph{define} $\apred$.
\end{definition}

The semantics of \slr\ formul{\ae} is given by the satisfaction
relation $(\universeGeneral,\struc) \models^\store_\asid \phi$ between
structures and formul{\ae}. This relation is parameterized by a
\emph{store} $\store : \vars \rightarrow \universeGeneral$ mapping the
free variables of a formula into elements of the universe and an SID
$\asid$. We write $\store[x \leftarrow u]$ for the store that maps $x$
into $u$ and agrees with $\store$ on all variables other than $x$. For
a term $\xi$, we denote by $(\struc,\store)(\xi)$ the value
$\struc(\xi)$ if $\xi$ is a constant, or $\store(\xi)$ if $\xi$ is a
first-order variable. The satisfaction relation is the least relation
that satisfies the following conditions:
\vspace*{-.5\baselineskip}
\[\begin{array}{lcl}
  (\universeGeneral,\struc) \models^\store_\asid \emp & \iff &
  \struc(\arel) = \emptyset \text{, for all } \arel\in\signature
  \\
  (\universeGeneral,\struc) \models^\store_\asid \xi \sim \chi & \iff &
  (\universeGeneral,\struc) \models^\store_\asid \emp \text{ and }
  (\struc,\store)(\xi) \sim (\struc,\store)(\chi) \text{, where} \sim \ \in\!\set{=,\neq}
  \\
  (\universeGeneral,\struc) \models^\store_\asid \arel(\xi_1, \ldots, \xi_{\arityof{\arel}}) & \iff &
  \struc(\arel) = \set{\tuple{(\struc,\store)(\xi_1), \ldots, (\struc,\store)(\xi_{\arityof{\arel}})}}\\
  && \quad \quad \quad\text{ and } \struc(\arel') = \emptyset, \text{for } \arel' \in \signature \setminus\set{\arel}
  \\
  (\universeGeneral,\struc) \models^\store_\asid \apred(\xi_1, \ldots, \xi_{\arityof{\apred}}) & \iff &
  (\universeGeneral,\struc) \models^\store_\asid \phi[x_1/\xi_1, \ldots, x_{\arityof{\apred}}/\xi_{\arityof{\apred}}],
  \\&& \quad \quad \quad
  \text{ for some } \apred(x_1, \ldots, x_{\arityof{\apred}}) \leftarrow \phi \in \asid
  \\
  (\universeGeneral,\struc) \models^\store_\asid \phi_1 * \phi_2 & \iff &
  \text{there exist structures } (\universeGeneral_1,\struc_1) \text{ and } (\universeGeneral_2,\struc_2)
  \text{, such that } \\
  && (\universeGeneral,\struc) = (\universeGeneral_1,\struc_1) \scomp (\universeGeneral_2,\struc_2)
  \text{ and } (\universeGeneral,\struc_i) \models^\store_\asid \phi_i \text{, for } i = 1,2
  \\
  (\universeGeneral,\struc) \models^\store_\asid \exists x ~.~ \phi & \iff &
  (\universeGeneral,\struc) \models^{\store[x\leftarrow u]}_\asid \phi \text{, for some } u \in \universeGeneral
\end{array}\]
Note that every structure $(\universeGeneral,\struc)$, such that
$(\universeGeneral,\struc) \models^\store_\asid \phi$, interprets each
relation symbol as a finite set of tuples, defined by a finite least
fixpoint iteration over the rules in $\asid$. In particular, the
assumption that each universe is infinite excludes the cases in which
a \slr\ formula becomes unsatisfiable because the universe does not
have enough elements to be assigned to the existentially quantified
variables during the unfolding of the rules.

If $\phi$ is a sentence, the satisfaction relation does not depend on
the store, in which case we write $(\universeGeneral,\struc)
\models_\asid \phi$ and say that $(\universeGeneral,\struc)$ is a
$\asid$-model of $\phi$. We denote by $\sidsem{\phi}{\asid}$ the set
of $\asid$-models of $\phi$.  We call $\sidsem{\phi}{\asid}$ an
\slr-\emph{definable} set.  By $\gsidsem{\phi}{\asid}{k}$ we denote
the set of guarded structures (Def. \ref{def:guarded}) of treewidth at
most $k$ from $\sidsem{\phi}{\asid}$.  We write $\sem{\text{\slr}}
\isdef \{\sidsem{\phi}{\asid} \mid \phi \text{ is a \slr\ formula},
\asid \text{ is a SID}\}$ and $\gsem{\text{\slr}}{k} \isdef
\{\gsidsem{\phi}{\asid}{k} \mid \phi \text{ is a \slr\ formula}, \asid
\text{ is a SID}\}$. Below we show that \slr-definable sets are unions
of isomorphic equivalence classes:

\begin{propositionE}\label{prop:sl-iso}
  Given isomorphic structures $(\universeGeneral,\struc)$ and
  $(\universeGeneral',\struc')$, for any sentence $\phi$ of \slr\ and
  any SID $\asid$, we have $(\universeGeneral,\struc) \models_\asid
  \phi \iff (\universeGeneral',\struc') \models_\asid \phi$.
\end{propositionE}
\begin{proofE}
   By induction on the definition of the satisfaction relation
   $\models_\asid$, we show that $(\universeGeneral,\struc)
   \models^{\store}_\asid \psi \iff (\universeGeneral',\struc')
   \models^{h \circ \store}_\asid \psi$, for any store $\store$ and
   any bijection $h : \universeGeneral \rightarrow \universeGeneral'$,
   such that, for any relation symbol $\arel \in \signature$, we have
   $\tuple{u_1, \ldots, u_n} \in \struc(\arel) \iff \tuple{h(u_1),
     \ldots, h(u_n)} \in \struc'(\arel)$ and, for any constant $\acst
   \in \signature$, we have $h(\struc(\acst))=\struc'(\acst)$. We
   consider the following cases: \begin{compactitem}
   \item $\phi = \emp$: for all $\arel\in\signature$, we have
     $\struc(\arel)=\emptyset \iff \struc'(\arel)=\emptyset$, by the
     existence of $h$.
   \item $\phi = \xi \sim \chi$, for $\sim \in \set{=,\neq}$: we
     prove only the case where $\xi\in\vars$ and
     $\chi\in\set{\acst_1,\ldots,\acst_\ncst}$, the other cases being
     similar. By the above point, for all $\sim \in \set{=,\neq}$,
     by the definition of $h$, we have: \begin{compactitem}
     \item $\struc(\arel)=\emptyset \iff \struc'(\arel)=\emptyset$, for
       all $\arel\in\signature$, and
     \item $\store(\xi) \sim \struc(\chi) \iff h(\store(\xi))
       \sim h(\struc(\chi)) \iff (h \circ \store)(\xi) \sim
       \struc'(\chi)$.
     \end{compactitem}
   \item $\phi = \arel(\xi_1, \ldots, \xi_{\arityof{\arel}})$:
     $\tuple{(\struc,\store)(\xi_1), \ldots,
     (\struc,\store)(\xi_{\arityof{\arel}})} \in \struc(\arel) \iff
     \langle h((\struc,\store)(\xi_1)), \ldots, \\
       h((\struc,\store)(\xi_{\arityof{\arel}})) \rangle \in
     \struc'(\arel)$. If $\xi_1, \ldots, \xi_{\arityof{\arel}} \in
     \vars$, the latter condition is $\langle(h\circ\store)(\xi_1),
       \ldots, \\ (h\circ\store)(\xi_{\arityof{\arel}}) \rangle \in
     \struc'(\arel)$. Else, if $\xi_1, \ldots, \xi_{\arityof{\arel}}
     \in \set{\acst_1, \ldots, \acst_\ncst}$, the condition is
     $\langle h(\struc(\xi_1)), \\ \ldots,
       h(\struc(\xi_{\arityof{\arel}})) \rangle \in \struc'(\arel)$. The general case
     $\xi_1, \ldots, \xi_{\arityof{\arel}} \in \vars \cup \set{\acst_1,
       \ldots, \acst_\ncst}$ is a combination of the above cases.
   \item $\phi = \apred(\xi_1, \ldots, \xi_{\arityof{\apred}})$: this
     case follows by the induction hypothesis.
   \item $\phi = \phi_1 * \phi_2$: $\struc \models^{\store}_\asid
     \phi_1 * \phi_2$ if and only if there exists disjoint and
     compatible structures $(\universeGeneral_1,\struc_1) \scomp (\universeGeneral_2,\struc_2) = \struc$, such
     that $(\universeGeneral_i,\struc_i) \models^{\store}_\asid \phi_i$, for all $i =
     1,2$. We consider the structures $(\universeGeneral'_1,\struc'_1)$ and $(\universeGeneral'_2,\struc'_2)$ as
     follows, for $i = 1,2$: \begin{compactitem}
     \item $\universeGeneral'_i = h(\universeGeneral_i)$,
     \item $\struc'_i(\arel) = \set{\tuple{h(u_1), \ldots,
         h(u_{\arityof{\arel}})} \mid \tuple{u_1, \ldots,
         u_{\arityof{\arel}}} \in \struc_i(\arel)}$, for all relation
       symbols $\arel \in \signature$,
     \item $\struc'_i(\acst) = h(\struc_i(\acst))$, for all constant
       symbols $\acst \in \signature$.
     \end{compactitem}
     Then $(\universeGeneral'_1,\struc'_1)$ and
     $(\universeGeneral'_2,\struc'_2)$ are locally disjoint and
     compatible and $(\universeGeneral',\struc') =
     (\universeGeneral'_1,\struc'_1) \scomp
     (\universeGeneral'_2,\struc'_2)$. Moreover, by the inductive
     hypothesis, we have $(\universeGeneral_i,\struc'_i)
     \models^{\store}_\asid \phi_i$, for all $i = 1,2$, leading to
     $(\universeGeneral,\struc') \models^{\store}_\asid \phi_1 *
     \phi_2$.
   \item $\phi = \exists x ~.~ \psi$: by the inductive hypothesis, we
     obtain $(\universeGeneral,\struc) \models^{\store[x\leftarrow
         u]}_\asid \psi \iff (\universeGeneral',\struc')
     \models^{h\circ(\store[x\leftarrow u])}_\asid \psi \iff
     (\universeGeneral',\struc') \models^{(h\circ\store)[x \leftarrow
         h(u)]}_\asid \psi \iff (\universeGeneral',\struc')
     \models^{h\circ\store}_\asid \exists x ~.~ \psi$.
   \end{compactitem}
 \end{proofE}

The other logic is the \emph{Weak Second Order Logic} (\sol) defined
using a set of \emph{second order variables} $\Vars = \set{X,\ldots}$,
in addition to first order variables $\vars$. We denote by
$\arityof{X}$ the arity of a second order variable $X$. Terms and
atoms are the same as in \slr. The formul{\ae} of \sol\ have the
following syntax:
\begin{flalign*}
  \psi := & ~\xi=\chi \mid \arel(\xi_1, \ldots, \xi_{\arityof{\arel}})
  \mid X(\xi_1, \ldots, \xi_{\arityof{X}})
  \mid \neg\psi \mid \psi \wedge \psi \mid \exists x ~.~ \psi \mid \exists X ~.~ \psi
\end{flalign*}
We write $\xi \neq \chi \isdef \neg \xi = \chi$, $\psi_1 \vee \psi_2
\isdef \neg(\neg\psi_1 \wedge \neg\psi_2)$, $\psi_1 \rightarrow \psi_2
\isdef \neg\psi_1 \vee \psi_2$, $\forall x ~.~ \psi \isdef \neg\exists
x ~.~ \neg\psi$ and $\forall X ~.~ \psi \isdef \neg\exists X ~.~
\neg\psi$. The Weak Monadic Second Order Logic (\mso) is the fragment
of \sol\ restricted to second-order variables of arity one. The Weak
Existential Second Order Logic (\eso) is the fragment of
\sol\ consisting of formul{\ae} of the form $\exists X_1 \ldots
\exists X_n ~.~ \phi$, where $\phi$ has only first order quantifiers.

The semantics of \sol\ is given by a relation $(\universeGeneral,\struc)
\Models^\store \psi$, where the store $\store : \vars \cup \Vars \rightarrow
\universeGeneral \cup \pow{\universeGeneral^+}$ maps each
first-order variable $x \in \vars$ to an element of the universe
$\store(x) \in \universeGeneral$ and each second-order variable $X \in \Vars$
to a finite relation $\store(X) \subseteq \universeGeneral^{\arityof{X}}$. The
satisfaction relation of \sol\ is defined inductively on the structure of
formul{\ae}:
\vspace*{-.5\baselineskip}
\[\begin{array}{lcl}
  (\universeGeneral,\struc) \Models^\store \xi=\chi & \iff & (\struc,\store)(\xi)=(\struc,\store)(\chi)
  \\
  (\universeGeneral,\struc) \Models^\store \arel(\xi_1, \ldots, \xi_{\arityof{\arel}}) & \iff &
  \tuple{(\struc,\store)(\xi_1), \ldots, (\struc,\store)(\xi_{\arityof{\arel}})} \in \struc(\arel)
  \\
  (\universeGeneral,\struc) \Models^\store X(\xi_1, \ldots, \xi_{\arityof{X}}) & \iff &
  \tuple{(\struc,\store)(\xi_1), \ldots, (\struc,\store)(\xi_{\arityof{X}})} \in \store(X)
  \\
  (\universeGeneral,\struc) \Models^\store \exists X ~.~ \psi & \iff &
  (\universeGeneral,\struc) \Models^{\store[X\leftarrow V]} \psi
  \text{, for some finite set } V \subseteq \universeGeneral^{\arityof{X}}
\end{array}\]
The semantics of negation, conjunction and first-order quantification
are standard and omitted for brevity. Note the difference between
equalities and relation atoms in \slr\ and \sol: in the former,
equalities (relation atoms) hold in an empty (singleton) structure,
whereas no such upper bounds on the cardinality of the model of an
atom occur in \sol.

However, \sol\ can express upper bounds on the cardinality of the
universe. Such formul{\ae} are unsatisfiable under the assumption that
the universe of each structure is infinite.  We chose to keep the
comparison between \slr\ and \sol\ simple and not consider the general
case of a finite universe, for the time being. A detailed study of
\seplog\ interpreted over finite universe heaps, with arbitrary
nesting of boolean and separating connectives but without inductive
definitions is given in \cite{DBLP:journals/tocl/EchenimIP20}.  We
plan to give a similar comparison in an extended version.

If $\phi$ is a sentence, we write $(\universeGeneral,\struc) \Models
\phi$ instead of $(\universeGeneral,\struc) \Models^\store \phi$ and
define $\sem{\phi} \isdef \{(\universeGeneral,\struc) \mid
(\universeGeneral,\struc) \Models \phi\}$ and $\gsem{\phi}{k}$ for the
restriction of $\sem{\phi}$ to guarded structures of treewidth at most
$k$. We call $\sem{\phi}$ an \solmso-\emph{definable} set.  We write
$\sem{\text{\solmso}} \isdef \{\sem{\phi} \mid \phi \text{ is a
  \solmso\ formula}\}$ and $\gsem{\text{\solmso}}{k} \isdef
\{\gsem{\phi}{k} \mid \phi \text{ is a \solmso\ formula}\}$.

The aim of this paper is comparing the expressive powers of \slr,
\mso\ and \sol, with respect to the properties that can be defined in
these logics.
We are concerned with the problems $\sem{\mathcal{L}_1} \subseteq
\sem{\mathcal{L}_2}$ and $\gsem{\mathcal{L}_1}{k} \subseteq
\sem{\mathcal{L}_2}$, where $\mathcal{L}_1$ and $\mathcal{L}_2$ are
any of the logics \slr, \mso\ and \sol, respectively. In particular,
for $\gsem{\mathcal{L}_1}{k} \subseteq \sem{\mathcal{L}_2}$, we
implicitly assume that $\mathcal{L}_1$ and $\mathcal{L}_2$ are sets of
formul{\ae} over the relational signature $\signature \cup
\set{\domsymb}$. Table \ref{tab:intro:expressiveness} summarizes our
results, with references to the sections in the paper where the
(non-trivial) proofs can be found, and the remaining open problems.

\section{$\gsem{\text{\slr}}{k} \not\subseteq \sem{\text{\mso}} \not\subseteq \sem{\text{\slr}}$}
\label{sec:mso-slr}

The argument that shows $\gsem{\text{\slr}}{k} \not\subseteq
\sem{\text{\mso}}$ is that
\mso\ cannot express the fact that the cardinality of a set is even
\cite[Proposition 6.2]{CourcelleI}. The \slr\ rules below state that
the cardinality of $\mathfrak{R}$ is even, for a predicate $\apred$ of
arity zero:
\begin{align*}
  \apred() \leftarrow \exists x \exists y ~.~ \mathfrak{R}(x) * \mathfrak{R}(y) * A()
  \hspace*{5mm} \apred() \leftarrow \emp
\end{align*}
Note that every model of $\apred()$ interprets $\mathfrak{R}$ as a set
with an even number of disconnected elements and every other relation
symbol by an empty set. The treewidth of such models is one, thus
$\gsem{\text{\slr}}{k} \not\subseteq \sem{\text{\mso}}$ for any
$k\geq1$, and we obtain $\sem{\text{\slr}} \not\subseteq
\sem{\text{\mso}}$, in general.

The argument for $\sem{\text{\mso}} \not\subseteq \sem{\text{\slr}}$
is that the set of cliques is \mso-definable (actually, even first
order definable) but not \slr-definable. First, the set
$\set{\clique{n} \mid n \in \nat}$ is defined by the following first
order formula in the signature of graph encodings
(Def. \ref{def:graph-struc}):
\begin{align*}
\forall x \forall y ~.~ \vertset(x) \wedge
\vertset(y) \wedge x \neq y \rightarrow \edgerel(x,y) \vee
\edgerel(y,x)
\end{align*}
Since this set is strictly treewidth-unbounded
(Prop. \ref{prop:strictly-unbounded}), it is sufficient to prove that
\slr\ cannot define strictly treewidth-unbounded sets. More precisely,
for each \slr\ sentence $\phi$ and SID $\asid$, we prove the existence
of an integer $W\geq1$, depending on $\phi$ and $\asid$ alone, such
that \begin{inparaenum}[(i)]
\item for each structure $(\universeGeneral,\struc) \in \sidsem{\phi}{\asid}$ there
  exists a structure $(\universeGeneral,\overline{\struc}) \in \sidsem{\phi}{\asid}$, of
  treewidth at most $W$, and
\item the function that maps $(\universeGeneral,\struc)$ into
  $(\universeGeneral,\overline{\struc})$ is locally co-finite (Lemma
  \ref{lemma:injective-model}).
\end{inparaenum}
Then each infinite \slr-definable set has an infinite
treewidth-bounded subset, i.e., it is not strictly treewidth-unbounded
(Prop. \ref{prop:slr-not-strictly-unbounded}).

A first ingredient of the proof is that each SID can be transformed
into an equivalent SID without equality constraints between variables:

\begin{definition}\label{def:normalized-sid}
  A rule $\apred(x_1, \ldots, x_{\arityof{\apred}}) \leftarrow \exists
  y_1 \ldots \exists y_n ~.~ \psi$, where $\psi$ is a quantifier-free
  formula, is \emph{normalized} iff no equality atom $x = y$ occurs in
  $\psi$, for distinct variables $x,y \in \set{x_1, \ldots,
    x_{\arityof{\apred}}} \cup \set{y_1, \ldots, y_n}$. An SID is
  \emph{normalized} iff it contains only normalized rules.
\end{definition}

\begin{lemmaE}[][category=four]\label{lemma:normalized-sid}
  Given an SID $\asid$, one can build a normalized SID $\asid'$ such
  that, for each structure $\struc$ and each predicate atom
  $\apred(\xi_1, \ldots, \xi_{\arityof{\apred}})$, we have
  $(\universeGeneral,\struc) \models_\asid \exists \xi_{i_1} \ldots
  \exists \xi_{i_n} ~.~ \apred(\xi_1, \ldots, \xi_{\arityof{\apred}})
  \iff (\universeGeneral,\struc) \models_{\asid'} \exists \xi_{i_1}
  \ldots \exists \xi_{i_n} ~.~ \apred(\xi_1, \ldots,
  \xi_{\arityof{\apred}})$, where $\set{\xi_{i_1}, \ldots, \xi_{i_n}}
  = \set{\xi_1, \ldots, \xi_{\arityof{\apred}}} \cap \vars$.
\end{lemmaE}
\begin{proofE}
  Let $\asid$ be an SID. For each predicate $\apred$ defined by a rule
  in $\asid$ and each partition $I_1, \ldots, I_k$ of
  $\interv{1}{\arityof{\apred}}$, we consider a fresh predicate
  $\apred_{I_1,\ldots,I_k}$ of arity $k\geq1$, not occurring in
  $\asid$. Let $\asid'$ be the SID obtained from $\asid$ by
  introducing, for each rule:
  \[\begin{array}{l}
    \apred(x_1, \ldots, x_{\arityof{\apred}}) \leftarrow \exists y_1 \ldots \exists y_m ~.~
    \phi * \Asterisk_{\ell=1}^h \bpred^\ell(z_{\ell,1}, \ldots,
    z_{\ell,\arityof{\bpred^\ell}}) \in \asid ~~(\dagger)
  \end{array}\]
  where $\phi$ is a quantifier- and predicate-free formula and for
  each equivalence relation $\approx$ on the set of variables
  $\set{x_1, \ldots, x_{\arityof{\apred}}} \cup \set{y_1,\ldots,y_m}$
  that is compatible with all equalities in $\phi$, i.e., $x = y$
  occurs in $\phi$ only if $x \approx y$, the following rules:
  \[\begin{array}{ll}
    \apred_{I_1,\ldots,I_k}(x_1, \ldots, x_k) \leftarrow \\
    \hspace*{5mm} \big(\exists y_{j_1} \ldots \exists y_{j_n} ~.~ \psi ~*
    \Asterisk_{\ell=1}^k \bpred^\ell_{J^\ell_1,\ldots,J^\ell_{s^\ell}}(z_{r_{\ell,1}}, \ldots,z_{r_{\ell,s^\ell}})\big)[x_{i_1}/x_1, \ldots, x_{i_k}/x_k] & (\ddagger) \\[1mm]
    \apred(x_1, \ldots, x_{\arityof{\apred}}) \leftarrow ~\apred_{I_1,\ldots,I_k}(x_{i_1}, \ldots, x_{i_k}) & (\star)
  \end{array}\]
  where:
  \begin{compactitem}
  \item $\approx$ induces the partitions $I_1, \ldots, I_k$ of
    $\interv{1}{\arityof{\apred}}$ and $J^\ell_1, \ldots,
    J^\ell_{s^\ell}$ of $\interv{1}{\arityof{\bpred^\ell}}$, for each
    $\ell \in \interv{1}{k}$,
  \item $x_{i_j}$ and $z_{r_{\ell,s^\ell}}$ are the first in their
    $\approx$-equivalence classes, respectively, in the total order
    $x_1 < \ldots < x_{\arityof{\apred}} < y_1 <
    \ldots < y_m$,
  \item $\psi$ is obtained from $\phi$ by replacing each variable $x$,
    such that $x \approx x_{i_j}$ with $x_{i_j}$, respectively each
    $z$, such that $z \approx z_{r_{\ell,s^j}}$ with
    $z_{r_{\ell,s^j}}$, and replacing the trivial equalities of the
    form $x = x$ by $\emp$,
  \item the quantifier prefix $\exists y_{j_1} \ldots \exists y_{j_n}$
    is the result of eliminating from $\exists y_1 \ldots \exists y_m$
    the variables that do not occur in $\fv{\psi} \cup
    \bigcup_{\ell=1}^h \set{z_{\ell,r_1}, \ldots,
      z_{r_{\ell,s^\ell}}}$.
  \end{compactitem}
  In particular, one can remove from $\asid'$ the rules containing
  unsatisfiable disequalities of the form $x \neq x$, obtained from
  the above transformations. We are left with proving the equivalence
  from the statement.

  \noindent
  ``$\Rightarrow$'' Assume that $(\universeGeneral,\struc)
  \models^\store_\asid \apred(\xi_1, \ldots, \xi_{\arityof{\apred}})$
  for a store $\store$, and let $\approx$ be the equivalence relation
  over $\set{\xi_1, \ldots, \xi_{\arityof{\apred}}}$, defined as
  $\xi_i \approx \xi_j \iff (\struc,\store)(\xi_i) = (\struc,\store)
  (\xi_j)$. We now prove by induction that $(\universeGeneral,\struc)
  \models^\store_\asid \apred(\xi_1, \ldots, \xi_{\arityof{\apred}})$
  implies $(\universeGeneral,\struc) \models^\store_{\asid'}
  \apred_{I_1, \ldots, I_k}(\xi_{i_1}, \ldots, \xi_{i_k})$, where
  $\set{I_1, \ldots, I_k}$ is the partition of
  $\interv{1}{\arityof{\apred}}$ induced by $\approx$ and the
  $\xi_{i_j}$ are minimal representatives of $I_j$ in some fixed total
  order. Since $(\universeGeneral,\struc) \models^\store_\asid
  \apred(\xi_1, \ldots, \xi_{\arityof{\apred}})$, there is a rule
  ($\dagger$) in $\asid$, a store $\store'$, that agrees with $\store$
  over $\xi_{1}, \ldots, \xi_{\arityof{\apred}}$ and structures
  $(\universeGeneral_0,\struc_0) \scomp \ldots \scomp
  (\universeGeneral_h,\struc_h) = (\universeGeneral,\struc)$, such
  that $(\universeGeneral_0,\struc_0) \models^{\store'}
  \phi\overline{s}$ and $(\universeGeneral_\ell,\struc_\ell)
  \models^{\store'}_\asid
  \bpred^\ell(z_{\ell,1},\ldots,z_{\ell,\arityof{\bpred^\ell}})\overline{s}$,
  for all $\ell \in \interv{1}{h}$, where $\overline{s} \isdef
  [x_{1}/\xi_{1}, \ldots, x_{n}/\xi_{n}]$ is the substitution that
  replaces the formal parameters by terms. Let $\approx'$ be the
  equivalence over $x_1 < \ldots < x_{\arityof{\apred}} < y_1 < \ldots
  < y_m$ defined as $x \approx y \iff \store'(x) = \store'(y)$. By the
  inductive hypothesis, we have $(\universeGeneral_\ell,\struc_\ell)
  \models^{\store'}_{\asid'} \bpred^\ell_{J_1, \ldots,
    r_{s^\ell}}(z_{J_{\ell,1}},\ldots,z_{r_{\ell,s^\ell}})\overline{s}$,
  where $z_{r_{\ell,1}},\ldots,z_{r_{\ell,s^\ell}}$ is the sequence of
  minimal representatives wrt $\approx'$.  Since $\approx' \ \supseteq
  \ \approx$, there exists a rule of type ($\ddagger$) in $\asid'$
  allowing to infer that $(\universeGeneral,\struc)
  \models^\store_{\asid'} \apred_{I_1, \ldots, I_k}(\xi_{i_1}, \ldots,
  \xi_{i_k})$. We finally obtain $(\universeGeneral,\struc)
  \models^\store_{\asid'} \apred(\xi_1, \ldots,
  \xi_{\arityof{\apred}})$, by a rule of type ($\star$).

  \noindent''$\Leftarrow$'' Assume that $(\universeGeneral,\struc)
  \models^\store_{\asid'} \apred(\xi_1, \ldots,
  \xi_{\arityof{\apred}})$.  Then, by a rule of type ($\star$) from
  $\asid'$, we must have $(\universeGeneral,\struc)
  \models^\store_{\asid'} \apred_{I_1, \ldots, I_k}(\xi_{i_1}, \ldots,
  \xi_{i_k})$. We now prove by induction that
  $(\universeGeneral,\struc) \models^\store_{\asid'} \apred_{I_1,
    \ldots, I_k}(\xi_{i_1}, \ldots, \xi_{i_k})$ implies that
  $(\universeGeneral,\struc) \models^{\store'}_{\asid} \apred(\xi_1,
  \ldots, \xi_{\arityof{\apred}})$, where $\store'$ is a store that
  maps each $\xi_j \in \vars$, such that $j \in I_j$, into
  $(\struc,\store)(\xi_{i_j})$. Let $(\universeGeneral,\struc)
  \models^{\store''}_{\asid'} (\psi * \Asterisk_{\ell=1}^h
  \bpred^\ell_{J^\ell_1,\ldots,J^\ell_{s^\ell}}(z_{r_{\ell,1}},
  \ldots, z_{r_{\ell,s^\ell}}))\overline{s}$ by a rule of type
  ($\ddagger$), where $\store''$ is a store that agrees with $\store$
  over $\xi_{i_1}, \ldots, \xi_{i_k}$ and $\overline{s} \isdef
  [x_1/\xi_{i_1}, \ldots, x_k/\xi_{i_k}]$ is a substitution. Then
  there exists structures $(\universeGeneral_0,\struc_0) \scomp \ldots
  \scomp (\universeGeneral_h,\struc_h) = (\universeGeneral,\struc)$,
  such that $(\universeGeneral_0,\struc_0) \models^{\store''}
  \psi\overline{s}$ and $(\universeGeneral_\ell,\struc_\ell)
  \models^{\store''}_{\asid'}
  \bpred^\ell_{J^\ell_1,\ldots,J^\ell_{s^\ell}}(z_{r_{\ell,1}},
  \ldots, z_{r_{\ell,s^\ell}})\overline{s}$, for all $\ell \in
  \interv{1}{h}$. By the definition of $\asid'$, there exists a rule
  of type ($\dagger$) in $\asid$ and a corresponding equivalence
  relation $\approx$ over $x_1 < \ldots < x_{\arityof{\apred}} < y_1 <
  \ldots < y_m$, which induces the partitions $\set{I_1, \ldots, I_k}$
  of $\interv{1}{\arityof{\apred}}$ and $\set{J^\ell_1, \ldots,
    J^\ell_{s^\ell}}$ of $\interv{1}{\arityof{\bpred^\ell}}$, for each
  $\ell \in \interv{1}{k}$, and $x_{i_j}$ and $z_{r_{\ell,s^\ell}}$
  are the first in their $\approx$-equivalence classes.  We can now
  choose a store $\store'''$, such that:
  \begin{compactitem}
  \item $(\struc,\store''')(\xi_j) =  (\struc,\store'')(\xi_{i_j})$,
    for all $j \in \interv{1}{k}$, and
  \item $(\struc,\store''')(z_j\overline{s}) = (\struc,\store'')(z_{r_{\ell,q}}\overline{s})$ if $j \in J^\ell_q$,
    for all $\ell\in\interv{1}{h}$ and $q \in    \interv{1}{s^\ell}$.
  \end{compactitem}
  By the definition of $\psi$, we have $(\universeGeneral_0,\struc_0)
  \models^{\store'''} \phi$. By the inductive hypothesis, we obtain
  $(\universeGeneral_\ell,\struc_\ell) \models^{\store'''}_\asid
  \bpred^\ell(z_{\ell,1}, \ldots, z_{\ell,\arityof{\bpred^\ell}})$,
  for all $\ell\in\interv{1}{h}$.  Hence $(\universeGeneral,\struc)
  \models^{\store'''}_\asid
  \apred(\xi_1,\ldots,\xi_{\arityof{\apred}})$, by a rule of type
  ($\dagger$).
\end{proofE}
\noindent A consequence is that, in the absence of equality
constraints, each existentially quantified variable instantiated by
the inductive definition of the satisfaction relation can be assigned
a distinct element of the universe. For instance, considering the
rules $\mathsf{fold\_ls}(x_1) \leftarrow \emp$ and
$\mathsf{fold\_ls}(x_1) \leftarrow \exists y ~.~ \heapsymb(x_1,y) *
\mathsf{fold\_ls}(y)$, the $\mathsf{fold\_ls}(x)$ formula defines an
infinite set of graphs whose edges are given by the interpretation of
a relation symbol $\heapsymb$, such that there exists an Eulerian path
visiting all edges exactly once, and all vertices possibly more than
once. Since there are no equality constraints, each model of
$\mathsf{fold\_ls}(x)$ can be expanded into an acyclic list that never
visits the same vertex twice, except at the endpoints. This graph has
treewidth two, if the endpoints coincide, and one otherwise.

Formally, we write $(\universeGeneral,\struc)
\imodels{\store}{\asid}{} \phi$ iff the satisfaction relation
$(\universeGeneral,\struc) \models^\store_\asid \phi$ can be
established by considering finite injective stores.
The definition of $\imodels{\store}{\asid}{}$ is the same as the one
of $\models_\asid^\store$ (\S\ref{sec:logics}), except for the cases
below:
\vspace*{-.2\baselineskip}
\begin{align*}
(\universeGeneral,\struc) \imodels{\store}{\asid}{} \phi_1 * \phi_2 \iff &
  \text{there exist structures } (\universeGeneral_1,\struc_1) \scomp (\universeGeneral_2,\struc_2) =
  (\universeGeneral,\struc) \text{, such that } \\[-1mm]
  & \universeGeneral_1 \cap \universeGeneral_2 = \store(\fv{\phi_1} \cap \fv{\phi_2}) \text{ and }
  (\universeGeneral_i,\struc_i) \imodels{\store\downharpoonright_{\fv{\phi_i}}}{\asid}{} \phi_i \text{, for } i = 1,2
\\
(\universeGeneral,\struc) \imodels{\store}{\asid}{} \exists x ~.~ \phi \iff &
(\universeGeneral,\struc) \imodels{\store[x\leftarrow u]}{\asid}{} \phi \text{, for some }
u \in \universeGeneral \setminus \store(\fv{\phi})
\end{align*}
For instance, we have $(\universeGeneral,\struc)
\imodels{\store}{\asid}{} \mathsf{fold\_ls}(x)$ only if
$\struc(\heapsymb)$ is a list of pairwise distinct elements.

\begin{lemmaE}[][category=four]\label{lemma:injective-model}
  Given a normalized SID $\asid$, a predicate atom
  $\apred(\xi_1,\ldots,\xi_{\arityof{\apred}})$, for each structure
  $(\universeGeneral,\struc)$ and a store $\store$, such that
  $(\universeGeneral,\struc) \models^{\store}_\asid
  \apred(\xi_1,\ldots,\xi_{\arityof{\apred}})$, there exists a
  structure $(\universeGeneral,\overline{\struc})$, such that
  $(\universeGeneral,\overline{\struc}) \imodels{\store}{\asid}{}
  \apred(\xi_1,\ldots,\xi_{\arityof{\apred}})$. Moreover, the function
  with domain
  $\sidsem{\apred(\xi_1,\ldots,\xi_{\arityof{\apred}})}{\asid}$ that
  maps $(\universeGeneral,\struc)$ into the set of structures
  isomorphic with $(\universeGeneral,\overline{\struc})$ is locally
  co-finite.
\end{lemmaE}
\begin{proofE} The structure $(\universeGeneral,\overline{\struc})$ is built inductively on the definition
  of the satisfaction relation $(\universeGeneral,\struc)
  \models^{\store}_\asid
  \apred(x_1,\ldots,x_{\arityof{\apred}})$. Since no existentially
  quantified variable is constrained by equality during this
  derivation, one can use the definition of
  $\imodels{\store}{\asid}{}$ instead, thus ensuring that
  $(\universeGeneral,\overline{\struc}) \imodels{\store}{\asid}{}
  \apred(x_1,\ldots,x_{\arityof{\apred}})$. Moreover, since the values
  of the existentially quantified variables are pairwise distinct in
  $(\universeGeneral,\overline{\struc})$, there are only finitely many
  non-isomorphic structures $(\universeGeneral,\struc')$, such that
  $\overline{\struc'} = \overline{\struc}$.
\end{proofE}
We show that the models defined on injective stores have bounded treewidth:

\begin{lemmaE}[][category=four]\label{lemma:injective-bounded}
  Given a normalized SID $\asid$ and a predicate atom $\apred(\xi_1,
  \ldots, \xi_{\arityof{\apred}})$, we have $\twof{\struc} \leq W$,
  for each structure $(\universeGeneral,\struc)$ and store $\store$,
  such that $(\universeGeneral,\struc) \imodels{\store}{\asid}{}
  \apred(\xi_1,\ldots,\xi_{\arityof{\apred}})$, where $W\geq1$ is a
  constant depending only on $\asid$.
\end{lemmaE}
\begin{proofE}
  Let $W$ be the maximal number of variables that occur free or bound
  in a rule from $\asid$. We build a tree decomposition $\tree =
  (\tnodes,\tedges,r,\alabel)$ of $(\universeGeneral,\struc)$, such
  that $\bigcup_{n \in \tnodes} \alabel(n) \subseteq
  \universeGeneral$, inductively on the definition of the satisfaction
  relation $(\universeGeneral,\struc) \imodels{\store}{\asid}{}
  \apred(\xi_1,\ldots,\xi_{\arityof{\apred}})$. Assume that
  $(\universeGeneral,\struc) \imodels{\store}{\asid}{}
  \apred(\xi_1,\ldots,\xi_{\arityof{\apred}})$ is the consequence of a
  rule $\apred(x_1,\ldots,x_{\arityof{\apred}}) \leftarrow \exists y_1
  \ldots \exists y_m ~.~ \psi * \Asterisk_{\ell=1}^k
  \bpred_\ell(z^\ell_1, \ldots, z^\ell_{\arityof{\bpred_\ell}})$ from
  $\asid$, where $\psi$ is a quantifier- and predicate-free formula,
  such that \((\universeGeneral,\struc) \imodels{\store'}{\asid}{}
  \exists y_1 \ldots \exists y_m ~.~ \psi * \Asterisk_{\ell=1}^k
  \bpred_\ell(z^\ell_1, \ldots, z^\ell_{\arityof{\bpred_\ell}})\) and
  $\store'(x_i)=(\struc,\store)(\xi_i)$, for all $i \in
  \interv{1}{\arityof{\apred}}$. Then there exists a store $\store''$
  that agrees with $\store'$ over $x_1, \ldots, x_{\arityof{\apred}}$,
  such that $\store''(y_1), \ldots, \store''(y_m)$ are pairwise
  distinct, and structures $(\universeGeneral_0,\struc_0) \scomp
  \ldots \scomp (\universeGeneral_k,\struc_k) =
  (\universeGeneral,\struc)$, such that: \begin{compactitem}
  \item $\universeGeneral_0 \cap \universeGeneral_\ell =
    \store''(\fv{\psi} \cap \set{z_1^\ell, \ldots,
      z_{\arityof{\bpred_\ell}}^\ell})$, for all $\ell \in
    \interv{1}{k}$,
  \item $\universeGeneral_i \cap \universeGeneral_j =
    \store''(\set{z_1^i, \ldots, z_{\arityof{\bpred_i}}^i} \cap
    \set{z_1^j, \ldots, z_{\arityof{\bpred_j}}^j})$, for all $1 \leq i
    < j \leq k$,
  \item $(\universeGeneral_0,\struc_0) \models^{\store''} \psi$, and
  \item $(\universeGeneral_\ell,\struc_\ell) \imodels{\store''}{\asid}{}
  \bpred_\ell(z^\ell_1, \ldots, z^\ell_{\arityof{\bpred_\ell}})$, for
  all $\ell \in \interv{1}{k}$.
  \end{compactitem}
  By the inductive hypothesis, there exists a tree decomposition
  $\tree_\ell = (\tnodes_\ell,\tedges_\ell,r_\ell,\alabel_\ell)$ of
  $\struc_\ell$, such that $\twof{\tree_\ell} \leq W$ and $\bigcup_{n
    \in \tnodes_\ell} \alabel(n) \subseteq \universeGeneral_\ell$, for
  each $\ell \in \interv{1}{k}$. We define the tree decomposition
  $\tree = (\tnodes,\tedges,r,\alabel)$ such that $\tree_1, \ldots,
  \tree_k$ are the immediate subtrees of the root and
  $\alabel(r)=\set{\store''(x_1), \ldots,
    \store''(x_{\arityof{\apred}})} \cup \set{\store''(y_1), \ldots,
    \store''(y_m)}$. Then, for each relation atom
  $\arel(z_1,\ldots,z_{\arityof{\arel}})$ that occurs in $\psi$, the
  set $\set{\store''(z_1), \ldots, \store''(z_{\arityof{\arel}})}$ is
  a subset of the label of the root, thus fulfilling point
  (\ref{it1:treewidth}) of Def. \ref{def:treewidth}. To check point
  (\ref{it2:treewidth}) of Def. \ref{def:treewidth}, let $u \in
  \alabel_i(n_i) \cap \alabel_j(n_j)$, where $n_i \in \tnodes_i$ and
  $n_j \in \tnodes_j$, for some $1 \leq i < j \leq k$. Since
  $(\universeGeneral_i,\struc_i) \imodels{\store''}{\asid}{}
  \bpred_i(z^i_1,\ldots,z^i_{\arityof{\bpred_i}})$,
  $(\universeGeneral_j,\struc_j) \imodels{\store''}{\asid}{}
  \bpred_j(z^j_1,\ldots,z^j_{\arityof{\bpred_j}})$ and
  $\universeGeneral_i \cap \universeGeneral_j = \store''(\set{z_1^i,
    \ldots, z_{\arityof{\bpred_i}}^i} \cap \set{z_1^j, \ldots,
    z_{\arityof{\bpred_j}}^j})$, we obtain that $u$ is not the image
  of an existentially quantified variable via $\store''$, hence $u =
  \store''(z)$, for some $z \in
  \set{z^i_1,\ldots,z^i_{\arityof{\bpred_i}}} \cap
  \set{z^j_1,\ldots,z^j_{\arityof{\bpred_j}}} \subseteq \set{\xi_1,
    \ldots, \xi_{\arityof{\apred}}} \cup \set{y_1, \ldots, y_m}$. Then
  $u \in \alabel(r)$, thus fulfilling point (\ref{it2:treewidth}) of
  Def. \ref{def:treewidth}. We have $\twof{\tree} =
  \max\set{\cardof{\alabel(r)}-1, \twof{\tree_1}, \ldots,
    \twof{\tree_k}} \leq W$, since $\cardof{\alabel(r)} \leq
  \arityof{\apred}+m \leq W$, by the definition of $\alabel(r)$, and
  $\twof{\tree_i} \leq W$, for all $i \in \interv{1}{k}$, by the
  inductive hypothesis.
\end{proofE}
Note that proving Lemmas \ref{lemma:injective-model} and
\ref{lemma:injective-bounded} for predicate atoms loses no generality,
because for each formula $\phi$, such that
$\fv{\phi}=\set{x_1,\ldots,x_n}$, we can consider a predicate symbol
$\apred_\phi$ of arity $n$ and extend the SID by the rule
$\apred_\phi(x_1, \ldots, x_{n}) \leftarrow \phi$.  The proof of
$\sem{\text{\mso}} \not\subseteq \sem{\text{\slr}}$ relies on the
following:

\begin{propositionE}[][category=four]\label{prop:slr-not-strictly-unbounded}
  Given a sentence $\phi$ and an SID $\asid$, $\sidsem{\phi}{\asid}$
  is either finite or it has an infinite subset of bounded treewidth.
\end{propositionE}
\begin{proofE}
Let $\asid'_\phi$ be the normalized SID, such that
$(\universeGeneral,\struc) \models_{\asid'_\phi} \apred_\phi()$, by
Lemma \ref{lemma:normalized-sid}. Then, there exists a structure
$(\universeGeneral,\overline{\struc})$ and a store $\store$, such that
$(\universeGeneral,\overline{\struc}) \imodels{\store}{\asid'_\phi}{}
\apred_\phi()$, by Lemma \ref{lemma:injective-model}. By Lemma
\ref{lemma:injective-bounded}, we also obtain
$\twof{\universeGeneral,\overline{\struc}} \leq W$, where $W$ depends
only of $\asid'_\phi$ and hence of $\asid$ and $\phi$. If
$\apred_\phi()$ has finitely many non-isomorphic models, there is
nothing to prove. Otherwise, consider an infinite sequence
$(\universeGeneral_1,\struc_1), (\universeGeneral_2,\struc_2), \ldots$
of non-isomorphic models of $\apred_\phi()$. Since, by Lemma
\ref{lemma:injective-model}, for any given $i\geq1$ there are only
finitely many non-isomorphic structures
$(\universeGeneral_j,\struc_j)$, such that $\overline{\struc}_i =
\overline{\struc}_j$, there exists an infinite subsequence
$(\universeGeneral_{i_1},\struc_{i_1}),
(\universeGeneral_{i_2},\struc_{i_2}), \ldots$, such that
$\twof{\universeGeneral_{i_j},\struc_{i_j}} \leq W$, for all $j \geq
1$.
\end{proofE}

\section{$\sem{\text{\slr}} \subseteq \sem{\text{\sol}}$}
\label{sec:slr-so}

Since \slr\ and \mso\ are incomparable, it is natural to ask for a
logic that subsumes both of them. In this section, we prove that
\sol\ is such a logic. Since \mso\ is a syntactic subset of \sol, we
have $\sem{\text{\mso}} \subseteq \sem{\text{\sol}}$ trivially. We
show that $\sem{\text{\slr}} \subseteq \sem{\text{\sol}}$ using the
fact that each model of a predicate atom in \slr\ is built according
to a \emph{finite unfolding tree} indicating the partial order in
which the rules of the SID are used in the inductive definition of the
satisfaction relation; in other words, unfolding trees are for SIDs
what derivation trees are for context-free grammars. More precisely,
any model of a \slr\ sentence can be decomposed into pairwise disjoint
substructures, each being the model of the quantifier- and
predicate-free subformula of a rule in the SID, such that there is a
one-to-one mapping between the nodes of the tree and the substructures
from the decomposition of the model. We use second order variables,
interpreted as finite relations, to define the unfolding tree and the
mapping between the nodes of the unfolding tree and the tuples in the
interpretation of the relation symbols from the model. These
second order variables are existentially quantified and the resulting
\sol\ formula describes the model, without the unfolding tree that
witnesses its construction according to the rules of the SID.

Let $\asid \isdef \set{\arule_1, \ldots, \arule_\nrule}$ be a given
SID. Without loss of generality, for each relation symbol $\arel \in
\signature$, we assume that there is at most one occurrence of an atom
$\arel(y_1, \ldots, y_{\arityof{\arel}})$ in each rule from
$\asid$. If this is not the case, we split the rule by introducing a
new predicate symbol for each relation atom with relation symbol
$\arel_i$, until the condition is satisfied.

\begin{definition}\label{def:unfolding-tree}
  An \emph{unfolding tree} for a predicate atom $\apred(\xi_1, \ldots,
  \xi_{\arityof{\apred}})$ is a $\asid$-labeled tree $\tree =
  (\tnodes,\tedges,r,\alabel)$, such that $\alabel(r)$ defines
  $\apred$ and, for each vertex $n \in \tnodes$, if $\bpred_1(z_{1,1},
  \ldots, z_{1,\arityof{\bpred_1}})$, $\ldots$, $\bpred_h(z_{h,1},
  \ldots, z_{h,\arityof{\bpred_h}})$ are the predicate atoms that
  occur in $\alabel(n)$, then $p_1, \ldots, p_h$ are the children of
  $n$ in $\tree$, such that $\alabel(p_\ell)$ defines $\bpred_\ell$,
  for all $\ell \in \interv{1}{h}$.
\end{definition}

\begin{myFiveTextE}
For a tree $\tree=(\tnodes,\tedges,r,\alabel)$ and a vertex $n \in
\tnodes$, we denote by $\subtree{\tree}{n}$ the subtree of $\tree$
whose root is $n$.  For a quantifier- and predicate-free formula
$\phi$, we denote by $\phi^n$ the formula in which every relation atom
$\arel(x_1, \ldots, x_{\arityof{\arel}})$ is annotated as
$\arel^n(x_1, \ldots, x_{\arityof{\arel}})$. Atoms $\arel^n(x_1,
\ldots, x_{\arityof{\arel}})$ (and consequently formulas $\phi^n$)
have the same semantics as atoms $\arel(x_1, \ldots,
x_{\arityof{\arel}})$ (resp. formulas $\phi$); these annotations serve
purely as explanatory devices in our construction
(Prop.~\ref{prop:slr-so}) that keep track of the node of the unfolding
tree where a relation atom was introduced.

\begin{definition}\label{def:characteristic-formula}
  An unfolding tree $\tree = (\tnodes,\tedges,r,\alabel)$ for a
  predicate atom $\apred(\xi_1, \ldots, \xi_{\arityof{\apred}})$ gives
  rise to a predicate-free formula defined inductively as follows:
  \[\begin{array}{l}
    \charform{n}{\tree}{\apred(\xi_1, \ldots, \xi_{\arityof{\apred}})} \isdef \\
  \left(\exists y_1 \ldots \exists y_m ~.~ \psi^r * \Asterisk_{\ell=1}^h
  \charform{p_\ell}{\subtree{\tree}{p_\ell}}{\bpred_\ell(z_{\ell,1},
    \ldots, z_{\ell,\arityof{\bpred_\ell}})}\right) [x_i/\xi_i]_{i \in \interv{1}{\arityof{\apred}}}
  \end{array}\]
  such that $\alabel(r)$ is the rule:
  \begin{align*}
    \apred(x_1, \ldots, x_{\arityof{\apred}}) \leftarrow \exists y_1
    \ldots \exists y_m ~.~ \psi * \Asterisk_{\ell=1}^h
    \bpred_\ell(z_{\ell,1}, \ldots, z_{\ell,\arityof{\bpred_\ell}})
  \end{align*}
  $\psi$ is a quantifier- and predicate-free formula, and $p_1$,
  $\ldots$, $p_h$ are the roots of the unfolding subtrees of
  $\bpred_1(z_{1,1}, \ldots, z_{1,\arityof{\bpred_1}})$, $\ldots$,
  $\bpred_h(z_{h,1}, \ldots, z_{h,\arityof{\bpred_h}})$ in $\tree$,
  respectively.
\end{definition}

The unfolding trees of a predicate atom describe the set of models of
that predicate atom. The following lemma is standard and we include it
for self-containment reasons:

\begin{lemma}\label{lemma:unfolding-tree}
  Given a structure $(\universeGeneral,\struc)$ and store $\store$, we have $(\universeGeneral,\struc)
  \models^\store_\asid \apred(\xi_1, \ldots, \xi_{\arityof{\apred}})$
  iff there exists an unfolding tree
  $\tree=(\tnodes,\tedges,r,\alabel)$ for $\apred(\xi_1, \ldots,
  \xi_{\arityof{\apred}})$, such that $(\universeGeneral,\struc) \models^\store
  \charform{r}{\tree}{\apred(\xi_1, \ldots, \xi_{\arityof{\apred}})}$.
\end{lemma}
\proof{
  We omit the annotations from $\phi^n$, in this proof and simply
  write $\phi$ because the annotations are not relevant for this
  proof. ``$\Rightarrow$'' By induction on the definition of the satisfaction
  relation $(\universeGeneral,\struc) \models^\store_\asid \apred(\xi_1, \ldots,
  \xi_{\arityof{\apred}})$. Assume the relation holds because:
  \[(\universeGeneral,\struc) \models^{\store'}_\asid \psi\overline{s} *
  \Asterisk_{\ell=1}^h \bpred_\ell(z_{\ell,1}, \ldots,
  z_{\ell,\arityof{\bpred_\ell}})\overline{s}\] for a rule $\arule :
  \apred(x_1, \ldots, x_{\arityof{\apred}}) \leftarrow \exists y_1
  \ldots \exists y_m ~.~ \psi * \Asterisk_{\ell=1}^h
  \bpred_\ell(z_{\ell,1}, \ldots, z_{\ell,\arityof{\bpred_\ell}})$
  from $\asid$, where $\psi$ is a quantifier- and predicate-free
  formula, $\overline{s} \isdef
  [x_1/\xi_1,\ldots,x_{\arityof{\apred}}/\xi_{\arityof{\apred}}]$ is a
  substitution and $\store'$ is a store that agrees with $\store$ over
  $\set{\xi_1, \ldots, \xi_{\arityof{\apred}}} \cap \vars$. Then there
  exist structures $(\universeGeneral_0,\struc_0) \scomp \ldots \scomp
  (\universeGeneral_h,\struc_h) = (\universeGeneral,\struc)$, such
  that $(\universeGeneral_0,\struc_0) \models^{\store'}
  \psi\overline{s}$ and $(\universeGeneral_\ell,\struc_\ell)
  \models^{\store'}_\asid \bpred_\ell(z_{\ell,1}, \ldots,
  z_{\ell,\arityof{\bpred_\ell}})\overline{s}$, for all $\ell \in
  \interv{1}{h}$. By the induction hypothesis, there exist unfolding
  trees $\tree_\ell = (\tnodes_\ell,\tedges_\ell,r_\ell,\alabel_\ell)$
  for $\bpred_\ell(z_{\ell,1}, \ldots,
  z_{\ell,\arityof{\bpred_\ell}})\overline{s}$, such that
  $(\universeGeneral_\ell,\struc_\ell) \models_\asid^{\store'}
  \charform{r_\ell}{\tree_\ell}{\bpred_\ell(z_{\ell,1}, \ldots,
    z_{\ell,\arityof{\bpred_\ell}})\overline{s}}$, for all $\ell \in
  \interv{1}{h}$. Then $\tree = (\tnodes,\tedges,r,\alabel)$ is
  defined as $\tnodes \isdef \set{r} \cup \bigcup_{\ell=1}^h
  \tnodes_\ell$, $\tedges \isdef \set{(r,r_\ell) \mid \ell \in
    \interv{1}{h}} \cup \bigcup_{\ell=1}^h \tedges_\ell$ and $\alabel
  = \set{(r,\arule)} \cup \bigcup_{\ell=1}^h \alabel_\ell$, assuming
  w.l.o.g. that $\tnodes_\ell \cap \tnodes_k = \emptyset$, for all $1
  \leq \ell < k \leq h$ and $r \not\in
  \bigcup_{\ell=1}^h\tnodes_\ell$. The check that
  $(\universeGeneral,\struc) \models^\store
  \charform{r}{\tree}{\apred(\xi_1, \ldots, \xi_{\arityof{\apred}})}$
  is routine.

  \vspace*{\baselineskip} \noindent''$\Leftarrow$'' By induction on
  the structure of $\tree$. Let $p_1, \ldots, p_h$ be the children of
  $r$ in $\tree$. By Def. \ref{def:unfolding-tree}, we have:
  \[((\universeGeneral,\struc) \models^{\store'} \psi\overline{s} *
  \Asterisk_{\ell=1}^h
  \charform{p_\ell}{\subtree{\tree}{p_\ell}}{\bpred_\ell(z_{\ell,1},
    \ldots, z_{\ell,\arityof{\bpred_\ell}})}\overline{s}\] where
  $\overline{s}\isdef[x_1/\xi_1,\ldots,x_{\arityof{\apred}}/\xi_{\arityof{\apred}}]$
  is a substitution, $\store'$ is a store that agrees with $\store$
  over $\set{\xi_1, \ldots, \xi_{\arityof{\apred}}} \cap \vars$ and
  $\alabel(r)=\apred(x_1, \ldots, x_{\arityof{\apred}}) \leftarrow
  \exists y_1 \ldots \exists y_m ~.~ \psi * \Asterisk_{\ell=1}^h
  \bpred_\ell(z_{\ell,1}, \ldots, z_{\ell,\arityof{\bpred_\ell}})$ is
  a rule from $\asid$. Then there exist structures
  $(\universeGeneral_0,\struc_0) \scomp \ldots \scomp
  (\universeGeneral_h,\struc_h) = (\universeGeneral,\struc)$, such
  that $(\universeGeneral_0,\struc_0) \models^{\store'}
  \psi\overline{s}$ and $(\universeGeneral_\ell,\struc_\ell)
  \models^{\store'}
  \charform{p_\ell}{\proj{\tree}{p_\ell}}{\bpred_\ell(z_{\ell,1},
    \ldots, z_{\ell,\arityof{\bpred_\ell}})}\overline{s}$, for all
  $\ell \in \interv{1}{h}$. Since $\subtree{\tree}{p_\ell}$ is an
  unfolding tree for $\bpred_\ell(z_{\ell,1}, \ldots,
  z_{\ell,\arityof{\bpred_\ell}})\overline{s}$, by the inductive
  hypothesis, we obtain $(\universeGeneral_\ell,\struc_\ell)
  \models^{\store'}_\asid \bpred_\ell(z_{\ell,1}, \ldots,
  z_{\ell,\arityof{\bpred_\ell}})\overline{s}$, for all $\ell \in
  \interv{1}{h}$. Then, we have $(\universeGeneral,\struc)
  \models^{\store'}_\asid \psi\overline{s} * \Asterisk_{\ell=1}^h
  \bpred_\ell(z_{\ell,1}, \ldots,
  z_{\ell,\arityof{\bpred_\ell}})\overline{s}$, leading to
  $(\universeGeneral,\struc) \models^{\store}_\asid \apred(\xi_1,
  \ldots, \xi_{\arityof{\apred}})$. \qed}

\vspace*{\baselineskip}\noindent
The formul{\ae} from the definition of $\mathfrak{F}$ are given below:
\end{myFiveTextE}

We build a \sol\ formula that defines the models of a relation atom
$\apred(\xi_1, \ldots, \xi_{\arityof{\apred}})$. As explained above,
this is without loss of generality. Let $\npred$ be the maximum number
of occurrences of predicate atoms in a rule from $\asid_\phi$.  We use
second order variables $Y_1, \ldots, Y_\npred$ of arity $2$, for the
edges of the unfolding tree and $X_1, \ldots, X_\nrule$ of arity $1$,
for the labels of the nodes in the unfolding tree, i.e., the rules of
$\asid$. First, we build a \sol\ formula $\mathfrak{T}(x,
\set{X_i}_{i=1}^\nrule,\set{Y_j}_{j=1}^\npred)$, as the conjunction of
\sol\ formul{\ae} that describe the following
facts:\begin{compactitem}
\item the root $x$ belongs to $X_i$, for some rule $\arule_i$ that defines $\apred$,
\item the sets $X_1, \ldots, X_\nrule$ are pairwise disjoint,
\item each vertex in $X_1 \cup \ldots \cup X_\nrule$ is reachable from $x$
  by a path with edges $Y_1, \ldots, Y_\npred$,
\item each vertex in $X_1 \cup \ldots \cup X_\nrule$, except for $x$, has
  exactly one incoming edge,
\item $x$ has no incoming edge,
\item each vertex from $X_i$ has exactly $h$ outgoing edges $Y_1,
  \ldots, Y_h$, each to a vertex from $X_{j_\ell}$, respectively, such
  that $\arule_{j_\ell}$ defines $\bpred_\ell$, for all $\ell \in
  \interv{1}{h}$, where $\bpred_1(z_{1,1}, \ldots,
  z_{1,\arityof{\bpred_1}}), \ldots,$ $\bpred_h(z_{h,1}, \ldots,
  z_{h,\arityof{\bpred_h}})$ are the predicate atoms that occur in
  $\arule_i$.
\end{compactitem}
Second, we build a \sol\ formula expressing the relationship between
the unfolding tree $\tree=(\tnodes,\tedges,r,\alabel)$ and the model.
The formula $\mathfrak{F}(\xi_1, \ldots, \xi_{\arityof{\apred}}, x,
\set{X_i}_{i=1}^\nrule, \set{Y_j}_{j=1}^\npred,
\set{\set{Z_{k,\ell}}_{\ell=1}^{\arityof{\arel_k}}}_{k=1}^\nrel)$ uses
second order variables $Z_{k,\ell}$, of arity $2$, that encode partial
functions mapping a tree node $n$ to the value of $\xi_\ell$ for the
(unique) atom $\arel_k(\xi_1, \ldots,\xi_{\arityof{\arel_i}})$ from
the rule $\alabel(n)$, in case such an atom exists. The formula
$\mathfrak{F}$ is the conjunction of following \sol-definable
facts\footnote{The exact \sol\ formul{\ae} are given in Appendix
  \ref{app:slr-so}.}: \begin{compactenum}[(i)]
\item\label{it1:slr-so} each second order variable $Z_{k,\ell}$
  denotes a functional binary relation,
\item\label{it2:slr-so} for each tree node labeled by a rule
  $\arule_i$ and each atom $\arel_k(\xi_1, \ldots,
  \xi_{\arityof{\arel_k}})$ occurring at that node, the
  interpretation of $\arel_k$ contains a tuple, whose elements are related to
  the node via $Z_{k,1}, \ldots, Z_{k,\arityof{\arel_k}}$, respectively,

  \begin{myFiveTextE}
    \vspace*{\baselineskip}
    {\raggedright(\text{\ref{it2:slr-so}})}
    $\begin{array}{l}
    \bigwedge_{i \in \interv{1}{\nrule}} \bigwedge_{\arel_k \text{
        occurs in } \arule_i} \forall y ~.~ X_i(y) \rightarrow
    \exists z_1 \ldots \exists z_{\arityof{\arel_k}} ~.~ \arel_k(z_1,\ldots,
    z_{\arityof{\arel_k}}) \wedge \bigwedge_{\ell \in
      \interv{1}{\arityof{\arel_k}}} Z_{k,\ell}(y,z_\ell)
    \end{array}$
  \end{myFiveTextE}
\item\label{it3:slr-so} for any (not necessarily distinct) rules
  $\arule_i$ and $\arule_j$ such that an atom with relation symbol
  $\arel_k$ occurs in both, the corresponding tuples from the
  interpretation of $\arel_k$ are distinct,

  \begin{myFiveTextE}
    \vspace*{\baselineskip}
    {\raggedright(\text{\ref{it3:slr-so}})}
    $\begin{array}{l}
      \bigwedge_{i, j \in \interv{1}{\nrule}} \bigwedge_{\arel_k \text{ occurs in } \arule_i, \arule_j}
      \forall y \forall y' \forall z_1 \forall z'_1 \ldots \forall z_{\arityof{\arel_k}} \forall z'_{\arityof{\arel_k}} ~.~ \\
      \Big(y \neq y' \wedge X_i(y) \wedge X_j(y') \wedge \bigwedge_{\ell\in\interv{1}{\arityof{\arel_k}}} (Z_{k,\ell}(y,z_\ell) \wedge Z_{k,\ell}(y',z'_\ell))\Big)
      \rightarrow \bigvee_{\ell\in\interv{1}{\arityof{\arel_k}}} z_\ell \neq z'_\ell
    \end{array}$
  \end{myFiveTextE}
\item\label{it4:slr-so} each tuple from the interpretation of
  $\arel_k$ must have been introduced by a relation atom with relation
  symbol $\arel_k$ that occurs in a rule $\arule_i$,

    \begin{myFiveTextE}
      \vspace*{\baselineskip}
    {\raggedright(\text{\ref{it4:slr-so}})}
    $\begin{array}{l}
      \bigwedge_{k \in \interv{1}{\nrel}}\forall z_1 \ldots \forall z_{\arityof{\arel_k}} ~.~ \arel_k(z_1, \ldots, z_{\arityof{\arel_k}}) \rightarrow \\
      \exists y ~.~ \bigvee_{\arel_k \text{ occurs in } \arule_i} \big(X_i(y) ~\wedge
      \bigwedge_{\ell \in \interv{1}{\arityof{\arel_k}}} Z_{k,\ell}(y,z_\ell)\big)
    \end{array}$
    \end{myFiveTextE}
\item\label{it5:slr-so} two terms $\xi_m$ and $\chi_n$ that occur in
  two relation atoms $\arel_k(\xi_1, \ldots, \xi_{\arityof{\arel_k}})$
  and $\arel_\ell(\chi_1, \ldots, \chi_{\arityof{\arel_\ell}})$ within
  rules $\arule_i$ and $\arule_j$, respectively, and are constrained
  to be equal (i.e., via equalities and parameter passing), must be
  equated,

    \begin{myFiveTextE}
      \vspace*{\baselineskip}
    {\raggedright(\text{\ref{it5:slr-so}})}
    $\begin{array}{l}
  \bigwedge_{k,\ell \in \interv{1}{\nrel}} \bigwedge_{\begin{array}{l}
      \scriptstyle{\arel_k \text{ occurs in } \arule_i} \\[-1mm]
      \scriptstyle{\arel_\ell \text{ occurs in } \arule_j}
  \end{array}} \bigwedge_{m \in \interv{1}{\arityof{\arel_k}}} \bigwedge_{n \in \interv{1}{\arityof{\arel_\ell}}}
  \forall y \forall y' \forall z' \forall z' ~.~ \\
  \big(X_i(y) \wedge X_j(y') \wedge \mathit{isEq}_{k,\ell,m,n}(y, y', \set{X_i}_{i=1}^\nrule,
  \set{Y_j}_{j=1}^\npred) \wedge~ Z_{k,m}(y,z) \wedge Z_{\ell,n}(y',z')\big) \rightarrow z = z'
    \end{array}$

    \vspace*{\baselineskip}\noindent The formula
    $\mathit{isEq}_{k,\ell,m,n}(y,y',\set{X_i}_{i=1}^\nrule,\set{Y_j}_{j=1}^\npred)$
    asserts that there is a path in the unfolding tree between the
    store values (i.e., vertices of the tree) of $y$ and $y'$, such
    that the $m$-th and $n$-th variables of the relation atoms
    $\arel_k(z_1, \ldots, z_{\arityof{\arel_k}})$ and
    $\arel_\ell(z'_1, \ldots, z'_{\arityof{\arel_\ell}})$ occurring in
    the labels of the vertices $y$ and $y'$, respectively, are bound
    to the same value.
    \end{myFiveTextE}
\item\label{it6:slr-so} a disequality $\xi \neq \chi$ that occurs in a
  rule $\arule_i$ is propagated throughout the tree to each pair of
  variables that occur within two relation atoms $\arel_k(\xi_1,
  \ldots, \xi_{\arityof{\arel_k}})$ and $\arel_\ell(\chi_1, \ldots,
  \chi_{\arityof{\arel_\ell}})$ in rules $\arule_{j_k}$ and
  $\arule_{j_\ell}$, respectively, such that $\xi$ is bound $\xi_r$
  and $\chi$ to $\chi_s$ by equalities and parameter passing,

  \begin{myFiveTextE}
    \vspace*{\baselineskip}
    {\raggedright(\text{\ref{it6:slr-so}})}
    $\begin{array}{l}
      \bigwedge_{\xi \neq \chi \text{ occurs in } \arule_i}
      \bigwedge_{k,\ell \in \interv{1}{\nrel}} \bigwedge_{\begin{array}{l}
          \scriptstyle{\arel_{k} \text{ occurs in } \arule_{j_k}} \\
          \scriptstyle{\arel_{\ell} \text{ occurs in } \arule_{j_\ell}}
      \end{array}} \bigwedge_{r \in \interv{1}{\arityof{\arel_k}}} \bigwedge_{s\in\interv{1}{\arityof{\arel_\ell}}} \\
      \forall y \forall y' \forall y'' \forall z' \forall z'' .
      \Big(X_i(y) \wedge X_{j_k}(y') \wedge X_{j_\ell}(y'') \wedge Z_{k,r}(y',z') ~\wedge \\
      Z_{\ell,s}(y'',z'') \wedge \mathit{varEq}_{\xi,k,r}(y, y', \set{X_i}_{i=1}^N, \set{Y_j}_{j=1}^M) ~\wedge
      \mathit{varEq}_{\chi,\ell,s}(y, y'', \set{X_i}_{i=1}^N)\Big) \rightarrow  z' \neq z''
      \end{array}$

  \vspace*{\baselineskip}\noindent The formula
  $\mathit{varEq}_{\xi,k,r}(x,y,\set{X_i}_{i=1}^\nrule,\set{Y_j}_{j=1}^\npred)$
  states that the variable $\xi$ occurring in the label of the
  unfolding tree vertex $x$ is bound to the variable $z_r$ that occurs
  in a relation atom $\arel_k(z_1, \ldots, z_{\arityof{\arel_i}})$ in
  the label of the vertex $y$.
  \end{myFiveTextE}
\item\label{it7:slr-so} each term in $\apred(\xi_1, \ldots,
  \xi_{\arityof{\apred}})$ that is bound to a variable from a relation
  atom $\arel_k(z_1, \ldots, z_{\arityof{\arel_k}})$ in the unfolding,
  must be equated to that variable.

  \begin{myFiveTextE}
    \vspace*{\baselineskip}
    {\raggedright(\text{\ref{it7:slr-so}})}
    $\begin{array}{l}
    \bigwedge_{\arule_i \text{ defines } \apred} \bigwedge_{j \in \interv{1}{\arityof{\apred}}} \forall y \forall z ~.~ \\
    \Big(X_i(x) \wedge \mathit{varEq}_{\xi_j,k,r}(x, y, \set{X_i}_{i=1}^\nrule,\set{Y_j}_{j=1}^\npred) \wedge Z_{k,r}(y,z)\Big)
    \rightarrow \xi_j = z
    \end{array}$
  \end{myFiveTextE}
\end{compactenum}

\begin{myFiveTextE}
\noindent The above formul{\ae}
$\mathit{isEq}_{k,\ell,m,n}(x,y,\set{X_i}_{i=1}^\nrule,\set{Y_j}_{j=1}^\npred)$
and
$\mathit{varEq}_{\xi,k,r}(x,y,\set{X_i}_{i=1}^\nrule,\set{Y_j}_{j=1}^\npred)$
can be written in \mso, using standard tree automata construction
techniques, similar to the definition of \mso\ formul{\ae} that track
parameters in an unfolding tree for \seplog, with edges definable by
\mso\ formul{\ae} over the signature of
\seplog\ \cite{DBLP:conf/cade/IosifRS13}.
\end{myFiveTextE}

Summing up, the \sol\ formula defining the models of the predicate
atom $\apred(\xi_1, \ldots, \xi_{\arityof{\apred}})$ with respect to
the SID $\asid$ is:
\vspace*{-.5\baselineskip}
\[\begin{array}{l}
\mathfrak{A}^\apred_\asid(\xi_1, \ldots, \xi_{\arityof{\apred}}) \isdef \exists x \exists \set{X_i}_{i=1}^\nrule \exists \set{Y_j}_{j=1}^\npred
\exists \set{Z_{1,\ell}}_{\ell=1}^{\arityof{\arel_1}} \ldots \exists \set{Z_{K,\ell}}_{\ell=1}^{\arityof{\arel_K}} ~.~ \\
\mathfrak{T}(x,\set{X_i}_{i=1}^\nrule,\set{Y_j}_{j=1}^\npred)
\wedge~ \mathfrak{F}(\xi_1, \ldots, \xi_{\arityof{\apred}},x, \set{X_i}_{i=1}^\nrule, \set{Y_j}_{j=1}^\npred,
\set{\set{Z_{k,\ell}}_{\ell=1}^{\arityof{\arel_k}}}_{k=1}^\nrel)
\end{array}\]
The correctness of the above construction is proved in the following
proposition, that also shows $\sem{\text{\slr}} \subseteq
\sem{\text{\sol}}$:

\begin{propositionE}[][category=five]\label{prop:slr-so}
  Given an SID $\asid$ and a predicate atom $\apred(\xi_1, \ldots,
  \xi_{\arityof{\apred}})$, for each structure
  $(\universeGeneral,\struc)$ and store $\store$, we have
  $(\universeGeneral,\struc) \models^{\store}_{\asid} \apred(\xi_1,
  \ldots, \xi_{\arityof{\apred}}) \iff (\universeGeneral,\struc)
  \Models^{\store} \mathfrak{A}^\apred_\asid(\xi_1, \ldots,
  \xi_{\arityof{\apred}})$.
\end{propositionE}
\begin{proofE}
  ``$\Rightarrow$'' By Lemma \ref{lemma:unfolding-tree}, there exists
  an unfolding tree $\tree = (\tnodes,\tedges,r,\alabel)$ of
  $\apred(\xi_1, \ldots, \xi_{\arityof{\apred}})$, such that
  $(\universeGeneral,\struc) \models^{\store}
  \charform{r}{\tree}{\apred(\xi_1, \ldots,
    \xi_{\arityof{\apred}})}$. Let $\charform{r}{\tree}{\apred(\xi_1,
    \ldots, \xi_{\arityof{\apred}})} = \exists y_1 \ldots \exists y_K
  ~.~ \Phi$, where $\Phi$ is a quantifier- and predicate-free formula.
  Note that, by Def. \ref{def:unfolding-tree}, no second order
  variables occur in $\charform{r}{\tree}{\apred(\xi_1, \ldots,
    \xi_{\arityof{\apred}})}$. Hence there exists a store $\store'$
  that agrees with $\store$ over $\xi_1, \ldots,
  \xi_{\arityof{\apred}}$, such that $(\universeGeneral,\struc) \models^{\store'}
  \Phi$. We define another store $\store''$, that agrees with $\store$
  and $\store'$ over $\xi_1, \ldots, \xi_{\arityof{\apred}}$ such
  that, moreover, we have: \begin{compactitem}
  \item $\store''(x)=r$,
  \item $\store''(X_i) = \set{n \in \tnodes \mid \alabel(n) =
    \arule_i}$, for all $i \in \interv{1}{\nrule}$,
  \item $\store''(Y_j) = \set{(n,m) \in \tnodes \times \tnodes \mid m
    \text{ is the $j$-th child of } n}$, for all $j \in
    \interv{1}{\npred}$; we consider that the order between the
    children of a vertex in an unfolding tree is the syntactic order
    of their corresponding predicate atoms, in the sense of
    Def. \ref{def:unfolding-tree},
  \item $\store''(Z_{k,\ell}) = \{(n,(\struc,\store')(\xi_\ell)) \mid n \in
    \tnodes,~ \arel^n_k(\xi_1,\ldots,\xi_{\arityof{\arel_k}}) \\ \text{
      occurs in } \Phi\}$, for all $k \in \interv{1}{\nrel}$ and $\ell
    \in \interv{1}{\arityof{\arel_k}}$.
  \end{compactitem}
  We have $(\universeGeneral,\struc) \Models^{\store''} \mathfrak{T}(x,
  \set{X_i}_{i=1}^\nrule, \set{Y_j}_{j=1}^\npred)$ because $\tree$ is
  an unfolding tree for $\apred(\xi_1,\ldots,\xi_{\arityof{\apred}})$,
  by Def. \ref{def:unfolding-tree}. The proof of \((\universeGeneral,\struc)
  \Models^{\store''} \mathfrak{F}(\xi_1, \ldots,
  \xi_{\arityof{\apred}},x, \set{X_i}_{i=1}^\nrule,
  \set{Y_j}_{j=1}^\npred,
  \set{\set{Z_{k,\ell}}_{\ell=1}^{\arityof{\arel_k}}}_{k=1}^\nrel)\)
  follows from $(\universeGeneral,\struc)\models^{\store'} \Phi$ and the definition of
  $\store''$, by the points (\ref{it1:slr-so}-\ref{it7:slr-so}) from
  the definition of $\mathfrak{F}$. We obtain $(\universeGeneral,\struc) \Models^{\store}
  \mathfrak{A}^\apred_\asid(\xi_1, \ldots, \xi_{\arityof{\apred}})$
  from the definition of $\mathfrak{A}^\apred_\asid$.

  \noindent''$\Leftarrow$'' There exists a store $\store'$ that agrees
  with $\store$ over $\xi_1, \ldots, \xi_{\arityof{\apred}}$, such
  that:
  \begin{align}
  (\universeGeneral,\struc) \Models^{\store'} & \mathfrak{T}(x, \set{X_i}_{i=1}^\nrule,
    \set{Y_j}_{j=1}^\npred) \label{eq:tree} \\
    (\universeGeneral,\struc) \Models^{\store'} &
    \mathfrak{F}(\xi_1, \ldots, \xi_{\arityof{\apred}},x,
    \set{X_i}_{i=1}^\nrule, \set{Y_j}_{j=1}^\npred,
    \set{\set{Z_{k,\ell}}_{\ell=1}^{\arityof{\arel_k}}}_{k=1}^\nrel) \label{eq:form}
  \end{align}
  By (\ref{eq:tree}) we obtain an unfolding tree $\tree =
  (\tnodes,\tedges,r,\alabel)$ for $\apred(\xi_1, \ldots,
  \xi_{\arityof{\apred}})$, such that: \begin{compactitem}
  \item $\tnodes = \bigcup_{i=1}^\nrule \store'(X_i)$,
  \item $\tedges = \bigcup_{j=1}^\npred \store'(Y_j)$,
  \item $\alabel(n) = \arule_i \iff n \in \store'(X_i)$, for all $n
    \in \tnodes$ and $i \in \interv{1}{\nrule}$.
  \end{compactitem}
  Let $\charform{r}{\tree}{\apred(\xi_1, \ldots,
    \xi_{\arityof{\apred}})} = \exists y_1 \ldots \exists y_K ~.~
  \Phi$, where $\Phi$ is a quantifier- and predicate-free formula
  (Def. \ref{def:unfolding-tree}). By Lemma
  \ref{lemma:unfolding-tree}, it is sufficient to show the existence
  of a store $\store''$ that agrees with $\store$ over $\xi_1, \ldots,
  \xi_{\arityof{\apred}}$, such that $(\universeGeneral,\struc)
  \models^{\store''} \Phi$. Let $f_{k,\ell}$ denote the partial
  mapping defined by $\store'(Z_{k,\ell})$, for each $k \in
  \interv{1}{\nrel}$ and $\ell \in \interv{1}{\arityof{\arel}}$, by
  point (\ref{it1:slr-so}) of the definition of $\mathfrak{F}$. For
  each $r \in \interv{1}{K}$, we define $\store''(y_r) \isdef
  f_{k,\ell}(n)$ if $y_r$ occurs in or is constrained to be equal to a
  term $\xi_\ell$ that occurs in an annotated relation atom
  $\arel^n_k(\xi_1, \ldots, \xi_{\arityof{\arel_k}})$ from
  $\Phi$. Note that there can be at most one such relation atom in
  $\Phi$, because of the assumption that in each rule from $\asid$ at
  most one relation atom $\arel_k(z_1,\ldots,z_{\arityof{\arel_k}})$
  occurs. Otherwise, if $y_r$ is not constrained in $\Phi$ to be equal
  to a term that occurs in a relation atom, $\store''(y_r)$ is given
  an arbitrary fresh value. Because the equalities and disequalities
  from $\Phi$ are taken care of by points (\ref{it5:slr-so}) and
  (\ref{it6:slr-so}) from the definition of $\mathfrak{F}$, it remains
  to check the satisfaction of the relation atoms from $\Phi$. To this
  end, we define a decomposition $(\universeGeneral,\struc) =
  \bigscomp_{n \in \tnodes,~ k \in \interv{1}{N}}
  (\universeGeneral_{n,k},\struc_{n,k})$ such that
  $(\universeGeneral_{n,k},\struc_{n,k}) \models^{\store''}
  \arel_k(\xi_1,\ldots,\xi_{\arityof{\arel_k}})$, for all relation
  atoms $\arel_k^n(\xi_1,\ldots,\xi_{\arityof{\arel_k}})$ from
  $\Phi$. Such a decomposition is possible due to points
  (\ref{it2:slr-so}-\ref{it4:slr-so}) from the definition of
  $\mathfrak{F}$.
\end{proofE}
We state as an open question whether the above formula can be written
in \eso, which would sharpen the comparison between \slr\ and \sol, as
\eso\ is known to be strictly less expressive than
\sol\ \cite{Immerman1999}. In particular, the problem is writing
$\mathfrak{F}$ in \eso.

\section{$\gsem{\text{\mso}}{k} \subseteq \sem{\text{\slr}}$}
\label{sec:k-mso-slr}

We prove that, for any \mso\ sentence $\phi$ and any integer $k \geq
1$, there exists an SID $\twformsid{k}{\phi}$ and a predicate
$\apred_{k,\phi}$ of arity zero, such that $\gsem{\phi}{k} =
\sidsem{\apred_{k,\phi}()}{\twformsid{k}{\phi}}$, i.e., the set of
guarded models of $\phi$ of treewidth at most $k$ corresponds to the
set of structures \slr-defined by the predicate atom
$\apred_{k,\phi}()$, when interpreted in the SID
$\twformsid{k}{\phi}$. Our proof leverages from a technique of
Courcelle \cite{CourcelleVII}, used to show that the models of bounded
treewidth of a given \mso\ sentence can be described by a finite set
of recursive equations, written using an algebra of operations on
structures. This result follows up in a long-standing line of work
(known as Feferman-Vaught theorems \cite{journals/apal/Makowsky04})
that reduces the evaluation of an \mso\ sentence on the result of an
algebraic operation to the evaluation of several related sentences in
the arguments of the respective operation.

\subsection{A Theorem of Courcelle}
\label{sec:courcelle}

We recall first a result of Courcelle \cite{CourcelleVII}, that
describes the structures of bounded treewidth, which satisfy a given
MSO formula $\phi$, by an effectively constructible set of recursive
equations. This set of equations uses two operations on structures,
namely $\glue$ and $\fgcst{j}$, that are lifted to sets of structures,
as usual. The result is developed in two steps. The first step builds
a generic set of equations, that characterizes all structures of
treewidth at most $k$. This set of equations is then refined, in the
second step, to describe only models of $\phi$. Because this result
applies to general (i.e., finite and infinite) structures
$(\universeGeneral,\struc)$, we do not require $\universeGeneral$ to
be infinite, for the purposes of this presentation. We consider a
fixed integer $k \geq 1$ and \mso\ sentence $\phi$ in the rest of this
section.

\paragraph{Operations on Structures}
Let $\signature_1$
and $\signature_2$
be two (possibly overlapping) signatures. The \emph{glueing} operation
$\glue : \strucof{\signature_1} \times \strucof{\signature_2}
\rightarrow \strucof{\signature_1\cup\signature_2}$ is the union of
structures with \emph{disjoint universes}, followed by fusion of the
elements denoted by constants. Formally, given $\astruc_i =
(\adomof{i},\interpof{i})$, for $i = 1,2$, such that $\adomof{1} \cap
\adomof{2} = \emptyset$, let $\sim$ be the least equivalence relation
on $\adomof{1} \cup \adomof{2}$ such that $\interpof{1}(\acst) \sim
\interpof{2}(\acst)$, for all $\acst \in \signature_1 \cap
\signature_2$. Let $[u]$ be the equivalence class of $u \in \adomof{1}
\cup \adomof{2}$ with respect to $\sim$ and lift this notation to
tuples and sets of tuples. Then $\glueof{\astruc_1}{\astruc_2} \isdef
(\adomof{}, \interpof{})$, where $\adomof{} \isdef \set{[u] \mid u \in
  \adomof{1} \cup \adomof{2}}$ and $\interpof{}$ is defined as
follows:
\begin{align*}
\struc(\arel) \isdef & \left\{\begin{array}{ll}
      \text{$[\struc_i(\arel)]$}, & \text{if } \arel \in \signature_i\setminus\signature_{3-i} \text{, for both } i = 1,2 \\
      \text{$[\struc_1(\arel)\cup\struc_2(\arel)]$}, & \text{if } \arel \in \signature_1 \cap \signature_2
\end{array}\right.
\end{align*}
Since we match isomorphic structures, the nature of the elements of
$\adomof{}$ (i.e., equivalence classes) is not important. The
\emph{forget} operation $\fgcst{j} : \strucof{\signature} \rightarrow
\strucof{\signature \setminus \set{\acst_j}}$ simply drops the
constant $\acst_j$ from the domain of its argument.

\paragraph{Structures of Bounded Treewidth}
Let $\signature = \{\arel_1, \ldots, \arel_\nrel, \acst_1, \ldots,
\acst_\ncst\}$ be a signature and $\ports = \{\acst_{\ncst+1}, \ldots,
\acst_{\ncst+k+1}\}$ be a set of constants disjoint from $\signature$,
called \emph{ports}. We consider variables $Y_i$, for all subsets
$\ports_i \subseteq \ports$, denoting sets of structures over the
signature $\signature \cup \ports_i$. The \emph{equation system}
$\tweq{k}$ is the set of recursive equations of the form $Y_0
\supseteq f(Y_1, \ldots, Y_n)$, where each $f$ is either $\glue$,
$\fgcst{\ncst + j}$, for any $j \in \interv{1}{k+1}$, or a singleton
relation of type $\arel_i$, consisting of a tuple
with at most $k+1$ distinct elements, for any $i \in
\interv{1}{\nrel}$. It is known that the set of structures of
treewidth at most $k$ is a component of the least solution of
$\tweq{k}$, in the domain of tuples of sets ordered by pointwise
inclusion \cite[Theorem 2.83]{courcelle_engelfriet_2012}.

\paragraph{Models of \mso\ Formul{\ae}}
The \emph{quantifier rank} $\qrof{\phi}$ of an \mso\ formula $\phi$ is
the maximal depth of nested quantifiers, i.e., $\qrof{\phi} \isdef 0$
if $\phi$ is an atom, $\qrof{\neg\phi_1} \isdef \qrof{\phi_1}$,
$\qrof{\phi_1 \wedge \phi_2} \isdef \max(\qrof{\phi_1},
\qrof{\phi_2})$ and $\qrof{\exists x ~.~ \phi_1} = \qrof{\exists X ~.~
  \phi_1} \isdef \qrof{\phi_1} + 1$. We denote by $\typesof{r}$ the
set of \mso\ sentences of quantifier rank at most $r$. This set is
finite, up to logical equivalence. For a structure $\astruc =
(\universeGeneral,\struc)$, we define its $r$-\emph{type} as
$\typeof{r}{\astruc} \isdef \set{\phi \in \typesof{r} \mid \astruc
  \Models \phi}$. We assume the sentences in $\typeof{r}{\astruc}$ to
use the signature over which $\astruc$ is defined; this signature will
be clear from the context in the following.

\begin{definition}\label{def:compatible}
An operation $f : \strucof{\signature_1} \times \ldots \times
\strucof{\signature_n} \rightarrow \strucof{\signature_{n+1}}$ is
(effectively) \mso-\emph{compatible}\footnote{Also referred to as
  \emph{smooth} operations in \cite{journals/apal/Makowsky04}.} iff,
for all structures $S_1, \ldots, S_n$, $\typeof{r}{f(\astruc_1,
  \ldots, \astruc_n)}$ depends only on (and can be effectively
computed from) $\typeof{r}{\astruc_1}, \ldots, \typeof{r}{\astruc_n}$
by an \emph{abstract operation} $\absof{f} :
\left(\pow{\typesof{r}}\right)^n \rightarrow \pow{\typesof{r}}$.
\end{definition}

The result of Courcelle establishes that glueing and forgetting of
constants are effectively \mso-compatible operations, with effectively
computable abstract operations $\absglue$ and $\absfgcst{\ncst + i}$,
for $i \in \interv{1}{k+1}$, see \cite[Lemmas 3.2 and
  3.3]{CourcelleVII}. As a consequence, one can build from $\tweq{k}$
a set of recursive equations $\abstweq{k}$ of the form $Y_0^{\tau_0} =
f(Y_1^{\tau_1}, \ldots, Y_n^{\tau_n})$, where $Y_0 = f(Y_1, \ldots,
Y_n)$ is an equation from $\tweq{k}$ and $\tau_0, \ldots, \tau_n$ are
$r$-types such that $\tau_0 = \absof{f}(\tau_1, \ldots, \tau_n)$.
Intuitively, each annotated variable $Y^\tau$ denotes the set of
structures whose $r$-type is $\tau$, from the $Y$-component of the
least solution of $\tweq{k}$.  Given some formula $\phi$ with
$\qrof{\phi}=r$, the set of models of $\phi$ of treewidth at most $k$
is the union of the $Y^\tau$-components of the least solution of
$\abstweq{k}$, such that $\phi \in \tau$ \cite[Theorem
  3.6]{CourcelleVII}.

\subsection{Encoding Types in \slr}
\label{sec:encoding-types-sids}

We begin explaining the proof for $\gsem{\text{\mso}}{k} \subseteq
\sem{\text{\slr}}$.  Instead of using the set of recursive equations
$\tweq{k}$ from the previous subsection, we give an SID $\twsid{k}$
that characterizes the guarded structures of bounded treewidth (Fig.
\ref{fig:sids}a). We use the separating conjunction to simulate the
glueing operation. The main problem is with the interpretation of the
separating conjunction, as composition of structures with possibly
overlapping universes (Def. \ref{def:composition}), that cannot be
glued directly. Our solution is to consider guarded structures
(Def. \ref{def:guarded}), where the unary relation symbol $\domsymb$
is used to enforce disjointness of the arguments of the composition
operation, in all but finitely many elements. Intuitively, $\domsymb$
``collects'' the values assigned to the existentially quantified
variables created by rule (\ref{rule:gen-exists}) of $\twsid{k}$ and
the top-level rule (\ref{rule:gen-top}) during the unraveling. This
ensures that
\begin{inparaenum}[(i)]
\item the variables of a predicate atom are mapped to pairwise
  distinct values and
\item the composition of two guarded structures is the same as glueing
  them.
\end{inparaenum}
Similar conditions have been used to define e.g., fragments of
\seplog\ with nice computational properties, such as the
\emph{establishment} condition used to ensure decidability of
entailments \cite{DBLP:journals/ipl/EchenimIP22}, or the
\emph{tightness} condition from
\cite[\S5.2]{AhrensBozgaIosifKatoen21}.

To alleviate the presentation, the SID $\twsid{k}$ defines only
structures $(\universeGeneral,\struc)\in\strucof{\signature,\domsymb}$ with
at least $k+1$ distinct elements in $\struc(\domsymb)$ (rule
\ref{rule:gen-top}) and $\struc(\arel)\neq\emptyset$ for at least one
relation symbol $\arel \in \signature$ (rule \ref{rule:gen-rel}). The
cases of structures such that $\cardof{\struc(\domsymb)} \leq k$ or
$\bigcup_{\arel\in\signature}\struc(\arel) = \emptyset$ can be dealt
with easily, by adding more rules to $\twsid{k}$. In the rest of this
section we show that $\twsid{k}$ defines all structures of $k$-bounded
treewidth (except for the mentioned corner cases).

\begin{figure}[t!]
  \vspace*{-\baselineskip}
  \begin{align}
    \apred(x_1, \ldots, x_{k+1}) \leftarrow & ~\apred(x_1, \ldots,
    x_{k+1}) * \apred(x_1, \ldots, x_{k+1}) \label{rule:gen-comp} \\[1mm]
    \apred(x_1, \ldots, x_{k+1}) \leftarrow & ~\exists y ~.~
    \domsymb(y) * \apred(x_1, \ldots,x_{k+1})[x_i/y] \text{ for all } i \in \interv{1}{k+1} \label{rule:gen-exists} \\[1mm]
    \apred(x_1, \ldots, x_{k+1}) \leftarrow &
    ~\arel(y_1,\ldots,y_{\arityof{\arel}}) \text{ for all } \arel\in\signature \text{ and } y_1,\ldots,y_{\arityof{\arel}} \in \set{x_1, \ldots, x_{k+1}} \label{rule:gen-rel} \\[1mm]
    \apred_k() \leftarrow & ~\exists x_1 \ldots \exists x_{k+1} ~.~ \domsymb(x_1) * \ldots * \domsymb(x_{k+1}) * \apred(x_1, \ldots, x_{k+1})  \label{rule:gen-top}
  \end{align}

  \vspace*{-.5\baselineskip}
  \centerline{\small(a)}

  \vspace*{-\baselineskip}
  \begin{align}
    \apred^\atype(x_1, \ldots, x_{k+1}) \leftarrow & ~\apred^{\atype_1}(x_1, \ldots, x_{k+1}) * \apred^{\atype_2}(x_1, \ldots, x_{k+1})
    \text{ where } \atype = \absglueof{\atype_1}{\atype_2}  \label{rule:ref-comp}  \\[1mm]
    \apred^\atype(x_1, \ldots, x_{k+1}) \leftarrow & \exists y ~.~ \domsymb(y) * \apred^{\atype_1}(x_1, \ldots, x_{k+1})[x_i/y] \text{ for all } i \in \interv{1}{k+1}, \text{where } \label{rule:ref-exists} \\[-1mm]
    &  \atype = \absglueof{\absfgcst{\ncst+i}(\atype_1)}{\rho_i} \text{ and } \rho_i \text{ is the type of some structure} \nonumber \\[-1mm]
    & \astruc \in \strucof{\set{\acst_{\ncst+i}},\domsymb} \text{ with singleton universe and } S \Models \domsymb(\acst_{\ncst+i}) \nonumber \\[1mm]
    \apred^\atype(x_1, \ldots, x_{k+1}) \leftarrow & ~\arel(y_1,\ldots,y_{\arityof{\arel}}) \text{ for some } y_1, \ldots, y_{\arityof{\arel}} \in \{x_1,\ldots,x_{k+1} \}, \text{ where} \label{rule:ref-rel} \\[-1mm]
    & \atype = \typeof{\qrof{\phi}}{\astruc}, ~\astruc \in \strucof{\signature\cup\set{\acst_{\ncst+1},\ldots,\acst_{\ncst+k+1}},\domsymb} \text{ and}  \nonumber \\[-1mm]
    & \astruc \Models \arel(y_1,\ldots,y_{\arityof{\arel}})  [x_1/\acst_{\ncst+1}, \ldots, x_{k+1}/\acst_{\ncst+k+1}] * \Asterisk_{i=1}^{k+1} \domsymb(\acst_{\ncst+i}) \nonumber \\[1mm]
    \apred_{k,\phi}() \leftarrow & \exists x_1 \ldots \exists x_{k+1} ~.~\domsymb(x_1) * \ldots * \domsymb(x_{k+1}) ~* \apred^\atype(x_1, \ldots, x_{k+1}) \label{rule:ref-top} \\[-1mm]
    &   \text{ for all } \atype \text{ such that } \phi\in\atype \nonumber
  \end{align}

  \vspace*{-.5\baselineskip}
  \centerline{\small(b)}
  \vspace*{-.5\baselineskip}
  \caption{The SID $\twsid{k}$ defining structures of treewidth at
    most $k$ (a) and its annotation $\twformsid{k}{\phi}$ defining the
    models of an \mso\ sentence $\phi$, of treewidth at most $k$ (b)}
  \label{fig:sids}
  \vspace*{-\baselineskip}
\end{figure}

\begin{mySixTextE}
We consider tree decompositions of a special form and show that this
restriction does not lose generality (see
Lemma~\ref{lemma:reduced-tree-decompositions} below).

\begin{definition}\label{def:reduced-td}
A tree decomposition $\tree=(\tnodes,\tedges,r,\alabel)$ of a
structure $(\universeGeneral,\struc)$ is said to be \emph{reduced} iff
the following hold: \begin{compactenum}
\item\label{it1:reduced-td} for each $\arel \in \signature$ and each $\tuple{u_1, \ldots,
  u_{\arityof{\arel}}} \in \struc(\arel)$ there exists a leaf $n \in
  \tnodes$ such that $\set{u_1, \ldots, u_{\arityof{\arel}}} \subseteq
  \alabel(n)$, called the \emph{witness} of $\tuple{u_1, \ldots,
    u_{\arityof{\arel}}} \in \struc(\arel)$,
\item\label{it2:reduced-td} a leaf witnesses exactly one tuple
  $\tuple{u_1, \ldots, u_{\arityof{\arel}}} \in \struc(\arel)$,
\item\label{it3:reduced-td} each node in $\tree$ has at most two
  children,
\item\label{it4:reduced-td} if $n \in \tnodes$ has two children $m_1,m_2 \in \tnodes$
  then $\alabel(n) = \alabel(m_1) = \alabel(m_2)$,
\item\label{it6:reduced-td} $\cardof{\alabel(n)} = k+1$, for all $n \in \tnodes$.
\item\label{it5:reduced-td} if $n \in \tnodes$ has one child $m \in \tnodes$ then either $\alabel(n) = \alabel(m)$
    or we have $\cardof{\alabel(n) \setminus \alabel(m)} = \cardof{\alabel(m)
  \setminus \alabel(n)} = 1$.
\end{compactenum}
\end{definition}

\begin{lemma}\label{lemma:reduced-tree-decompositions}
  A structure $(\universeGeneral,\struc)$ has a tree decomposition of
  width $k$ iff it has a reduced tree decomposition of width $k$.
\end{lemma}
\proof{
  Let $\tree=(\tnodes,\tedges,r,\alabel)$ be the original tree
  decomposition. We perform the following actions, corresponding to
  the cases from Def. \ref{def:reduced-td}, in this
  order: \begin{compactitem}
  \item[(\ref{it1:reduced-td})] For each non-leaf node $n_0\in\tnodes$,
    witnessing a tuple $\tuple{u_1, \ldots, u_{\arityof{\arel}}} \in
    \struc(\arel)$, we add a leaf $m_1$ directly to $n_0$, labeled
    with $\alabel(n_0)$.
  \item[(\ref{it2:reduced-td})] for each leaf $n_0\in\tnodes$, that
    witnesses $k\geq2$ tuples, we append $k$ children $m_1, \ldots,
    m_k$, each labeled by $\alabel(n_0)$.
  \item[(\ref{it3:reduced-td})] Let $n_0 \in \tnodes$ be a node with
    children $n_1, \ldots, n_\ell \in \tnodes$, for some $\ell\geq2$.
    We introduce nodes $m_1, \ldots, m_{\ell-1}$ labeled with
    $\alabel(n_0)$, delete the edges from $n_0$ to $n_1, \ldots,
    m_\ell$ and add edges $m_i$ to $n_i$ and $m_{i+1}$, for all $i \in
    \interv{1}{\ell-2}$ then edges from $m_{\ell-1}$ to $n_{\ell-1}$
    and $n_\ell$. This step is repeated for each node with more than
    two children in $\tree$.
  \item[(\ref{it4:reduced-td})] For each node $n$ with children
    $n_1, n_2$, we relabel add nodes $n_1', n_2'$ with $\alabel(n) =
    \alabel(n_1')=\alabel(n_2')$, delete the edges from $n$ to $n_1,n_2$, and add edges from $n$ to $n_1',n_2'$ as well as edges from $n_1'$ to $n_1$ and from $n_2'$ to $n_2$.
  \item[(\ref{it6:reduced-td})] For each node $n \in \tnodes$, we add
    $k+1-\cardof{\alabel(n)}$ fresh elements from $\universeGeneral$.
    Note that the added elements appear exactly once in the tree decomposition, and, hence, the conditions of a tree decomposition remain satisfied.
  \item[(\ref{it5:reduced-td})]
    Let $n$ be a node whose only child is $n'$ with $\alabel(n) \neq \alabel(n')$.
    If $\alabel(n)$ and $\alabel(n')$ differ by more than one element we add a chain of nodes between $n$ and $n'$ such that the sets of labels of two neighboring nodes in the chain differ by exactly one element.
  \end{compactitem}

  \qed}

We show that the set $\sidsem{\apred_k()}{\twsid{k}}$ contains only
structures of treewidth at most $k$.  Let $(\universeGeneral,\struc)
\in \strucof{\signature,\domsymb}$ be a structure such that
$(\universeGeneral,\struc) \models_{\twsid{k}} \apred_k()$.  By rule
(\ref{rule:gen-top}), there exists a store $\store$ with $\store(x_i)
= \vertex_i$, for all $1 \le i \le k+1$, and a structure
$(\universeGeneral,\struc') \in \strucof{\signature,\domsymb}$ that
agrees with $(\universeGeneral,\struc)$ on $\signature$ such that
$(\universeGeneral,\struc') \models^\store_{\twsid{k}} \apred(x_1,
\ldots, x_{k+1})$ and $\struc'(\domsymb) = \struc(\domsymb) \setminus
\set{\vertex_1,\ldots,\vertex_{k+1}}$.  The rest of the proof goes by
induction on the definition of the satisfaction relation for \slr\ and
is given in the following lemma:

\begin{lemma}
\label{lem:sid-implies-tree-decomposition}
Let $(\universeGeneral,\struc) \in \strucof{\signature,\domsymb}$ be a structure and
$\store$ be a store, such that $(\universeGeneral,\struc) \models^\store_{\twsid{k}}
\apred(x_1,\ldots,x_{k+1})$ and $\store(x_i) \not\in
\struc(\domsymb)$, for all $i \in \interv{1}{k+1}$.  Then, there exists a
reduced tree decomposition $\tree = (\tnodes,\tedges,r,\alabel)$ of
$\struc$, such that $\twof{\tree}=k$,
$\alabel(r)=\set{\store(x_1),\ldots,\store(x_{k+1})}$ and $\alabel(n)
\subseteq \struc(\domsymb) \cup
\{\store(x_1),\ldots,\store(x_{k+1})\}$, for all $n \in \tnodes$.
\end{lemma}
\proof{ We prove the claim by induction on the number of rule
  applications. The claim clearly holds for the base case, by rule
  (\ref{rule:gen-rel}). We now consider the rule
  (\ref{rule:gen-comp}), i.e., we assume that
  $(\universeGeneral,\struc) \models^\store_{\twsid{k}}
  \apred(x_1,\ldots,x_{k+1}) * \apred(x_1,\ldots,x_{k+1})$. Then,
  there exist structures $(\universeGeneral_1,\struc_1)$ and
  $(\universeGeneral_2,\struc_2)$, such that
  $(\universeGeneral_i,\struc_i,) \models^\store_{\twsid{k}}
  \apred(x_1,\ldots,x_{k+1})$, for $i = 1,2$ and
  $(\universeGeneral_1,\struc_1) \scomp (\universeGeneral_2,\struc_2)
  = (\universeGeneral,\struc)$.  We note that the latter implies that
  $\struc_1(\domsymb) \cap \struc_2(\domsymb) = \emptyset$
  ($\dagger$). We now apply the inductive hypothesis and obtain
  reduced tree decompositions $\tree_i$ for
  $(\universeGeneral_i,\struc_i)$ whose respective roots are labelled
  by $\{\store(x_1),\ldots,\store(x_{k+1})\}$ and $\alabel(n)
  \subseteq \struc(\domsymb) \cup
  \{\store(x_1),\ldots,\store(x_{k+1})\}$, for all nodes $n$ of
  $\tree_i$ and all $i = 1,2$. We obtain a reduced tree decomposition
  for $(\universeGeneral,\struc) $ by composing $\tree_1$ and
  $\tree_2$ with a fresh root node labelled by
  $\{\store(x_1),\ldots,\store(x_{k+1})\}$. Note that $\tree$ is
  indeed a tree decomposition,note t because the only elements that
  may appear in labels of both $\tree_1$ and $\tree_2$ must belong to
  $\{\store(x_1),\ldots,\store(x_{k+1})\}$, by ($\dagger$). Clearly,
  we have $\alabel(n) \subseteq \struc_1(\domsymb) \cup
  \struc_2(\domsymb) \cup \{\store(x_1),\ldots,\store(x_{k+1})\} =
  \struc(\domsymb) \cup \{\store(x_1),\ldots,\store(x_{k+1})\}$, for
  all nodes $n$ of the resulting tree decomposition $\tree$. We now
  consider the rule (\ref{rule:gen-exists}), i.e., we assume that
  $(\universeGeneral,\struc) \models^\store_{\twsid{k}} \exists y ~.~
  \domsymb(y) * \apred(x_1, \ldots,x_{k+1})[x_i/y]$. Then, there is an
  element $\vertex \not\in \set{\store(x_1), \ldots, \store(x_{k+1})}$
  such that $(\universeGeneral,\struc') \models^{\store[x_i \leftarrow
      \vertex])}_{\twsid{k}} \apred(x_1, \ldots,x_{k+1})$, where the
  structure $(\universeGeneral,\struc')$ agrees with
  $(\universeGeneral,\struc)$, except that $\struc(\domsymb) =
  \struc'(\domsymb) \setminus \set{\vertex}$.  We now apply the
  inductive hypothesis and obtain a reduced tree decomposition
  $\tree_1$ for $\struc'$ whose root is labelled by
  $(\{\store(x_1),\ldots,\store(x_{k+1})\} \setminus \{\store(x_i)\})
  \cup \{\vertex\}$ and $\alabel(n) \subseteq ((\struc'(\domsymb) \cup
  \{\store(x_1),\ldots,\store(x_{k+1})\}) \setminus \{\store(x_i)\})
  \cup \{\vertex\}$, for all nodes $n$ of $\tree_1$ ($\ddagger$).  We
  can now obtain a reduced tree decomposition $\tree$ for $\struc$ by
  composing $\tree_1$ with an additional root labelled by
  $\{\store(x_1),\ldots,\store(x_{k+1})\}$.  Note that $\tree$ is
  indeed a tree decomposition because $\store(x_i) \not\in \alabel(n)$
  for every node $n$ of $\tree_1$, because of ($\ddagger$), $\vertex
  \neq \store(x_i)$ and the assumption that $\store(x_i) \not\in
  \struc(\domsymb)$.  We have $\alabel(n) \subseteq \struc'(\domsymb)
  \cup \{\store(x_1),\ldots,\store(x_{k+1})\} \cup \{\vertex\} =
  \struc(\domsymb) \cup \{\store(x_1),\ldots,\store(x_{k+1})\}$ for
  all nodes $n$ of the resulting tree decomposition $\tree$. \qed}

Dually, we prove that any structure of treewidth at most $k$ belongs
to $\sidsem{\apred_k()}{\twsid{k}}$. To this end, Lemma
\ref{lem:tree-decomposition-implies-sid} below shows that, for each
such structure $(\universeGeneral,\struc)$ there exists a store $\store$, such that
$\store(x_i) \not\in \struc(\domsymb)$, for all $i \in
\interv{1}{k+1}$ and $\struc \models^\store_{\twsid{k}}
\apred(x_1,\ldots,x_{k+1})$, and conclude by an application of rule
(\ref{rule:gen-top}).

\begin{lemma}\label{lem:tree-decomposition-implies-sid}
  Let $(\universeGeneral,\struc) \in \strucof{\signature,\domsymb}$ be a structure with
  $\twof{\struc} \le k$ witnessed by some reduced tree decomposition
  $\tree=(\tnodes,\tedges,r,\alabel)$, with
  $\alabel(r)=\{\vertex_1,\ldots,\vertex_{k+1}\}$ and
  $\struc(\domsymb) = \bigcup_{n \in \tnodes} \alabel(n) \setminus
  \{\vertex_1,\ldots,\vertex_{k+1}\}$, and let $\store$ be a store
  with $\store(x_i) = \vertex_i$, for all $i \in \interv{1}{k+1}$.
  Then, $(\universeGeneral,\struc) \models^\store_{\twsid{k}} \apred_k(x_1, \ldots,
  x_{k+1})$.
\end{lemma}
\proof{
  The proof goes by induction on the structure of $\tree$. The claim
  clearly holds for the base case, where $\tree$ consists of a single
  leaf, by rule (\ref{rule:gen-rel}). Consider first the case where
  the root of $\tree$ has two children. The subtrees $\tree_1$ and
  $\tree_2$ rooted in the two children induce substructures $(\universeGeneral,\struc_1)$
  and $(\universeGeneral,\struc_2)$ of $(\universeGeneral,\struc)$, where $\tuple{v_1, \ldots, v_{\arityof{\arel}}} \in \struc_i(\arel)$ iff $\tuple{v_1, \ldots,
  v_{\arityof{\arel}}}$ is witnessed by some leaf of $\tree_i$, for $i
  = 1,2$. Because $\tree$ is a reduced tree decomposition, there is at
  most one leaf that witnesses a tuple $\tuple{v_1, \ldots,
  v_{\arityof{\arel}}} \in \struc_i(\arel)$.  Hence, we have $\struc
  = \struc_1 \comp \struc_2$.
  From the inductive hypothesis we
  obtain $(\universeGeneral,\struc_i) \models^\store_{\twsid{k}} \apred(x_1, \ldots,
  x_{k+1})$, for $i = 1,2$.  Hence $(\universeGeneral,\struc_1 \comp \struc_2)
  \models^\store_{\twsid{k}} \apred(x_1, \ldots, x_{k+1}) *
  \apred(x_1, \ldots, x_{k+1})$, thus $(\universeGeneral,\struc)
  \models^\store_{\twsid{k}} \apred(x_1, \ldots, x_{k+1})$, by rule
  (\ref{rule:gen-comp}).
  Consider now the case where the root of
  $\tree$ has a single child which is not a leaf and consider the
  subtree $\tree_1$ rooted at this child.
  Then, there is an element
  $\vertex \not\in \set{\vertex_1, \ldots, \vertex_{k+1}}$, such that
  the root of $\tree_1$ is labeled by
  $\{\vertex_1,\ldots,\vertex_{i-1},\vertex,\vertex_{i+1},\ldots,\vertex_{k+1}\}$.
  Let $(\universeGeneral,\struc')$ be the structure that agrees with $(\universeGeneral,\struc)$ except that
  we have $\struc'(\domsymb) = \struc(\domsymb) \setminus
  \{\vertex\}$.  By the inductive hypothesis, we obtain $(\universeGeneral,\struc')
  \models^{\store[x_i \leftarrow \vertex]}_{\twsid{k}} \apred(x_1,
  \ldots, x_{k+1})$.  Hence, $(\universeGeneral,\struc) \models^\store_{\twsid{k}}
  \exists y ~.~ \domsymb(y) * \apred(x_1,\ldots,x_{k+1})[x_i / y]$,
  thus $(\universeGeneral,\struc) \models^\store_{\twsid{k}} \apred(x_1,\ldots,x_{k+1})$,
  by rule (\ref{rule:gen-exists}). \qed}
\end{mySixTextE}

The main property of $\twsid{k}$ is stated below:

\begin{mySixTextE}
The main property of $\twsid{k}$ now follows from Lemmas
\ref{lem:sid-implies-tree-decomposition} and
\ref{lem:tree-decomposition-implies-sid}:
\end{mySixTextE}

\begin{lemmaE}[][category=six]\label{lemma:k-sid}
  For any guarded structure $(\universeGeneral,\struc) \in \strucof{\signature,\domsymb}$, such that $\cardof{\struc(\domsymb)}
  \geq k+1$ and $\struc(\arel)\neq\emptyset$, for at least some
  $\arel\in\signature$, we have $\twof{\struc} \leq k$ iff
  $(\universeGeneral,\struc) \models_{\twsid{k}} \apred_k()$.
\end{lemmaE}
\begin{proofE}
``$\Rightarrow$'' If $(\universeGeneral,\struc)$ has tree decomposition of width at most
$k$, then it also has a reduced tree decomposition
$\tree=(\tnodes,\tedges,r,\alabel)$ of width $k$, by
Lemma~\ref{lemma:reduced-tree-decompositions}.
Let
$\alabel(r)=\set{\vertex_1, \ldots, \vertex_{k+1}}$ and $\store$ be a
store such that $\store(x_i) = \vertex_i$, for all
$i \in \interv{1}{k+1}$.  By
Lemma~\ref{lem:tree-decomposition-implies-sid}, we get that
$(\universeGeneral,\struc) \models^\store_{\twsid{k}} \apred(x_1, \ldots,
x_{k+1}) * \domsymb(x_1) * \cdots * \domsymb(x_{k+1})$.
Thus, $(\universeGeneral,\struc) \models_{\twsid{k}} \apred_k()$, by rule
(\ref{rule:gen-top}).

``$\Leftarrow$''
  By rule
  (\ref{rule:gen-top}),
  there exists a store $\store$ with $\store(x_i) = \vertex_i$, for all $1 \le i \le k+1$, and a structure $(\universeGeneral,\struc') \in \strucof{\signature,\domsymb}$ that agrees with $(\universeGeneral,\struc)$ on $\signature$ such that $(\universeGeneral,\struc') \models^\store_{\twsid{k}} \apred(x_1, \ldots, x_{k+1})$ and $\struc'(\domsymb) = \struc(\domsymb) \setminus \set{\vertex_1,\ldots,\vertex_{k+1}}$.
  By Lemma \ref{lem:sid-implies-tree-decomposition}, there
  exists a reduced tree decomposition $\tree$ of $(\universeGeneral,\struc')$ of width
  $k$.
  Thus $\twof{\struc} \leq k$, because $(\universeGeneral,\struc')$ agrees with $(\universeGeneral,\struc)$ on $\signature$ and we have $\struc'(\domsymb) = \struc(\domsymb) \setminus \set{\vertex_1,\ldots,\vertex_{k+1}}$.
\end{proofE}

We remark that the encoding of $\glue$ and $\fgcst{j}$ used in the
definition of $\twsid{k}$ can be used to show that any inductive set
of structures, i.e., a set defined by finitely many recursive
equations written using $\glue$ and $\fgcst{j}$, can be also defined
in \slr. This means that \slr\ is at least as expressive than the
inductive sets, which are always of bounded treewidth.

The second step of our construction is the annotation of the rules in
$\twsid{k}$ with $\qrof{\phi}$-types, in order to obtain an SID
$\twformsid{k}{\phi}$ (Fig. \ref{fig:sids}b) describing the models of
an \mso\ sentence $\phi$, of treewidth at most $k$. We consider the
set of ports $\ports = \set{\acst_{\ncst+1}, \ldots,
  \acst_{\ncst+k+1}}$ disjoint from $\signature$. The encoding of the
store values of the variables $x_1,\ldots,x_{k+1}$ in a given
structure is defined below:

\begin{definition}\label{def:encode}
Let $\signature = \set{\arel_1, \ldots, \arel_\nrel, \acst_1, \ldots,
  \acst_\ncst}$ be a signature, $\ports = \set{\acst_{\ncst+1},
  \ldots, \acst_{\ncst+k+1}}$ be a set of constants not in
$\signature$, and let $(\universeGeneral,\struc) \in
\strucof{\signature,\domsymb}$ be a structure.  Let $\store$ be a
store mapping $x_1, \ldots, x_{k+1}$ to elements of
$\universeGeneral\setminus\struc(\domsymb)$. 
Then, $\encof{k+1}{(\universeGeneral,\struc)}{\store} \in
\strucof{\signature \cup \ports,\domsymb}$ is a structure with
universe $\universeGeneral$ that agrees with
$(\universeGeneral,\struc)$ over $\signature$, maps each
$\acst_{\ncst+i}$ to $\store(x_i)$, for $i \in \interv{1}{k+1}$ and
maps $\domsymb$ to $\struc(\domsymb) \cup
\set{\store(x_1),\ldots,\store(x_{k+1})}$.
\end{definition}

The correctness of our construction relies on the fact that the
composition acts like glueing, for structures with universe
$\universeGeneral$, whose sets of elements involved in the
interpretation of some relation symbol may only overlap at the
interpretation of the ports from $\ports$:

\begin{lemmaE}[][category=six]\label{lemma:abstract-glue}
  For an integer $r\geq0$, a store $\store$ and locally disjoint
  compatible structures $(\universeGeneral_1,\struc_1), (\universeGeneral_2,\struc_2)
  \in \strucof{\signature \cup \ports,\domsymb}$, such that
  $\Rel{\struc_1} \cap \Rel{\struc_2} \subseteq
  \{\struc_1(\acst_{\ncst+1}),\ldots,\struc_1(\acst_{\ncst+k+1})\}$
  and $(\struc_1(\domsymb)\cup\struc_2(\domsymb)) \cap
  \set{\store(x_i) \mid i \in \interv{1}{m}} = \emptyset$, we have:
  \vspace*{-1.5\baselineskip}

  $$
  \typeof{r}{\encof{k+1}{(\universeGeneral_1,\struc_1) \scomp (\universeGeneral_2,\struc_2)}{\store}} =
  \absglueof{\typeof{r}{\encof{k+1}{(\universeGeneral_1,\struc_1)}{\store}}}
  {\typeof{r}{\encof{k+1}{(\universeGeneral_2,\struc_2)}{\store}}}
  $$
  \vspace*{-2\baselineskip}
\end{lemmaE}
\begin{proofE}
  Let us consider the structures $(\universeGeneral_1,\struc_1') = \encof{k+1}{(\universeGeneral_1,\struc_1)}{\store}$ and
  $(\universeGeneral_2,\struc_2') = \encof{k+1}{(\universeGeneral_2,\struc_2)}{\store}$.
  In order to apply
  glueing, we will now consider two structures isomorphic to
  $(\universeGeneral_1,\struc_1')$ and $(\universeGeneral_1,\struc_2')$, respectively.
  We note that
  $\Dom{\struc_1'} \cap \Dom{\struc_2'} \subseteq \{\struc_2'(\acst_{\ncst_1}),\ldots,\struc_2'(\acst_{\ncst+k+1})\}$
  because of the assumption $\Rel{\struc_1} \cap \Rel{\struc_2} \subseteq \{\store(x_1),\ldots,\store(x_{k+1})\}$).
  We further note that $\struc_1'(\acst_{\ncst_i})
  = \struc_2'(\acst_{\ncst_i})$, for all
  $i \in \interv{1}{\ncst+k+1}$, because $\struc_1$ and $\struc_2$ are
  compatible and the interpretation of the additional constants
  $\acst_{\ncst+1},\ldots,\acst_{\ncst+k+1}$ has been chosen w.r.t. the same store $\store$.
  Because $\universeGeneral_1$ and $\universeGeneral_2$ are infinite (recall that we require all universes to be infinite), we can now choose some partitioning $\universeGeneral_1' \uplus \universeGeneral_2' = \universeGeneral_1 \cup \universeGeneral_2$, such that $\universeGeneral_1'$ and $\universeGeneral_2'$ are infinite,  $\Dom{\struc_1'} \subseteq \universeGeneral_1$ and
  $\Dom{\struc_2'} \setminus \{\struc_2'(\acst_1),\ldots,\struc_2'(\acst_{\ncst+k+1})\} \subseteq \universeGeneral_2$.
  We can now choose a structure $(\universeGeneral_2,\struc''_2)$ that is isomorphic to
  $(\universeGeneral_2,\struc_2')$ and that agrees with $(\universeGeneral_2,\struc_2')$ except for
  $\acst_1,\ldots\acst_{\ncst+k+1}$, whose interpretation is chosen as
  $\struc''_2(\acst_1), \ldots, \struc''_2(\acst_{\ncst+k+1}) \in \universeGeneral_2 \setminus
  (\Dom{\struc_1'} \cup \Dom{\struc_2'})$.
  Then,
  $(\universeGeneral_1',\struc_1')$ [resp. $(\universeGeneral_2',\struc''_2)$] is
  isomorphic to $(\universeGeneral_1,\struc_1')$
  [resp. $(\universeGeneral_2,\struc_2')$]. In particular, they have the same
  type, i.e., $\typeof{r}{\universeGeneral_1',\struc_1'}
  = \typeof{r}{\universeGeneral_1,\struc_1'}$ and
  $\typeof{r}{\universeGeneral_2',\struc''_2}
  = \typeof{r}{\universeGeneral_2,\struc_2'}$.  Moreover, we have that
  $\glueof{(\universeGeneral_1',\struc_1')}{(\universeGeneral_2',\struc''_2)}$ and $\encof{k+1}{(\universeGeneral_1,\struc_1) \scomp (\universeGeneral_2,\struc_2)}{\store}$  are isomorphic.
  In particular, they have the same
  type, i.e.,
  $$\typeof{r}{\encof{k+1}{(\universeGeneral_1,\struc_1) \scomp (\universeGeneral_2,\struc_2)}{\store}}
  = \\ \typeof{r}{\glueof{(\universeGeneral_1',\struc_1')}{(\universeGeneral_2',\struc''_2)}}.$$
  We compute:
  \vspace*{-.5\baselineskip}
  \[\begin{array}{l} \typeof{r}{\encof{k+1}{(\universeGeneral_1,\struc_1) \scomp (\universeGeneral_2,\struc_2)}{\store}}
  = \\ \typeof{r}{\glueof{(\universeGeneral_1',\struc_1')}{(\universeGeneral_2',\struc''_2)}}
  = \\ \absglueof{\typeof{r}{\universeGeneral_1',\struc_1'}}{\typeof{r}{\universeGeneral_2',\struc''_2}}
  = \\ \absglueof{\typeof{r}{\universeGeneral_1,\struc_1'}}{\typeof{r}{\universeGeneral_2,\struc'_2}}
  = \\ \absglueof{\typeof{r}{\encof{k+1}{(\universeGeneral_1,\struc_1)}{\store}}}{\typeof{r}{\encof{k+1}{(\universeGeneral_2,\struc_2)}{\store}}} \end{array}\]
\end{proofE}

\begin{mySixTextE}
We prove that the set
$\sidsem{\apred_{k,\phi}()}{\twformsid{k}{\phi}}$ contains only
structures whose $\qrof{\phi}$-type contains $\phi$, i.e., models of
$\phi$.  Let $(\universeGeneral,\struc) \in
\strucof{\signature,\domsymb}$ be a structure such that
$(\universeGeneral,\struc) \models_{\twformsid{k}{\phi}}
\apred_{k,\phi}$.  By rule (\ref{rule:ref-top}), there exists a store
$\store$ with $\store(x_i) = \vertex_i$, for all $1 \le i \le k+1$,
and a structure $(\universeGeneral,\struc') \in
\strucof{\signature,\domsymb}$ that agrees with
$(\universeGeneral,\struc)$ on $\signature$ such that
$(\universeGeneral,\struc') \models^\store_{\twformsid{k}{\phi}}
\apred(x_1, \ldots, x_{k+1})$ and $\struc'(\domsymb) =
\struc(\domsymb) \setminus \set{\vertex_1,\ldots,\vertex_{k+1}}$.  The
rest of the proof uses induction on the definition of the satisfaction
relation for \slr:

\begin{lemma}\label{lem:refined-SID-is-sound}
Let $k\geq1$ be an integer, $\phi$ be an \mso\ sentence,
$(\universeGeneral,\struc) \in \strucof{\signature,\domsymb}$ be a structure and
$\store$ be a store such that $(\universeGeneral,\struc) \models^\store_{\twformsid{k}{\phi}} \apred^\atype(x_1, \ldots, x_{k+1})$,
$\store(x_i) \not\in \struc(\domsymb)$ for all $i \in
\interv{1}{k+1}$, and $\store(x_i) \neq \store(x_j)$ for all $i \neq j$.
Then,
$\typeof{\qrof{\phi}}{\encof{k+1}{(\universeGeneral,\struc)}{\store}} =
\atype$.
\end{lemma}
\proof{
  We prove the claim by induction on the number of rule applications.
  The claim clearly holds for the base case, by a rule of type
  (\ref{rule:ref-rel}). We now consider a rule of type
  (\ref{rule:ref-comp}), i.e., we have $(\universeGeneral,\struc)
  \models^\store_{\twformsid{k}{\phi}} \apred^{\atype_1}(x_1, \ldots,
  x_{k+1}) * \apred^{\atype_2}(x_1, \ldots, x_{k+1})$, such that
  $\atype = \absglueof{\atype_1}{\atype_2}$. Then, there are
  structures $(\universeGeneral_1,\struc_1)$ and $(\universeGeneral_2,\struc_2)$ with $(\universeGeneral_i,\struc_i)
  \models^\store_{\twformsid{k}{\phi}} \apred^{\atype_i}(x_1, \ldots,
  x_{k+1})$, for $i = 1,2$, and $\struc_1 \comp \struc_2 = \struc$.
  By the
  inductive hypothesis, we have that
  $\typeof{\qrof{\phi}}{\encof{k+1}{(\universeGeneral_i,\struc_i)}{\store}}
  = \atype_i$, for $i=1,2$.  Because every derivation of
  $\twformsid{k}{\phi}$ is also a derivation of $\twsid{k}$,
  obtained by removing the type annotations from the rules in
  $\twformsid{k}{\phi}$, we get that
  $(\universeGeneral_i,\struc_i) \models^\store_{\twsid{k}} \apred(x_1, \ldots,
  x_{k+1})$, for $i=1,2$.  By
  Lemma~\ref{lem:sid-implies-tree-decomposition} we have that
  $\Rel{\struc_i} \subseteq \struc_i(\domsymb) \cup \{\store(x_1), \ldots, \store(x_{k+1})\}$,
  for $i=1,2$. Since $\struc = \struc_1 \comp \struc_2$, we have that
  $\struc_1(\domsymb) \cap \struc_2(\domsymb) = \emptyset$.
  We compute
  \vspace*{-.5\baselineskip}
  \[\begin{array}{l} \typeof{\qrof{\phi}}{\encof{k+1}{(\universeGeneral,\struc)}{\store}}
  = \\ \absglueof{\typeof{\qrof{\phi}}{\encof{k+1}{(\universeGeneral_1,\struc_1)}{\store}}}{\typeof{\qrof{\phi}}{\encof{k+1}{(\universeGeneral_2,\struc_2)}{\store}}}
  = \\ \absglueof{\atype_1}{\atype_2} = \atype \end{array}\]

  \vspace*{-.5\baselineskip}\noindent
  by Lemma~\ref{lemma:abstract-glue}.
  We now consider rules of type
  (\ref{rule:ref-exists}), i.e., we have $(\universeGeneral,\struc)
  \models^\store_{\twformsid{k}{\phi}} \exists y ~.~ \domsymb(y) *
  \apred^{\atype_1}(x_1, \ldots,x_{k+1})[x_i/y]$, for some $i \in
  \interv{1}{k+1}$, such that $\atype =
  \absglueof{\absfgcst{\ncst+i}(\atype_1)}{\rho_i}$ for the type $\rho_i$ of some structure $\astruc \in  \strucof{\set{\acst_{\ncst+i}}}$ with a singleton universe and $\astruc \Models \domsymb(\acst_{\ncst+i})$.
  Then, there is an element $\vertex \in
  \universeGeneral$, such that $(\universeGeneral,\struc')
  \models^{\store[x_i \leftarrow \vertex]}_{\twformsid{k}{\phi}} \apred^{\atype_1}(x_1,
  \ldots,x_{k+1})$, where the structure $(\universeGeneral,\struc')$ agrees with
  $(\universeGeneral,\struc)$, except that $\domsymb$ does not hold for $\vertex$ in
  $(\universeGeneral,\struc')$.
  By the inductive hypothesis,
  $\typeof{\qrof{\phi}}{\encof{k+1}{(\universeGeneral,\struc')}{\store[x_i        \leftarrow \vertex]}} = \atype_1$.
  Because every derivation of $\twformsid{k}{\phi}$ is also a derivation of
  $\twsid{k}$, obtained by removing the type annotations
  from the rules in $\twformsid{k}{\phi}$, we get that
  $(\universeGeneral,\struc')
  \models^{\store[x_i \leftarrow \vertex]}_{\twsid{k}} \apred(x_1, \ldots,
  x_{k+1})$.
  By Lemma~\ref{lem:sid-implies-tree-decomposition} we have that
  $\Rel{\struc'} \subseteq \struc'(\domsymb) \cup \{\store(x_1),
  \ldots, \store(x_{k+1})\} \setminus \{\store(x_i)\} \cup \{u\}$.
  Because of $\store(x_i) \neq u$ (due to the assumption $\store(x_i)
  \not\in \struc(\domsymb)$) and because of $\store(x_i) \neq
  \store(x_j)$ for all $i \neq j$, we get that
  $\encof{k+1}{(\universeGeneral,\struc)}{\store} =
  \glueof{\fgcst{\ncst+i}(\encof{k+1}{(\universeGeneral,\struc')}{\store[x_i
        \leftarrow \vertex]})}{\astruc'}$, where $\astruc' \in
  \strucof{\set{\acst_{\ncst+i}}}$ is the structure with
  singleton universe $\{\store(x_i)\}$ and $\astruc' \Models \domsymb(\acst_{\ncst+i})$.  Because $\astruc$ is
  isomorphic to $\astruc'$, we obtain:
  \[\begin{array}{l} \typeof{\qrof{\phi}}{\encof{k+1}{(\universeGeneral,\struc)}{\store}} = \\
  \typeof{\qrof{\phi}}{ \glueof{\fgcst{\ncst+i}(\encof{k+1}{(\universeGeneral,\struc')}{\store[x_i \leftarrow \vertex]})}{\astruc'}} = \\
  \absglueof{\typeof{\qrof{\phi}}{\fgcst{\ncst+i}(\encof{k+1}{(\universeGeneral,\struc')}{\store[x_i \leftarrow \vertex]})}}{\typeof{\qrof{\phi}}{\astruc'}} = \\
  \absglueof{\absfgcst{\ncst+i}(\typeof{\qrof{\phi}}{\encof{k+1}{(\universeGeneral,\struc')}{\store[x_i \leftarrow \vertex]}})}{\typeof{\qrof{\phi}}{\astruc}} = \\
  \absglueof{\absfgcst{\ncst+i}(\atype_1)}{\rho_i} = \atype \end{array}\] \qed
}

Second, we prove the dual statement, namely that each structure of
treewidth at most $k$ and of $\qrof{\phi}$-type that contains $\phi$
must belong to $\sidsem{\apred_{k,\phi}()}{\twformsid{k}{\phi}}$. To
this end, Lemma \ref{lem:refined-SID-is-complete} below shows that,
for each structure $(\universeGeneral,\struc)$ of treewidth at most $k$ and
$\qrof{\phi}$-type $\atype$, there exists a store $\store$, such that
$\store(x_i) \not\in\struc(\domsymb)$, for all $i \in \interv{1}{k+1}$
and $(\universeGeneral,\struc) \models^\store_{\twformsid{k}{\phi}}
\apred^\atype(x_1,\ldots,x_{k+1})$, and we conclude by an application
of the rule (\ref{rule:ref-top}):

\begin{lemma}\label{lem:refined-SID-is-complete}
Let $k\geq1$ be an integer, $\atype$ be some $r$-type, $(\universeGeneral,\struc) \in
\strucof{\signature,\domsymb}$ be a structure of treewidth
$\twof{\struc} \le k$, witnessed by some reduced tree decomposition
$\tree=(\tnodes,\tedges,r,\alabel)$, and $\store$ be a store with
$\alabel(r)=\{\store(x_1),\ldots,\store(k+1)\}$ and $\store(x_i) \neq
\store(x_j)$ for all $i \neq j$, such that $\struc(\domsymb) =
\bigcup_{n \in \tnodes} \alabel(n) \setminus
\{\vertex_1,\ldots,\vertex_{k+1}\}$ and
$\typeof{r}{\encof{k+1}{(\universeGeneral,\struc)}{\store}} = \atype$.  Then, we have
$(\universeGeneral,\struc) \models^\store_{\twformsid{k}{\phi}} \apred^\atype(x_1,
\ldots, x_{k+1})$.
\end{lemma}
\proof{
  The proof proceeds by induction on the structure of $\tree$.
  Clearly the claim holds for the base case, by a rule of type
  (\ref{rule:ref-rel}). For the inductive step, we assume first that
  the root of $\tree$ has two children. The subtrees $\tree_1$ and
  $\tree_2$ rooted in the two children induce substructures $(\universeGeneral,\struc_1)$
  and $(\universeGeneral,\struc_2)$ of $(\universeGeneral,\struc)$, where $\tuple{v_1, \ldots,
    v_{\arityof{\arel}}} \in \struc_i(\arel)$ if and only if
  $\tuple{v_1, \ldots, v_{\arityof{\arel}}}$ is witnessed by some leaf
  of the respective subtree.  Because $\tree$ is a reduced tree
  decomposition, there is at most one leaf that witnesses a tuple
  $\tuple{v_1, \ldots, v_{\arityof{\arel}}} \in \struc_i(\arel)$.
  Hence, we have $\struc = \struc_1 \comp \struc_2$.  Because the only elements that can
  appear as labels in $\tree_1$ and $\tree_2$ are
  $\store(x_1),\ldots,\store(x_{k+1})$ (as these are the labels of the root of the tree decomposition), we get
  that $\Rel{\struc_1} \cap \Rel{\struc_2} \subseteq
  \set{\store(x_1),\ldots,\store(x_{k+1})}$.  Let $\atype_i =
  \typeof{r}{\encof{k+1}{(\universeGeneral,\struc_i)}{\store}}$, for $i=1,2$.
  From the inductive hypothesis, we obtain
  $(\universeGeneral,\struc_i) \models^\store_{\twformsid{k}{\phi}}
  \apred^{\atype_i}(x_1, \ldots, x_{k+1})$, for $i=1,2$.
  Let $\atype
  \isdef \absglueof{\atype_1}{\atype_2}$.  By
  Lemma~\ref{lemma:abstract-glue}, we obtain:
  \[\begin{array}{l}
  \typeof{\qrof{\phi}}{\encof{k+1}{(\universeGeneral,\struc)}{\store}} = \typeof{\qrof{\phi}}{\encof{k+1}{(\universeGeneral,\struc_1) \scomp (\universeGeneral,\struc_2)}{\store}} = \\
  \absglueof{\typeof{\qrof{\phi}}{\encof{k+1}{(\universeGeneral,\struc_1)}{\store}}}{\typeof{\qrof{\phi}}{\encof{k+1}{(\universeGeneral,\struc_2)}{\store}}} = \\
  \absglueof{\atype_1}{\atype_2}  = \atype
  \end{array}\]
  Hence, we now get that $(\universeGeneral,\struc) \models^\store_{\twformsid{k}{\phi}}
  \apred^\atype(x_1, \ldots, x_{k+1})$, by a rule of type
  (\ref{rule:ref-comp}).
  We now assume that the root of $\tree$ has a
  single child which is not a leaf. We consider the subtree $\tree_1$
  rooted at this child. Then, there is an element $\vertex \not\in
  \set{\store(x_1), \ldots, \store(x_{k+1})}$, such that the root of
  $\tree_1$ is labeled by $\set{\store(x_1), \ldots,\store(x_{i-1}),
    \vertex,\store(x_{i+1}), \ldots, \store(x_{k+1})}$.  Let $\atype_1
  \isdef \typeof{r}{\encof{k+1}{(\universeGeneral,\struc)}{\store[x_i \leftarrow
        \vertex]}}$ and let $(\universeGeneral,\struc')$ be the structure that agrees
  with $(\universeGeneral,\struc)$ except that we have $\struc'(\domsymb) =
  \struc(\domsymb) \setminus \{\vertex\}$.  By the inductive
  hypothesis, we obtain that $(\universeGeneral,\struc') \models^{\store[x_i \leftarrow
      \vertex]}_{\twformsid{k}{\phi}} \apred^{\atype_1}(x_1, \ldots,
  x_{k+1})$.  Let $\atype \isdef
  \absglueof{\absfgcst{\ncst+i}(\atype_1)}{\rho_i}$ for the type
  $\rho_i$ of the structure $\astruc \in
  \strucof{\set{\acst_{\ncst+i}}}$ with singleton universe
  $\{\store(x_i)\}$ and $\astruc \Models \domsymb(\acst_{\ncst+i})$.
  Because of $\store(x_i) \neq u$ and $\store(x_i) \neq \store(x_j)$
  for all $i \neq j$, we have $\encof{k+1}{(\universeGeneral,\struc)}{\store} =
  \glueof{\fgcst{\ncst+i}(\encof{k+1}{(\universeGeneral,\struc')}{\store[x_i \leftarrow \vertex]})}{\astruc}$.  We compute:
  \[\begin{array}{l}
  \typeof{r}{\encof{k+1}{(\universeGeneral,\struc)}{\store}} = \\
  \typeof{r}{ \glueof{\fgcst{\ncst+i}(\encof{k+1}{(\universeGeneral,\struc')}{\store[x_i \leftarrow \vertex]})}{\astruc}} = \\
  \absglueof{\typeof{r}{\fgcst{\ncst+i}(\encof{k+1}{(\universeGeneral,\struc')}{\store[x_i \leftarrow \vertex]})}}{\typeof{r}{\astruc}} = \\
  \absglueof{\absfgcst{\ncst+i}(\typeof{r}{\encof{k+1}{(\universeGeneral,\struc')}{\store[x_i \leftarrow \vertex]}})}{\typeof{r}{\astruc}} = \\
  \absglueof{\absfgcst{\ncst+i}(\atype_1)}{\rho_i} = \atype
  \end{array}\]

  \vspace*{-.5\baselineskip}\noindent
  Hence, we get
  that $(\universeGeneral,\struc) \models^\store_{\twformsid{k}{\phi}}
  \exists y ~.~ \domsymb(y) * \apred_\tau(x_1, \ldots,x_{k+1})[x_i/y]$,
  i.e., that $(\universeGeneral,\struc)
  \models^\store_{\twformsid{k}{\phi}} \apred^\atype(x_1, \ldots, x_{k+1})$
  by a rule of type (\ref{rule:ref-exists}). \qed}
\end{mySixTextE}

Finally, the main property of $\twformsid{k}{\phi}$ is stated and proved below:

\begin{mySixTextE}
The main property of the $\twformsid{k}{\phi}$ SID is proved below, by an application of Lemma
\ref{lem:refined-SID-is-sound} and \ref{lem:refined-SID-is-complete}:
\end{mySixTextE}

\begin{propositionE}[][category=six]\label{prop:kform-sid}
   For any $k\geq1$, \mso\ sentence $\phi$, and guarded structure  $(\universeGeneral,\struc) \in \strucof{\signature,\domsymb}$, the following are equivalent:
  \begin{inparaenum}[(1)]
  \item\label{it2:kform-sid} $(\universeGeneral,\struc) \Models \phi$ and $\twof{\struc}
    \leq k$, and
  \item\label{it1:kform-sid} $(\universeGeneral,\struc) \models_{\twformsid{k}{\phi}} \apred_{k,\phi}()$.
  \end{inparaenum}
\end{propositionE}
\begin{proofE}
  ``(\ref{it2:kform-sid}) $\Rightarrow$ (\ref{it1:kform-sid})'' Since
  $\twof{\universeGeneral,\struc} \leq k$, there exists a reduced tree decomposition
  $\tree=(\tnodes,\tedges,r,\alabel)$ of width $k$, by
  Lemma~\ref{lemma:reduced-tree-decompositions}.  Let $\store$ be a
  store such that $\alabel(r)
  = \set{\store(x_1), \ldots, \store(x_{k+1})}$ and
  $\store(x_i) \neq \store(x_j)$ for all $i \neq
  j \in \interv{1}{k+1}$.
  Let $(\universeGeneral,\struc')$ be the structure that agrees with $(\universeGeneral,\struc)$ on $\signature$, and for which  $\struc'(\domsymb) = \struc(\domsymb) \setminus \set{\store(x_1),\ldots,\store(x_{k+1})}$.
  We observe that $\encof{k+1}{(\universeGeneral,\struc')}{\store} = (\universeGeneral,\struc)$.
  Thus, we obtain
  $(\universeGeneral,\struc) \models^\store_{\twformsid{k}{\phi}} \domsymb(x_1) * \ldots * \domsymb(x_{k+1}) * \apred^\atype(x_1,\ldots,x_{k+1})$,
  by Lemma~\ref{lem:refined-SID-is-complete}, for
  $\atype \isdef \typeof{\qrof{\phi}}{\universeGeneral,\struc} =  \typeof{\qrof{\phi}}{\encof{k+1}{(\universeGeneral,\struc')}{\store}}$.
  Since, moreover, we have assumed that $(\universeGeneral,\struc) \Models \phi$, we have
  $\phi \in \atype \isdef \typeof{\qrof{\phi}}{\universeGeneral,\struc}$.
  Then, we obtain $(\universeGeneral,\struc) \models_{\twformsid{k}{\phi}} \apred_{k,\phi}()$,
  by a rule of type (\ref{rule:ref-top}).

  ``(\ref{it1:kform-sid}) $\Rightarrow$ (\ref{it2:kform-sid})''
  Since $(\universeGeneral,\struc) \models_{\twformsid{k}{\phi}} \apred_{k,\phi}()$, there
  exists a store $\store$ with $\store(x_i) \neq \store(x_j)$ for all $i \neq
  j \in \interv{1}{k+1}$, such that
  $(\universeGeneral,\struc) \models^\store_{\twformsid{k}{\phi}} \domsymb(x_1) * \ldots * \domsymb(x_{k+1}) * \apred^\atype(x_1, \ldots,
  x_{k+1})$, for some type $\atype$ with $\phi \in \atype$, by a rule of type (\ref{rule:ref-top}).
  Let $(\universeGeneral,\struc')$ be the structure that agrees with $(\universeGeneral,\struc)$ on $\signature$, and for which  $\struc'(\domsymb) = \struc(\domsymb) \setminus \set{\store(x_1),\ldots,\store(x_{k+1})}$.
  We observe that $\encof{k+1}{(\universeGeneral,\struc')}{\store} = \struc$.
  By Lemma~\ref{lem:refined-SID-is-sound}, we obtain  $\typeof{\qrof{\phi}}{\encof{k+1}{(\universeGeneral,\struc')}{\store}} = \atype$.
  With $\encof{k+1}{(\universeGeneral,\struc')}{\store} = (\universeGeneral,\struc)$ and $\phi \in \atype$ we then obtain $(\universeGeneral,\struc) \Models \phi$.
  Moreover, we have $\twof{\universeGeneral,\struc} \leq k$, by
  Lemma \ref{lemma:k-sid}, because each derivation of $(\universeGeneral,\struc') \models_{\twformsid{k}{\phi}} \apred_{k,\phi}()$
  corresponds to a derivation of
  $(\universeGeneral,\struc) \models_{\twsid{k}} \apred_{k}()$, obtained by removing the
  type annotations from the rules in $\twformsid{k}{\phi}$.
\end{proofE}

The above result shows that \slr\ can define the guarded models
$(\universeGeneral,\struc) \in \strucof{\signature,\domsymb}$ of a
given \mso\ formula whose treewidth is bounded by a given integer. We
do not know, for the moment, if this result holds on unguarded
structures as well.

The above construction of the SID $\twformsid{k}{\phi}$ is effectively
computable, except for the rule~(\ref{rule:ref-rel}), where one needs
to determine the type of a structure $\astruc =
(\universeGeneral,\struc)$ with infinite universe. However, we prove
in the following that determining this type can be reduced to
computing the type of a finite structure, which amounts to solving
finitely many \mso\ model checking problems on finite structures, each
of which being \pspace-complete \cite{DBLP:conf/stoc/Vardi82}.  Given
an integer $n\geq0$ and a structure $\astruc =
(\universeGeneral,\struc) \in \strucof{\signature}$, we define the
finite structure $\astruc^n = (\Dom{\struc} \cup \set{\vertex_1,
  \ldots, \vertex_n}, \struc)$, for pairwise distinct elements
$\vertex_1, \ldots, \vertex_n \in \universeGeneral \setminus
\Dom{\struc}$.  Then, for any quantifier rank $r$, the structures
$\astruc$ and $\astruc^{2^r}$ have the same $r$-type:

\begin{lemmaE}[][category=six]\label{lemma:single-type}
  Given $r\geq0$ and $\astruc = (\universeGeneral,\struc) \in
  \strucof{\signature}$, we have $\typeof{r}{\astruc} = \typeof{r}{\astruc^{2^r}}$.
\end{lemmaE}
\begin{proofE}
  We consider the $r$-round MSO Ehrenfeucht-Fra\"{i}ss\'{e} game: In
  every round, Spoiler picks a vertex $\vertex$ or a set of vertices
  $\vertexSet$ in $\astruc$ or $\astruc^{2^r}$ and Duplicator answers
  with a vertex or a set of vertices in the other structure.
  Duplicator wins iff after $r$-rounds the substructures induced by
  the selected vertices are isomorphic.  We now sketch a winning
  strategy for Duplicator (as the argument is standard, we leave the
  details to the reader): For vertices in $\Dom{\struc}$ the
  Duplicator plays the same vertices in the respective other structure
  (this applies to single vertices as well as sets of vertices).  For
  the vertices that do not belong to $\Dom{\struc}$ the Duplicator
  selects vertices that do not belong to $\Dom{\struc}$ in the other
  structure such that the selected vertices belong to the same subsets
  played in the previous rounds; care has to be taken for the involved
  cardinalities, if playing in $\astruc^{2^r}$ the Duplicator limits
  himself to at most $2^{r-k-1}$ vertices that belong to any
  combination of previously played subsets or the complement of these
  subsets, where $k$ is the number of previously played rounds; this
  choice is sufficient such that after $r$-rounds the substructures
  induced by the selected vertices are isomorphic.
\end{proofE}

As a final remark, we notice that the idea used to prove
$\gsem{\text{\mso}}{k} \subseteq \sem{\text{\slr}}$ can be extended to
show also $\gsem{\text{\ccmso}}{k} \subseteq \sem{\text{\slr}}$, where
$\ccmso$ denotes the extension of $\mso$ with cardinality constraints
$\cardof{X}_{p,q}$ stating that the cardinality of a set of vertices
$X$ equals $p$ modulo $q$, for some constants $0 \leq p < q$. This is
because glueing and forgetting constants are $\ccmso$-compatible
operations \cite[Lemma 4.5, 4.6 and 4.7]{CourcelleI}.

\section{The Remaining Cases}
\label{sec:remaining}

We discuss the results from Table \ref{tab:intro:expressiveness}, that
are not already covered by \S\ref{sec:mso-slr}, \S\ref{sec:slr-so} and
\S\ref{sec:k-mso-slr}.

\paragraph{$\gsem{\text{\sol}}{k} \not\subseteq \sem{\text{\mso}}$}
Since $\sem{\slr} \subseteq \sem{\sol}$ and$\gsem{\text{\slr}}{k}
\not\subseteq \sem{\text{\mso}}$, we obtain that $\gsem{\sol}{k}
\not\subseteq \sem{\text{\mso}}$. Moreover, $\sem{\text{\sol}}
\not\subseteq \sem{\text{\mso}}$ follows from the fact that our
counterexample for $\gsem{\text{\slr}}{k} \not\subseteq
\sem{\text{\mso}}$ involves only structures of treewidth one.

\paragraph{$\gsem{\text{\slr}}{k} \subseteq \sem{\text{\sol}}$}
By applying the translation of \slr\ to \sol\ from \S\ref{sec:slr-so}
to $\twsid{k}$ (Fig. \ref{fig:sids}a) and to a given SID $\asid$
defining a predicate $\apred$ of zero arity, respectively, and taking
the conjunction of the results with the \sol\ formula defining guarded
structures\footnote{$\bigwedge_{\arel\in\signature} \forall x_1 \ldots
  \forall x_{\arityof{\arel}} ~.~ \arel(x_1, \ldots,
  x_{\arityof{\arel}}) \rightarrow \bigwedge_{i \in
    \interv{1}{\arityof{\arel}}} \domsymb(x_i)$.}, we obtain an
\sol\ formula that defines the set $\gsidsem{\apred()}{\asid}{k}$,
thus proving that $\gsem{\text{\slr}}{k} \subseteq \sem{\text{\sol}}$.

\paragraph{$\gsem{\text{\solmso}}{k} \subseteq \sem{\solmso}$}
For each given $k \geq 1$, there exists an \mso\ formula $\theta_k$
that defines the structures of treewidth at most $k$ \cite[Proposition
5.11]{courcelle_engelfriet_2012}. This is a consequence of the Graph
Minor Theorem proved by Robertson and Seymour
\cite{DowneyFellowsBook}, combined with the fact that bounded
treewidth graphs are closed under taking minors and that the property
of having a given finite minor is \mso-definable\footnote{The proof of
Robertson and Seymour does not build $\theta_k$, see
\cite{10.5555/1347082.1347153} for an effective proof.}. Then, for any given
\solmso\ formula $\phi$, the \solmso\ formula $\phi \wedge \theta_k$
defines the models of $\phi$ of treewidth at most $k$.

\paragraph{Open Problems}
The following problems from Table \ref{tab:intro:expressiveness} are
currently open: $\gsem{\text{\slr}}{k} \subseteq \sem{\text{\slr}}$
and $\gsem{\text{\sol}}{k} \subseteq \sem{\text{\slr}}$, both
conjectured to have a negative answer.  In particular, the difficulty
concerning $\gsem{\text{\slr}}{k} \subseteq \sem{\text{\slr}}$ is
that, in order to ensure treewidth boundedness, it seems necessary to
force the composition of structures to behave like glueing (see the
definition of $\twsid{k}$ in Fig. \ref{fig:sids}a), which seems
difficult without the additional relation symbol $\domsymb$.

Since $\gsem{\text{\mso}}{k} \subseteq \sem{\text{\slr}}$ but
$\sem{\text{\mso}} \not\subseteq \sem{\text{\slr}}$, we naturally ask
for the existence of a fragment of \slr\ that describes only
\mso-definable families of structures of bounded treewidth. In
particular, \cite[\S6]{DBLP:journals/corr/abs-2202-09637} defines a
fragment of \slr\ that has bounded-treewidth models and is
\mso-definable. However, in general, since \slr\ can define
context-free sets of guarded graphs (the grammar in Figure \ref{fig:sids}a can be adapted to encode Hyperedge Replacement (HR) grammars~\cite{journals/tcs/Courcelle94}), the \mso-definability of a \slr-definable set is
undecidable, as a consequence of the undecidability of the
recognizability of context-free languages \cite{Greibach68}. 
On the other hand, the treewidth-boundedness of a \slr-definable set is an
open problem, that we conjecture decidable.

A possible direction for future work is also adding Boolean connectives to \slr.
Here, one might study an \slr\ variant that supports Boolean connectives in a top-level logic but not within the inductive definitions, similar to the \seplog\ studied in~\cite{DBLP:conf/lpar/KatelaanZ20,tocl-guarded-separation-logic}.
Adding Boolean connectives within the inductive definitions appears more difficult, as one will need to impose syntactic restitutions such as positive occurrences of predicate atoms in the right hand side of definitions or stratification of negation in order to ensure well-definedness. 
\section{Conclusions}

We have compared the expressiveness of \slr, \mso\ and \sol, in
general and for models of bounded treewidth. Interestingly, we found
that \slr\ and \mso\ are, in general, incomparable and subsumed by
\sol, whereas the models of bounded treewidth of \mso\ can be defined
by \slr, modulo augmenting the signature with a unary relation symbol
used to store the elements that occur in the original
structure.

\bibliography{refs}

\begin{thebibliography}{10}

\bibitem{AbitebouldBunemanSuciu00}
Serge Abiteboul, Peter Buneman, and Dan Suciu.
\newblock {\em Data on the Web: From Relations to Semistructured Data and XML}.
\newblock Morgan Kaufmann, 2000.

\bibitem{ACZEL1977739}
Peter Aczel.
\newblock An introduction to inductive definitions.
\newblock In {\em Handbook of Mathematical Logic}, volume~90 of {\em Studies in
  Logic and the Foundations of Mathematics}, pages 739--782. Elsevier, 1977.
\newblock URL:
  \url{https://www.sciencedirect.com/science/article/pii/S0049237X08711200},
  \href {https://doi.org/10.1016/S0049-237X(08)71120-0}
  {\path{doi:10.1016/S0049-237X(08)71120-0}}.

\bibitem{10.5555/1347082.1347153}
Isolde Adler, Martin Grohe, and Stephan Kreutzer.
\newblock Computing excluded minors.
\newblock In {\em Proceedings of the Nineteenth Annual ACM-SIAM Symposium on
  Discrete Algorithms}, SODA '08, page 641–650, USA, 2008. Society for
  Industrial and Applied Mathematics.

\bibitem{AhrensBozgaIosifKatoen21}
Emma Ahrens, Marius Bozga, Radu Iosif, and Joost{-}Pieter Katoen.
\newblock Reasoning about distributed reconfigurable systems.
\newblock {\em Proc. {ACM} Program. Lang.}, 6({OOPSLA2}):145--174, 2022.
\newblock \href {https://doi.org/10.1145/3563293} {\path{doi:10.1145/3563293}}.

\bibitem{10.1007/978-3-642-00596-1_6}
Timos Antonopoulos and Anuj Dawar.
\newblock Separating graph logic from mso.
\newblock In Luca de~Alfaro, editor, {\em Foundations of Software Science and
  Computational Structures}, pages 63--77, Berlin, Heidelberg, 2009. Springer
  Berlin Heidelberg.

\bibitem{10.1145/2933575.2934508}
Miko\l{}aj Boja\'{n}czyk and Micha\l{} Pilipczuk.
\newblock Definability equals recognizability for graphs of bounded treewidth.
\newblock In {\em Proceedings of the 31st Annual ACM/IEEE Symposium on Logic in
  Computer Science}, LICS '16, page 407–416, New York, NY, USA, 2016.
  Association for Computing Machinery.
\newblock \href {https://doi.org/10.1145/2933575.2934508}
  {\path{doi:10.1145/2933575.2934508}}.

\bibitem{DBLP:conf/cade/BozgaBI22}
Marius Bozga, Lucas Bueri, and Radu Iosif.
\newblock Decision problems in a logic for reasoning about reconfigurable
  distributed systems.
\newblock In Jasmin Blanchette, Laura Kov{\'{a}}cs, and Dirk Pattinson,
  editors, {\em Automated Reasoning - 11th International Joint Conference,
  {IJCAR} 2022, Haifa, Israel, August 8-10, 2022, Proceedings}, volume 13385 of
  {\em Lecture Notes in Computer Science}, pages 691--711. Springer, 2022.
\newblock \href {https://doi.org/10.1007/978-3-031-10769-6\_40}
  {\path{doi:10.1007/978-3-031-10769-6\_40}}.

\bibitem{DBLP:journals/corr/abs-2202-09637}
Marius Bozga, Lucas Bueri, and Radu Iosif.
\newblock Decision problems in a logic for reasoning about reconfigurable
  distributed systems.
\newblock {\em CoRR}, abs/2202.09637, 2022.
\newblock URL: \url{https://arxiv.org/abs/2202.09637}, \href
  {https://arxiv.org/abs/2202.09637} {\path{arXiv:2202.09637}}.

\bibitem{DBLP:conf/concur/BozgaBI22}
Marius Bozga, Lucas Bueri, and Radu Iosif.
\newblock On an invariance problem for parameterized concurrent systems.
\newblock In Bartek Klin, Slawomir Lasota, and Anca Muscholl, editors, {\em
  33rd International Conference on Concurrency Theory, {CONCUR} 2022, September
  12-16, 2022, Warsaw, Poland}, volume 243 of {\em LIPIcs}, pages 24:1--24:16.
  Schloss Dagstuhl - Leibniz-Zentrum f{\"{u}}r Informatik, 2022.
\newblock \href {https://doi.org/10.4230/LIPIcs.CONCUR.2022.24}
  {\path{doi:10.4230/LIPIcs.CONCUR.2022.24}}.

\bibitem{DBLP:journals/tcs/BozgaIS23}
Marius Bozga, Radu Iosif, and Joseph Sifakis.
\newblock Verification of component-based systems with recursive architectures.
\newblock {\em Theor. Comput. Sci.}, 940(Part):146--175, 2023.
\newblock \href {https://doi.org/10.1016/j.tcs.2022.10.022}
  {\path{doi:10.1016/j.tcs.2022.10.022}}.

\bibitem{DBLP:journals/iandc/BrocheninDL12}
R{\'{e}}mi Brochenin, St{\'{e}}phane Demri, and {\'{E}}tienne Lozes.
\newblock On the almighty wand.
\newblock {\em Inf. Comput.}, 211:106--137, 2012.

\bibitem{DBLP:conf/csl/BrotherstonFPG14}
James Brotherston, Carsten Fuhs, Juan Antonio~Navarro P{\'{e}}rez, and Nikos
  Gorogiannis.
\newblock A decision procedure for satisfiability in separation logic with
  inductive predicates.
\newblock In Thomas~A. Henzinger and Dale Miller, editors, {\em Joint Meeting
  of the Twenty-Third {EACSL} Annual Conference on Computer Science Logic
  {(CSL)} and the Twenty-Ninth Annual {ACM/IEEE} Symposium on Logic in Computer
  Science (LICS), {CSL-LICS} '14, Vienna, Austria, July 14 - 18, 2014}, pages
  25:1--25:10. {ACM}, 2014.
\newblock \href {https://doi.org/10.1145/2603088.2603091}
  {\path{doi:10.1145/2603088.2603091}}.

\bibitem{DBLP:conf/popl/BrotherstonGKR16}
James Brotherston, Nikos Gorogiannis, Max~I. Kanovich, and Reuben Rowe.
\newblock Model checking for symbolic-heap separation logic with inductive
  predicates.
\newblock In Rastislav Bod{\'{\i}}k and Rupak Majumdar, editors, {\em
  Proceedings of the 43rd Annual {ACM} {SIGPLAN-SIGACT} Symposium on Principles
  of Programming Languages, {POPL} 2016, St. Petersburg, FL, USA, January 20 -
  22, 2016}, pages 84--96. {ACM}, 2016.
\newblock \href {https://doi.org/10.1145/2837614.2837621}
  {\path{doi:10.1145/2837614.2837621}}.

\bibitem{CalcagnoDistefanoOHearnYang11}
Cristiano Calcagno, Dino Distefano, Peter~W. O’Hearn, and Hongseok Yang.
\newblock Compositional shape analysis by means of bi-abduction.
\newblock {\em J. ACM}, 58(6), December 2011.
\newblock \href {https://doi.org/10.1145/2049697.2049700}
  {\path{doi:10.1145/2049697.2049700}}.

\bibitem{CalcagnoGardnerHague05}
Cristiano Calcagno, Philippa Gardner, and Matthew Hague.
\newblock From separation logic to first-order logic.
\newblock In {\em Foundations of Software Science and Computational
  Structures}, pages 395--409, Berlin, Heidelberg, 2005. Springer Berlin
  Heidelberg.

\bibitem{CalcagnoOHearnYan07}
Cristiano Calcagno, Peter~W. O'Hearn, and Hongseok Yang.
\newblock Local action and abstract separation logic.
\newblock In {\em 22nd {IEEE} Symposium on Logic in Computer Science {(LICS}
  2007), 10-12 July 2007, Wroclaw, Poland, Proceedings}, pages 366--378. {IEEE}
  Computer Society, 2007.
\newblock \href {https://doi.org/10.1109/LICS.2007.30}
  {\path{doi:10.1109/LICS.2007.30}}.

\bibitem{DBLP:conf/fsttcs/CalcagnoYO01}
Cristiano Calcagno, Hongseok Yang, and Peter~W. O'Hearn.
\newblock Computability and complexity results for a spatial assertion language
  for data structures.
\newblock In Ramesh Hariharan, Madhavan Mukund, and V.~Vinay, editors, {\em
  {FST} {TCS} 2001: Foundations of Software Technology and Theoretical Computer
  Science, 21st Conference, Bangalore, India, December 13-15, 2001,
  Proceedings}, volume 2245 of {\em Lecture Notes in Computer Science}, pages
  108--119. Springer, 2001.

\bibitem{Cardelli2002Spatial}
Luca Cardelli, Philippa Gardner, and Giorgio Ghelli.
\newblock {A Spatial Logic for Querying Graphs}.
\newblock In Peter Widmayer, Francisco~Triguero Ruiz, Rafael~Morales Bueno,
  Matthew Hennessy, Stephan Eidenbenz, and Ricardo Conejo, editors, {\em
  Proceedings of the 29\textsuperscript{th} International Colloquium on
  Automata, Languages and Programming ({ICALP'02})}, volume 2380 of {\em
  Lecture Notes in Computer Science}, pages 597--610. Springer, July 2002.
\newblock \href {https://doi.org/10.1007/3-540-45465-9_51}
  {\path{doi:10.1007/3-540-45465-9_51}}.

\bibitem{10.1145/325694.325742}
Luca Cardelli and Andrew~D. Gordon.
\newblock Anytime, anywhere: Modal logics for mobile ambients.
\newblock In {\em Proceedings of the 27th ACM SIGPLAN-SIGACT Symposium on
  Principles of Programming Languages}, POPL '00, page 365–377, New York, NY,
  USA, 2000. Association for Computing Machinery.
\newblock \href {https://doi.org/10.1145/325694.325742}
  {\path{doi:10.1145/325694.325742}}.

\bibitem{DBLP:journals/logcom/CollinsonMP14}
Matthew Collinson, Kevin McDonald, and David~J. Pym.
\newblock A substructural logic for layered graphs.
\newblock {\em J. Log. Comput.}, 24(4):953--988, 2014.
\newblock \href {https://doi.org/10.1093/logcom/exu002}
  {\path{doi:10.1093/logcom/exu002}}.

\bibitem{DBLP:conf/concur/CookHOPW11}
Byron Cook, Christoph Haase, Jo{\"{e}}l Ouaknine, Matthew~J. Parkinson, and
  James Worrell.
\newblock Tractable reasoning in a fragment of separation logic.
\newblock In Joost{-}Pieter Katoen and Barbara K{\"{o}}nig, editors, {\em
  {CONCUR} 2011 - Concurrency Theory - 22nd International Conference, {CONCUR}
  2011, Aachen, Germany, September 6-9, 2011. Proceedings}, volume 6901 of {\em
  Lecture Notes in Computer Science}, pages 235--249. Springer, 2011.
\newblock \href {https://doi.org/10.1007/978-3-642-23217-6\_16}
  {\path{doi:10.1007/978-3-642-23217-6\_16}}.

\bibitem{CourcelleI}
Bruno Courcelle.
\newblock The monadic second-order logic of graphs. i. recognizable sets of
  finite graphs.
\newblock {\em Information and Computation}, 85(1):12--75, 1990.
\newblock URL:
  \url{https://www.sciencedirect.com/science/article/pii/089054019090043H},
  \href {https://doi.org/10.1016/0890-5401(90)90043-H}
  {\path{doi:10.1016/0890-5401(90)90043-H}}.

\bibitem{CourcelleVII}
Bruno Courcelle.
\newblock The monadic second-order logic of graphs {VII:} graphs as relational
  structures.
\newblock {\em Theor. Comput. Sci.}, 101(1):3--33, 1992.
\newblock \href {https://doi.org/10.1016/0304-3975(92)90148-9}
  {\path{doi:10.1016/0304-3975(92)90148-9}}.

\bibitem{journals/tcs/Courcelle94}
Bruno Courcelle.
\newblock Monadic second-order definable graph transductions: {A} survey.
\newblock {\em Theor. Comput. Sci.}, 126(1):53--75, 1994.
\newblock \href {https://doi.org/10.1016/0304-3975(94)90268-2}
  {\path{doi:10.1016/0304-3975(94)90268-2}}.

\bibitem{courcelle_engelfriet_2012}
Bruno Courcelle and Joost Engelfriet.
\newblock {\em Graph Structure and Monadic Second-Order Logic: A
  Language-Theoretic Approach}.
\newblock Encyclopedia of Mathematics and its Applications. Cambridge
  University Press, 2012.
\newblock \href {https://doi.org/10.1017/CBO9780511977619}
  {\path{doi:10.1017/CBO9780511977619}}.

\bibitem{Dawar2007Expressiveness}
Anuj Dawar, Philippa Gardner, and Giorgio Ghelli.
\newblock {Expressiveness and Complexity of Graph Logic}.
\newblock {\em Information and Computation}, 205(3):263--310, February 2007.
\newblock \href {https://doi.org/10.1016/j.ic.2006.10.006}
  {\path{doi:10.1016/j.ic.2006.10.006}}.

\bibitem{DBLP:conf/birthday/2008montanari}
Pierpaolo Degano, Rocco~De Nicola, and Jos{\'{e}} Meseguer, editors.
\newblock {\em Concurrency, Graphs and Models, Essays Dedicated to Ugo
  Montanari on the Occasion of His 65th Birthday}, volume 5065 of {\em Lecture
  Notes in Computer Science}. Springer, 2008.
\newblock \href {https://doi.org/10.1007/978-3-540-68679-8}
  {\path{doi:10.1007/978-3-540-68679-8}}.

\bibitem{DBLP:journals/tocl/DemriD16}
St{\'{e}}phane Demri and Morgan Deters.
\newblock Expressive completeness of separation logic with two variables and no
  separating conjunction.
\newblock {\em {ACM} Trans. Comput. Log.}, 17(2):12, 2016.

\bibitem{DBLP:journals/tocl/DemriLM21}
St{\'{e}}phane Demri, {\'{E}}tienne Lozes, and Alessio Mansutti.
\newblock The effects of adding reachability predicates in quantifier-free
  separation logic.
\newblock {\em {ACM} Trans. Comput. Log.}, 22(2):14:1--14:56, 2021.
\newblock \href {https://doi.org/10.1145/3448269} {\path{doi:10.1145/3448269}}.

\bibitem{DBLP:books/daglib/0030488}
Reinhard Diestel.
\newblock {\em Graph Theory, 4th Edition}, volume 173 of {\em Graduate texts in
  mathematics}.
\newblock Springer, 2012.

\bibitem{10.1007/978-3-642-38856-9_10}
Cezara Dr{\u{a}}goi, Constantin Enea, and Mihaela Sighireanu.
\newblock Local shape analysis for overlaid data structures.
\newblock In Francesco Logozzo and Manuel F{\"a}hndrich, editors, {\em Static
  Analysis}, pages 150--171, Berlin, Heidelberg, 2013. Springer Berlin
  Heidelberg.

\bibitem{DBLP:books/daglib/0082516}
Heinz{-}Dieter Ebbinghaus and J{\"{o}}rg Flum.
\newblock {\em Finite model theory}.
\newblock Perspectives in Mathematical Logic. Springer, 1995.

\bibitem{DBLP:journals/tocl/EchenimIP20}
Mnacho Echenim, Radu Iosif, and Nicolas Peltier.
\newblock The bernays-sch{\"{o}}nfinkel-ramsey class of separation logic with
  uninterpreted predicates.
\newblock {\em {ACM} Trans. Comput. Log.}, 21(3):19:1--19:46, 2020.

\bibitem{EchenimIosifPeltier21b}
Mnacho Echenim, Radu Iosif, and Nicolas Peltier.
\newblock {Decidable Entailments in Separation Logic with Inductive
  Definitions: Beyond Establishment}.
\newblock In Christel Baier and Jean Goubault-Larrecq, editors, {\em 29th EACSL
  Annual Conference on Computer Science Logic (CSL 2021)}, volume 183 of {\em
  Leibniz International Proceedings in Informatics (LIPIcs)}, pages
  20:1--20:18, Dagstuhl, Germany, 2021. Schloss Dagstuhl--Leibniz-Zentrum
  f{\"u}r Informatik.
\newblock URL: \url{https://drops.dagstuhl.de/opus/volltexte/2021/13454}, \href
  {https://doi.org/10.4230/LIPIcs.CSL.2021.20}
  {\path{doi:10.4230/LIPIcs.CSL.2021.20}}.

\bibitem{EchenimIosifPeltier21}
Mnacho Echenim, Radu Iosif, and Nicolas Peltier.
\newblock Unifying decidable entailments in separation logic with inductive
  definitions.
\newblock In Andr{\'{e}} Platzer and Geoff Sutcliffe, editors, {\em Automated
  Deduction - {CADE} 28 - 28th International Conference on Automated Deduction,
  Virtual Event, July 12-15, 2021, Proceedings}, volume 12699 of {\em Lecture
  Notes in Computer Science}, pages 183--199. Springer, 2021.
\newblock \href {https://doi.org/10.1007/978-3-030-79876-5\_11}
  {\path{doi:10.1007/978-3-030-79876-5\_11}}.

\bibitem{DBLP:journals/ipl/EchenimIP22}
Mnacho Echenim, Radu Iosif, and Nicolas Peltier.
\newblock Entailment is undecidable for symbolic heap separation logic
  formul{\ae} with non-established inductive rules.
\newblock {\em Inf. Process. Lett.}, 173:106169, 2022.
\newblock \href {https://doi.org/10.1016/j.ipl.2021.106169}
  {\path{doi:10.1016/j.ipl.2021.106169}}.

\bibitem{DBLP:conf/lics/FigueiraL15}
Diego Figueira and Leonid Libkin.
\newblock Path logics for querying graphs: Combining expressiveness and
  efficiency.
\newblock In {\em 30th Annual {ACM/IEEE} Symposium on Logic in Computer
  Science, {LICS} 2015, Kyoto, Japan, July 6-10, 2015}, pages 329--340. {IEEE}
  Computer Society, 2015.
\newblock \href {https://doi.org/10.1109/LICS.2015.39}
  {\path{doi:10.1109/LICS.2015.39}}.

\bibitem{DBLP:series/txtcs/FlumG06}
J{\"{o}}rg Flum and Martin Grohe.
\newblock {\em Parameterized Complexity Theory}.
\newblock Texts in Theoretical Computer Science. An {EATCS} Series. Springer,
  2006.
\newblock \href {https://doi.org/10.1007/3-540-29953-X}
  {\path{doi:10.1007/3-540-29953-X}}.

\bibitem{Greibach68}
S.~Greibach.
\newblock A note on undecidable properties of formal languages.
\newblock {\em Math. Systems Theory}, 2:1--6, 1968.
\newblock \href {https://doi.org/10.1007/BF01691341}
  {\path{doi:10.1007/BF01691341}}.

\bibitem{Immerman1999}
Neil Immerman.
\newblock {\em Second-Order Logic and Fagin's Theorem}, pages 113--124.
\newblock Springer New York, New York, NY, 1999.
\newblock \href {https://doi.org/10.1007/978-1-4612-0539-5_8}
  {\path{doi:10.1007/978-1-4612-0539-5_8}}.

\bibitem{DBLP:conf/cade/IosifRS13}
Radu Iosif, Adam Rogalewicz, and Jir{\'{\i}} Sim{\'{a}}cek.
\newblock The tree width of separation logic with recursive definitions.
\newblock In Maria~Paola Bonacina, editor, {\em Automated Deduction - {CADE-24}
  - 24th International Conference on Automated Deduction, Lake Placid, NY, USA,
  June 9-14, 2013. Proceedings}, volume 7898 of {\em Lecture Notes in Computer
  Science}, pages 21--38. Springer, 2013.
\newblock \href {https://doi.org/10.1007/978-3-642-38574-2\_2}
  {\path{doi:10.1007/978-3-642-38574-2\_2}}.

\bibitem{Ishtiaq00bias}
Samin~S. Ishtiaq and Peter~W. O'Hearn.
\newblock {BI} as an assertion language for mutable data structures.
\newblock In Chris Hankin and Dave Schmidt, editors, {\em Conference Record of
  {POPL} 2001: The 28th {ACM} {SIGPLAN-SIGACT} Symposium on Principles of
  Programming Languages, London, UK, January 17-19, 2001}, pages 14--26. {ACM},
  2001.

\bibitem{JansenKatelaanMathejaNollZuleger17}
Christina Jansen, Jens Katelaan, Christoph Matheja, Thomas Noll, and Florian
  Zuleger.
\newblock Unified reasoning about robustness properties of symbolic-heap
  separation logic.
\newblock In {\em {European Symposium on Programming (ESOP)}}, volume 10201 of
  {\em Lecture Notes in Computer Science}, pages 611--638. Springer, 2017.

\bibitem{10.1145/582153.582161}
Neil~D. Jones and Steven~S. Muchnick.
\newblock A flexible approach to interprocedural data flow analysis and
  programs with recursive data structures.
\newblock In {\em Proceedings of the 9th ACM SIGPLAN-SIGACT Symposium on
  Principles of Programming Languages}, POPL '82, page 66–74, New York, NY,
  USA, 1982. Association for Computing Machinery.
\newblock \href {https://doi.org/10.1145/582153.582161}
  {\path{doi:10.1145/582153.582161}}.

\bibitem{DBLP:conf/cade/KatelaanJW18}
Jens Katelaan, Dejan Jovanovic, and Georg Weissenbacher.
\newblock A separation logic with data: Small models and automation.
\newblock In Didier Galmiche, Stephan Schulz, and Roberto Sebastiani, editors,
  {\em Automated Reasoning - 9th International Joint Conference, {IJCAR} 2018,
  Held as Part of the Federated Logic Conference, FloC 2018, Oxford, UK, July
  14-17, 2018, Proceedings}, volume 10900 of {\em Lecture Notes in Computer
  Science}, pages 455--471. Springer, 2018.
\newblock \href {https://doi.org/10.1007/978-3-319-94205-6\_30}
  {\path{doi:10.1007/978-3-319-94205-6\_30}}.

\bibitem{DBLP:conf/lpar/KatelaanZ20}
Jens Katelaan and Florian Zuleger.
\newblock Beyond symbolic heaps: Deciding separation logic with inductive
  definitions.
\newblock In Elvira Albert and Laura Kov{\'{a}}cs, editors, {\em {LPAR} 2020:
  23rd International Conference on Logic for Programming, Artificial
  Intelligence and Reasoning, Alicante, Spain, May 22-27, 2020}, volume~73 of
  {\em EPiC Series in Computing}, pages 390--408. EasyChair, 2020.
\newblock \href {https://doi.org/10.29007/vkmj} {\path{doi:10.29007/vkmj}}.

\bibitem{10.1007/978-3-540-27864-1_26}
Viktor Kuncak and Martin Rinard.
\newblock Generalized records and spatial conjunction in role logic.
\newblock In {\em Static Analysis}, pages 361--376, Berlin, Heidelberg, 2004.
  Springer Berlin Heidelberg.

\bibitem{DBLP:journals/corr/cs-LO-0410073}
Viktor Kuncak and Martin~C. Rinard.
\newblock On spatial conjunction as second-order logic.
\newblock {\em CoRR}, cs.LO/0410073, 2004.
\newblock URL: \url{http://arxiv.org/abs/cs.LO/0410073}.

\bibitem{PhD-lozes}
{\'E}tienne Lozes.
\newblock {\em Expressivit{\'e} des logiques spatiales}.
\newblock Th{\`e}se de doctorat, Laboratoire de l'Informatique du
  Parall{\'e}lisme, ENS Lyon, France, November 2004.
\newblock URL:
  \url{http://www.lsv.ens-cachan.fr/Publis/PAPERS/PS/PhD-lozes.ps}.

\bibitem{journals/apal/Makowsky04}
Johann~A. Makowsky.
\newblock Algorithmic uses of the feferman-vaught theorem.
\newblock {\em Ann. Pure Appl. Log.}, 126(1-3):159--213, 2004.

\bibitem{DBLP:phd/hal/Mansutti20}
Alessio Mansutti.
\newblock {\em Logiques de s{\'{e}}paration : complexit{\'{e}},
  expressivit{\'{e}}, calculs. (Reasoning with separation logics : complexity,
  expressive power, proof systems)}.
\newblock PhD thesis, University of Paris-Saclay, France, 2020.
\newblock URL: \url{https://tel.archives-ouvertes.fr/tel-03094373}.

\bibitem{tocl-guarded-separation-logic}
Christoph Matheja, Jens Pagel, and Florian Zuleger.
\newblock A decision procedure for guarded separation logic complete entailment
  checking for separation logic with inductive definitions.
\newblock {\em {ACM} Trans. Comput. Log.}, 24(1):1:1--1:76, 2023.
\newblock \href {https://doi.org/10.1145/3534927} {\path{doi:10.1145/3534927}}.

\bibitem{PymOHearn99}
Peter~W. O'Hearn and David~J. Pym.
\newblock The logic of bunched implications.
\newblock {\em Bull. Symb. Log.}, 5(2):215--244, 1999.

\bibitem{DowneyFellowsBook}
M.~R.~Fellows R.~G.~Downey.
\newblock {\em Parameterized Complexity}.
\newblock Springer New York, NY, 1999.
\newblock \href {https://doi.org/10.1007/978-1-4612-0515-9}
  {\path{doi:10.1007/978-1-4612-0515-9}}.

\bibitem{Reynolds02}
John~C. Reynolds.
\newblock Separation logic: {A} logic for shared mutable data structures.
\newblock In {\em 17th {IEEE} Symposium on Logic in Computer Science {(LICS}
  2002), 22-25 July 2002, Copenhagen, Denmark, Proceedings}, pages 55--74.
  {IEEE} Computer Society, 2002.
\newblock \href {https://doi.org/10.1109/LICS.2002.1029817}
  {\path{doi:10.1109/LICS.2002.1029817}}.

\bibitem{Seese91}
D.~Seese.
\newblock The structure of the models of decidable monadic theories of graphs.
\newblock {\em Annals of Pure and Applied Logic}, 53(2):169--195, 1991.
\newblock URL:
  \url{https://www.sciencedirect.com/science/article/pii/016800729190054P},
  \href {https://doi.org/10.1016/0168-0072(91)90054-P}
  {\path{doi:10.1016/0168-0072(91)90054-P}}.

\bibitem{DBLP:books/daglib/0080654}
Dirk van Dalen.
\newblock {\em Logic and structure {(3.} ed.)}.
\newblock Universitext. Springer, 1994.

\bibitem{DBLP:conf/stoc/Vardi82}
Moshe~Y. Vardi.
\newblock The complexity of relational query languages (extended abstract).
\newblock In {\em Proceedings of the 14th Annual {ACM} Symposium on Theory of
  Computing, May 5-7, 1982, San Francisco, California, {USA}}, pages 137--146.
  {ACM}, 1982.

\end{thebibliography}

\appendix

\section{Technical Material from Section \ref{sec:mso-slr}}
\label{app:slr-mso}
\printProofs[four]

\section{Technical Material from Section \ref{sec:slr-so}}
\label{app:slr-so}
\printProofs[five]

\section{Technical Material from Section \ref{sec:k-mso-slr}}
\label{app:k-mso-slr}
\printProofs[six]

\end{document}